\documentclass{article}
\usepackage{jheppub}
\pdfoutput=1

\usepackage{amsmath}
\usepackage{amssymb}
\usepackage{latexsym}
\usepackage{amsfonts}
\usepackage{graphicx}
\usepackage{epstopdf}
\usepackage{mdwlist}
\usepackage{enumitem}
\usepackage{stackengine}
\usepackage{verbatim}

\usepackage{array,multirow}

\usepackage{color}%
\def\hhref#1{\href{http://arxiv.org/abs/#1}{arXiv:#1}} 

\def\XXint#1#2#3{{\setbox0=\hbox{$#1{#2#3}{\int}$}
     \vcenter{\hbox{$#2#3$}}\kern-.5\wd0}}

\newcommand{\be}{\begin{equation}}
\newcommand{\ee}{\end{equation}} 
\def\bea{\begin{align}}
\def\ena{\end{align}}
\def\beqa{\begin{eqnarray}}
\def\enqa{\end{eqnarray}}

\begin{document}

\title{Transmutation of a Trans-series: The Gross-Witten-Wadia Phase Transition}

\author{Anees Ahmed and}
\author{Gerald~V.~Dunne}
\affiliation{Department of Physics, University of Connecticut, Storrs CT 06269, USA}
\emailAdd{anees.ahmed@uconn.edu}
\emailAdd{gerald.dunne@uconn.edu}
\date{\today}

\abstract{We study the change in the resurgent asymptotic properties of a trans-series in two parameters, a coupling $g^2$ and a gauge index $N$, as a system passes through a large $N$ phase transition, using the universal example of the Gross-Witten-Wadia third-order phase transition in the unitary matrix model. This transition is well-studied in the immediate vicinity of the transition point, where it is characterized by a double-scaling limit Painlev\'e II equation, and also away from the transition point using the pre-string {\it difference} equation.
Here we present a complementary analysis of the transition at all coupling and all finite $N$, in terms of a {\it differential} equation, using the explicit Tracy-Widom mapping of the Gross-Witten-Wadia partition function to a solution of a Painlev\'e III equation.
This mapping provides a simple method to generate trans-series expansions in all parameter regimes, and to study their transmutation as the parameters are varied. 
For example, at any finite $N$ the 
weak coupling expansion is divergent, with a non-perturbative trans-series completion; on the other hand, the strong coupling expansion is convergent, and yet there is still a non-perturbative trans-series completion. We show how the different instanton terms `condense' at the transition point to match with the double-scaling limit trans-series. We also define a {\it uniform} large $N$ strong-coupling expansion (a non-linear analogue of uniform WKB), which is much more precise than the conventional large $N$ expansion through the transition region, and apply it to the evaluation of Wilson loops.
}

\maketitle

\section{Introduction} 

The third-order phase transition of the Gross-Witten-Wadia (GWW) unitary matrix model \cite{gw,wadia,Wadia:1980cp} at $N=\infty$ is a universal model of large $N$ phase transitions, which appear in many fields, ranging from large dynamical systems, to quantum field theory, statistical physics, 2d gravity, string theory, matrix models, and random matrices \cite{may,migdal,brezin-wadia,2dgravity,Rossi:1996hs,Moshe:2003xn,Verbaarschot:2000dy,forrester-book,oxford,Szabo:2011eg,marcos-lectures,marcos-book}. Two dimensional gauge theory provides a particularly explicit realization \cite{Neuberger:1980qh,Witten:1991we,Witten:1992xu,Douglas:1993iia,matytsin,Forrester:2010ah}. There are also mathematical applications in combinatorial problems \cite{subsequences,baik}.
The immediate vicinity of the transition has a universal nature \cite{dotsenko,schehr}, characterized by the Tracy-Widom distribution \cite{Tracy:1992rf,Tracy-Widom:1994,tracy-widom2} and the associated Painlev\'e II equation. A natural approach to these systems is asymptotic analysis for large $N$, combined with an expansion in another parameter such as a coupling. Recently, modern ideas of resurgent asymptotic analysis with trans-series \cite{Ecalle:1981,Costin:2009,Aniceto:2013fka,Dorigoni:2014hea} have been applied to such physical systems, studying large $N$ and/or strong and weak coupling asymptotics \cite{marino-matrix,Aniceto:2011nu,Hatsuda:2013oxa,Aniceto:2014hoa,Nishigaki:2014ija,Codesido:2014oua,Russo:2014bda,Grassi:2014zfa,Basar:2015xna,Couso-Santamaria:2015wga,Hatsuda:2015owa,Dunne:2016qix,Honda:2016mvg,Honda:2016vmv,Couso-Santamaria:2016vcc,Gukov:2016njj}.
The related interpretation in terms of complex saddle points has also been studied recently for the Gross-Witten-Wadia (GWW) unitary matrix model \cite{Buividovich:2015oju,Alvarez:2016rmo,Okuyama:2017pil,Alfinito:2017hsh}.
\begin{table}[htb]
\centering
 \medskip
\begin{tabular}{|c|l|l|}\hline
\multirow{1}{*}{~} & \qquad\qquad Weak coupling & \qquad\qquad Strong coupling\\ \hline 
\multirow{4}{*}{\stackanchor{
Fixed $N$;}
{expansion in coupling $g^2$}
} 
    & \qquad\qquad $g^2\ll N$ & \qquad\qquad $g^2\gg N$ \\
    \cline{2-3}
    & $\bullet$ divergent (non-alternating) & $\bullet$ convergent \\ 
    \cline{2-3}
    & $\bullet$ trans-series completion & $\bullet$ trans-series completion \\
     \cline{2-3}
    & $\bullet$ imaginary trans-series parameter & $\bullet$ real trans-series parameter \\\hline
\multirow{4}{*}{
\stackanchor{\stackanchor{Large $N$  't Hooft limit:}{$N\to\infty$, $t\equiv N g^2/2$ fixed;}}{expansion in $1/N$}}
    & \qquad\qquad $t\ll 1$ & \qquad\qquad $t\gg 1$ \\
    \cline{2-3}
    & $\bullet$ divergent (non-alternating) & $\bullet$ divergent (alternating) \\\cline{2-3}
    & $\bullet$ trans-series completion & $\bullet$ trans-series completion \\\cline{2-3}
    & $\bullet$ imaginary trans-series parameter & $\bullet$ real trans-series parameter \\\hline
\multirow{4}{*}{
\stackanchor{\stackanchor{Double-scaling limit:}{$N\to\infty$, $t\sim 1+\kappa/N^{2/3}$;}}{expansion in $\kappa$}} 
    &  \qquad\qquad $\kappa\leq 0$ &  \qquad\qquad $\kappa \geq 0$ \\
    \cline{2-3}
    & $\bullet$ divergent (non-alternating) & $\bullet$ divergent (alternating) \\\cline{2-3}
    & $\bullet$ trans-series completion & $\bullet$ trans-series completion \\\cline{2-3}
    & $\bullet$  imaginary trans-series parameter & $\bullet$ real trans-series parameter \\\hline
\end{tabular}
 \medskip
 \caption{The basic structure of the weak-coupling and strong-coupling expansions of the free energy $\ln Z(g^2, N)$ in three physical limits. At fixed $N$, the weak coupling expansion is divergent, with a trans-series completion having a pure imaginary trans-series parameter. In contrast, at fixed $N$ the strong coupling expansion is convergent; nevertheless it has a non-perturbative trans-series completion as an instanton expansion: see Section \ref{sec:delta-strong}. In the 't Hooft large $N$ limit, both weak-coupling and strong-coupling expansions are divergent, with non-perturbative trans-series completions, but the form of the trans-series is quite different on either side of the Gross-Witten-Wadia phase transition, which occurs at 't Hooft parameter $t=1$, where $t\equiv N g^2/2$. In the double-scaling limit, the trans-series match smoothly to those of the 't Hooft limit, but also to the fixed N expansions, in a way discussed in Section \ref{sec:matching}.
 }
    \label{tab:summary}
\end{table}

This paper addresses the question of how a trans-series in two parameters, $g^2$ and $N$, rearranges itself at weak- and strong-coupling, and at large and small values of the parameter $N$. The trans-series expression for the partition function (or the free energy, or the specific heat, or a Wilson loop) involves both perturbative and non-perturbative contributions, in a unified self-consistent form that encodes the full analytic structure.
Often, perturbative coupling expansions may be convergent in one regime (weak or strong coupling), but divergent in the other \cite{LeGuillou:1990nq}. Divergent expansions are naturally related to non-perturbative contributions beyond perturbation theory, but we can ask how such non-perturbative terms arise for a convergent expansion, as indeed occurs in the GWW model. This behavior becomes considerably richer when another parameter, $N$, is introduced.
A generic phenomenon is the possible appearance of a phase transition in the $N\to \infty$ limit at a particular critical value of a scaled ``'t Hooft'' parameter $t=N g^2/2$. The large $N$ expansion is divergent on either side of the transition, where the trans-series has very different structure. The immediate vicinity of the transition point is described by a double-scaling limit ($N\to \infty$ and $t\to t_c$ in a correlated way such that $N$ is scaled out), in terms of a solution to the Painlev\'e II equation. This universal behavior is captured in a simple and explicit form by the Gross-Witten-Wadia unitary matrix model. In order to probe more finely the full rearrangement of the trans-series, we study this transition keeping the full $N$ dependence. This has been done long ago by Marino using the pre-string equation \cite{marino-matrix}, and a detailed study of finite $N$ hermitean matrix models appeared in \cite{Couso-Santamaria:2015wga}. Here we use a different approach, using the explicit connection of the GWW unitary matrix model to the Painlev\'e III equation for all $N$ and all coupling.

The essential trans-series structure in various limits is sketched in Table \ref{tab:summary}, and in slightly more detail in equations (\ref{eq:lnz-general-weak})--(\ref{eq:lnz-general-strong}) and (\ref{eq:ds-general-weak})--(\ref{eq:ds-general-strong}). The weak coupling perturbative expansion, for fixed $N$, is divergent and becomes a trans-series when the associated non-perturbative terms are included. In contrast, the strong coupling expansion, for fixed $N$, is convergent. Nevertheless, it also has a non-perturbative completion as a trans-series, as shown in Section \ref{sec:delta-strong}. When expressed as large $N$ expansions, these are divergent in both weak and strong coupling, although the form of the expansion is radically different in these two  coupling regimes, passing through the GWW phase transition. For example, in the large $N$ 't Hooft limit, the free energy, $\ln Z(t, N)$ has the following trans-series structure (for details see Section \ref{sec:delta} below):
\begin{eqnarray}
\ln Z(t, N)\bigg|_{\rm weak}
&\sim& \ln \left(\frac{G(N+1)}{(2\pi)^{N/2}}\right)+\frac{N^2}{t}-\frac{N^2}{2}\ln \left(\frac{N}{t}\right) +\frac{1}{8}\ln(1-t) \nonumber\\
&&+\sum_{n=0}^\infty \frac{f_n^{(k), {\rm weak}}(t)}{N^{2n}}
+\sum_{k=1}^\infty P_{\rm weak}^{(k)}(t)\left(\frac{i\, e^{-N\, S_{\rm weak}(t)}}{\sqrt{2\pi N}}\right)^{k} \sum_{n=0}^\infty \frac{f_n^{(k), {\rm weak}}(t)}{N^n}
\label{eq:lnz-general-weak}
\\
\ln Z(t, N)\bigg|_{\rm strong}
&\sim& \frac{N^2}{4t^2} +\sum_{k=1}^\infty P_{\rm strong}^{(k)}(t)\left(\frac{e^{- N\, S_{\rm strong}(t)}}{\sqrt{2\pi N}}\right)^{2k} \sum_{n=0}^\infty \frac{f_n^{(k), {\rm strong}}(t)}{N^n} 
\label{eq:lnz-general-strong}
\end{eqnarray}
At weak coupling, the perturbative expansion is in powers of $1/N^2$, while higher instanton terms have fluctuation expansions in powers of $1/N$. All these perturbative expansions are factorially divergent and non-alternating in sign, and correspondingly the trans-series parameter appearing in the instanton sum has an imaginary part (in fact, with the appropriate boundary conditions it is pure imaginary). At strong coupling, there is only one perturbative term, and all the fluctuation expansions in the instanton sum are in powers of $1/N$, and  the fluctuations coefficients are factorially divergent and alternating in sign. Correspondingly the trans-series parameter is real, with a value fixed by boundary conditions.
These large $N$ expansions break down at the transition point, because the actions $S_{\rm weak}(t)$ and $S_{\rm strong}(t)$ vanish there, and also because the
prefactors $P^{(k)}(t)$ diverge as $t\to 1$. This can be improved by introducing a {\it uniform} large $N$ expansion at strong coupling, expanding in instanton factors of ${\rm Ai}\left(\left(\frac{3}{2}\,N\, S_{\rm strong}(t)\right)^{2/3}\right)$ and ${\rm Ai}^\prime\left(\left(\frac{3}{2}\,N\, 
S_{\rm strong}(t)\right)^{2/3}\right)$, instead of the usual exponential instanton factors $e^{- N\, S_{\rm strong}(t)}$. Then the prefactors and fluctuation terms are smooth all the way through the transition point, and provide a much better approximation all the way into the double-scaling region. This is a non-linear extension of the familiar uniform WKB approximation (see Sections \ref{sec:uniform} and  \ref{sec:wilson}).

In the double-scaling limit, the parameters $t$ and $N$ are scaled together as $t\sim 1+\kappa/N^{2/3}$, which effectively zooms in on a narrow window, of width $1/N^{2/3}$, of the transition point. The associated Hastings-McLeod solution of the Painlev\'e II equation has trans-series structure \cite{marino-matrix} (see Section \ref{sec:matching} below):
\begin{eqnarray}
W(\kappa)\bigg|_{\rm weak}
&\sim & \sqrt{-\kappa} \sum_{n=0}^\infty \frac{f_{n, {\rm double}}^{(0), {\rm weak}}}{(-\kappa)^{3n}}
+\sum_{k=1}^\infty P_{\rm double}^{(k), {\rm weak}}(\kappa)\left(\frac{i\, e^{-\frac{4}{3}(-\kappa)^{3/2}}}{\sqrt{2\pi} (-\kappa)^{1/4}}\right)^{k} \sum_{n=0}^\infty \frac{f_{n, {\rm double}}^{(k), {\rm  weak}}}{(-\kappa)^{3n/2}}
\label{eq:ds-general-weak}
\\
W(\kappa)\bigg|_{\rm strong}
&\sim & \sum_{k=0}^\infty P_{\rm double}^{(k), {\rm strong}}(\kappa)\left(\frac{ e^{-\frac{2\sqrt{2}}{3}\,\kappa^{3/2}}}{\sqrt{2\pi} \,\kappa^{1/4}}\right)^{k} \sum_{n=0}^\infty \frac{f_{n, {\rm double}}^{(k), {\rm strong}}}{\kappa^{3n/2}}
\label{eq:ds-general-strong}
\end{eqnarray}
Notice the factor of $\sqrt{2}$ difference in the instanton factor exponents at strong and weak coupling.
These double-scaling limit solutions match to the weak-coupling and strong-coupling large $N$ expansions in (\ref{eq:lnz-general-weak})--(\ref{eq:lnz-general-strong}). But the expansions (\ref{eq:ds-general-weak}, \ref{eq:ds-general-strong}) also blow up at the transition point, where $\kappa\to 0$, because the exponential factors become of order 1, and the prefactors diverge. As above, this can  be cured by a uniform approximation, expressing the double-scaling strong-coupling expansion as a trans-series in powers of Airy functions ${\rm Ai}$ and ${\rm Ai}^\prime$. See for example the Wilson loop expressions in Section \ref{sec:wilson}.

In Section \ref{sec:z} we describe the basic structure of the trans-series expansions for the partition function $Z$, using the resurgent trans-series expansions for the individual entries of the Toeplitz determinant representation of $Z$. While this reveals the essential form, it is not sufficiently explicit to study the large $N$ limit.
Sections \ref{sec:pIII} and \ref{sec:simpler} show how the partition function can be obtained from simpler functions, such as the expectation value $\Delta\equiv \langle \det\, U\rangle$, which satisfy  Painlev\'e III type equations with respect to the inverse coupling, with $N$ appearing as a parameter in the equation. These Painlev\'e III equations make it straightforward to generate trans-series expansions in any parameter regime. Section \ref{sec:delta} develops the detailed trans-series expansions for $\langle \det\, U\rangle$, and the corresponding expansions for other physical quantities such as the partition function, the free energy and the specific heat. The connection to the double-scaling limit, in terms of the coalescence of the Painlev\'e III equation to the Painlev\'e II equation is described in Section \ref{sec:matching}. Section \ref{sec:wilson} studies new uniform large $N$ expansions for winding Wilson loops in the GWW model.

\section{Towards trans-series expansions of the partition function}
\label{sec:z} 
The Gross-Witten-Wadia (GWW) model \cite{gw,wadia,Wadia:1980cp} is a unitary matrix model whose partition function is defined by an integral over $N\times N$ unitary matrices:
\begin{eqnarray}
Z(g^2, N)=\int_{U(N)} DU \, \exp\left[\frac{1}{g^2} {\rm tr}
\left(U+U^\dagger\right)\right]
\label{eq:z}
\end{eqnarray}
This matrix integral can be evaluated as a Toeplitz determinant, for any $N$ and any coupling $g^2$ \cite{Bars:1979xb,Rossi:1996hs,forrester-book,borodin}:
\begin{eqnarray}
Z(x, N) =\det \left[ I_{j-k} \left(x\right)\right]_{j, k =1, \dots , N}
\label{eq:z1}
\end{eqnarray}
where $I_j(x)$ is the modified Bessel function, evaluated at twice the inverse coupling
\begin{eqnarray}
x\equiv \frac{2}{g^2}
\label{eq:x}
\end{eqnarray}
We will also use a ``'t Hooft'' parameter, $t$, when studying the large $N$ limit:
\begin{eqnarray}
t\equiv \frac{N g^2}{2}\equiv \frac{N}{x}
\label{eq:t}
\end{eqnarray}
The GWW model is known to have a third order phase transition at the critical value $t_c=1$, in the $N\to\infty$ limit \cite{gw,wadia,Wadia:1980cp}.

Other important functions considered in this paper include the expectation value \cite{Rossi:1982vw}
\begin{eqnarray}
\Delta(x, N) \equiv \langle \det U \rangle = \frac{\det \left[ I_{j-k+1} \left(x\right)\right]_{j, k =1, \dots , N}}{\det \left[ I_{j-k} \left(x\right)\right]_{j, k =1, \dots , N}}
\label{eq:delta-def}
\end{eqnarray}
and the (normalized) winding Wilson loops \cite{Green:1980bg,Okuyama:2017pil,Alfinito:2017hsh}
\begin{eqnarray}
{\mathcal W}_p(x, N)\equiv \frac{1}{N} \langle {\rm tr} \left( U^p \right) \rangle
\label{eq:winding}
\end{eqnarray}

We first discuss the basic trans-series structure of the weak-coupling and strong-coupling expansions for $Z(x, N)$ at fixed finite $N$, which follow from the large and small $x$ expansions of the Bessel functions appearing in the determinant expression (\ref{eq:z1}). These `brute-force' expansions reveal much of the essential structure, but also illustrate that more sophisticated methods are needed in order to probe the full details, in particular in the strong-coupling regime.

\subsection{Coupling expansions of the partition function $Z(x, N)$, at fixed $N$}
\label{sec:z-expansions}

The partition function is a function of two variables, the inverse coupling $x=2/g^2$ and the integer $N$ of the $U(N)$ group. We first consider expansions in the coupling, for fixed but arbitrary $N$.

\subsubsection{Weak coupling expansion of the partition function $Z(x, N)$, at fixed $N$}
\label{sec:z-weak1}
At weak coupling, $x\to+\infty$, we need the large argument asymptotics of the modified Bessel functions \cite{dlmf:bessel-large}. At fixed index $j$, the large $x$ resurgent asymptotic expansion of the modified Bessel function involves two exponential terms:\footnote{Since the Bessel function satisfies a linear second order equation there are only two saddle terms \cite{Costin:2009}.}
\begin{eqnarray}
I_j(x) \sim \frac{e^{x}}{\sqrt{2 \pi x}} \sum_{n=0}^\infty(-1)^n \frac{\alpha_n(j)}{x^n} \pm i e^{ij\pi} \frac{e^{-x}}{\sqrt{2 \pi x}} \sum_{n=0}^\infty \frac{\alpha_n(j)}{x^n}, \qquad \left|{\rm arg}(x) - \frac{\pi}{2}\right| <\pi
\label{eq:bessel}
\end{eqnarray}
where the fluctuation coefficients are
\begin{eqnarray}
\alpha_n(j) = \frac{1}{8^n n!}\prod_{l=1}^n \big(4 j^2 - (2l-1)^2\big)
=\frac{1}{8^n n!} \frac{(2j+2n-1)!!}{(2j-2n-1)!!}
\end{eqnarray}
Note that we need to keep {\it both} exponential terms in (\ref{eq:bessel}) in order to have direct access to the non-perturbative terms in the weak-coupling expansion.\footnote{The same strategy can be applied to the analysis of Wilson loop expectation values at finite $N$ \cite{Okuyama:2017pil,Alfinito:2017hsh}. We discuss Wilson loops in Section \ref{sec:wilson} below.}

The fluctuation factors about each of the two exponentials in (\ref{eq:bessel}) are related by resurgence. First, they are clearly related since they only differ from one another by an alternating sign. Second, they exhibit the generic Berry-Howls \cite{berry-howls,Aniceto:2013fka} type of resurgence relation connecting the large order growth of the coefficients about one non-perturbative term to the low orders of the expansion coefficients about the other non-perturbative term. Indeed,  at large order $n$ (for a fixed but arbitrary Bessel index $j$) we have the remarkable relation:
\begin{eqnarray}
\alpha_n(j)\sim \frac{\cos(j \pi)}{\pi} 
\frac{(-1)^n (n-1)!}{2^n} \left(\alpha_0(j)
+ \frac{2\, \alpha_1(j)}{(n-1)}
+ \frac{2^2\, \alpha_2(j)}{(n-1)(n-2)}
+ \frac{2^3\, \alpha_3(j)}{(n-1)(n-2)(n-3)}
+\dots\right)
\nonumber\\
\label{eq:self-resurgence}
\end{eqnarray}
Notice that the large order coefficients are factorially divergent in the expansion order $n$, with leading and sub-leading coefficients that are expressed in terms of their own low order terms: $\alpha_0(j)$, $\alpha_1(j)$, $\alpha_2(j)$, \dots, for any Bessel index $j$. This is a strong form of ``self-resurgence''.
Furthermore, notice that when the Bessel index $j$ is a half-odd-integer, the large-order  expression (\ref{eq:self-resurgence}) vanishes, consistent with the fact that in this case the asymptotic expansions in (\ref{eq:bessel}) truncate. Physically, this is an illustrative example of a cancellation due to interference between different saddle contributions \cite{Unsal:2012zj,Dunne:2016jsr}.

We now consider how this kind of large-order/low-order resurgence behavior manifests itself in the partition function $Z(x, N)$ in (\ref{eq:z1}), which is a determinant involving Bessel $I_j(x)$ functions with indices ranging from $j=0$ to $j=N-1$. Thus, the partition function consists structurally of a sum of products of individual Bessel functions, with varying indices. It is therefore clear that since each Bessel function has two different exponential terms in its resurgent asymptotic expansion (\ref{eq:bessel}), the expanded determinant has an expansion involving $(N+1)$ different exponential terms:\footnote{This sum over $(N+1)$ instanton terms is consistent with the fact that $Z(x, N)$ satisfies a linear differential equation of order $(N+1)$. For example, $Z(x, 1)=I_0(x)$ satisfies the Bessel equation, while for $N=2$ we have $Z(x, 2)=I_0^2(x)-I_1^2(x)$, which satisfies the third order equation: $x^2 Z^{\prime\prime\prime}+5x Z^{\prime\prime}+(3-4x^2)Z^\prime - 4x Z=0$.}
\begin{eqnarray}
Z(x, N)\sim Z_0(x, N) \sum_{k=0}^N Z^{(k)}(x,N) e^{-2 k x} \sum_{n=0}^\infty  \frac{a_n^{(k)}(N)}{x^n}
\label{eq:z-transseries}
\end{eqnarray}
The leading piece, corresponding to taking for each Bessel function the leading growing exponential factor, $\frac{e^{x}}{\sqrt{2 \pi x}}$, together with its first fluctuation correction, is easily deduced as \cite{Rossi:1996hs}
\begin{eqnarray}
Z_0(x,N)=\frac{G(N+1)}{(2\pi)^{N/2}} e^{N x} x^{-N^2/2}
\label{eq:z0}
\end{eqnarray}
where $G$ is the Barnes G-function \cite{dlmf:barnes}. For a given integer $N$ there is only a finite number of "instanton" terms in the trans-series (\ref{eq:z-transseries}).\footnote{Of course, the corresponding trans-series for $\ln Z$, and hence for the free energy, specific heat, and Wilson loop expectation values, have an infinite number of instanton terms.} Also notice that because of the simple relation between the expansion coefficients for the fluctuations about the two different exponentials in (\ref{eq:bessel}), there is a symmetry between the fluctuation coefficients $a_n^{(k)}(N)$ and $a_n^{(N-k)}(N)$ in the $k$-instanton and $(N-k)$-instanton sectors of $Z(x, N)$. The prefactors $Z^{(k)}(x,N)$ in (\ref{eq:z-transseries}) are chosen so that the leading fluctuation coefficients are unity: $a_0^{(k)}=1$ for all $k$.
More explicitly we can write the trans-series as
\begin{eqnarray}
Z(x,N) &\sim& Z_0(x, N) \left[\sum_{n=0}^\infty \frac{a_n^{(0)}(N)}{x^n} + i\frac{(4x)^{N-1}}{\Gamma(N)} e^{-2 x}\sum_{n=0}^\infty \frac{a_n^{(1)}(N)}{x^n}
+\frac{(4x)^{2(N-2)}}{\Gamma(N)\Gamma(N-1)} e^{-4x}\sum_{n=0}^\infty \frac{a_n^{(2)}(N)}{x^n} +
\right.
\nonumber\\
&&\left.
\ldots + \xi_N \frac{G(N+1)}{\prod_{i=0}^{N-1} \Gamma(N-i)} e^{-2 N x}\sum_{n=0}^\infty \frac{a_n^{(N)}(N)}{x^n}\right]
\label{eq:zweak}
\end{eqnarray}
where $\xi_N$ is 1 ($i$) when $N$ is even (odd), respectively. By brute-force expansion of the Toeplitz determinant (\ref{eq:z1}) in the weak-coupling limit, $x\to \infty$, for various values of $N$, we can deduce the early coefficients in these expansions. For example, the first few terms of the zero-instanton, one-instanton and two-instanton fluctuation terms are:
\begin{eqnarray}
\Gamma^{(0)}(x, N)\equiv \sum_{n=0}^\infty \frac{a_n^{(0)}(N)}{x^n}&=&1+\frac{N}{8}\frac{1}{x} + \frac{9N^2}{128} \frac{1}{x^2} + \frac{3N(17N^2+8)}{1024} \frac{1}{x^3} + \ldots 
\label{eq:zweaka}\\
\Gamma^{(1)}(x, N)\equiv \sum_{n=0}^\infty \frac{a_n^{(1)}(N)}{x^n}&=&1-\frac{(N-2)(2N-3)}{8}\frac{1}{x} + \frac{(4 N^4-36 N^3+129 N^2-220 N+132)}{128}  \frac{1}{x^2} + \ldots 
\nonumber\\
\label{eq:zweakb}\\
\Gamma^{(2)}(x, N)\equiv \sum_{n=0}^\infty \frac{a_n^{(2)}(N)}{x^n}&=&1-\frac{(N-4)(4N-9)}{8}\frac{1}{x} + ...
\label{eq:zweakc}
\end{eqnarray}

\begin{figure}[htb]
    \centering
    \includegraphics[scale=1.3]{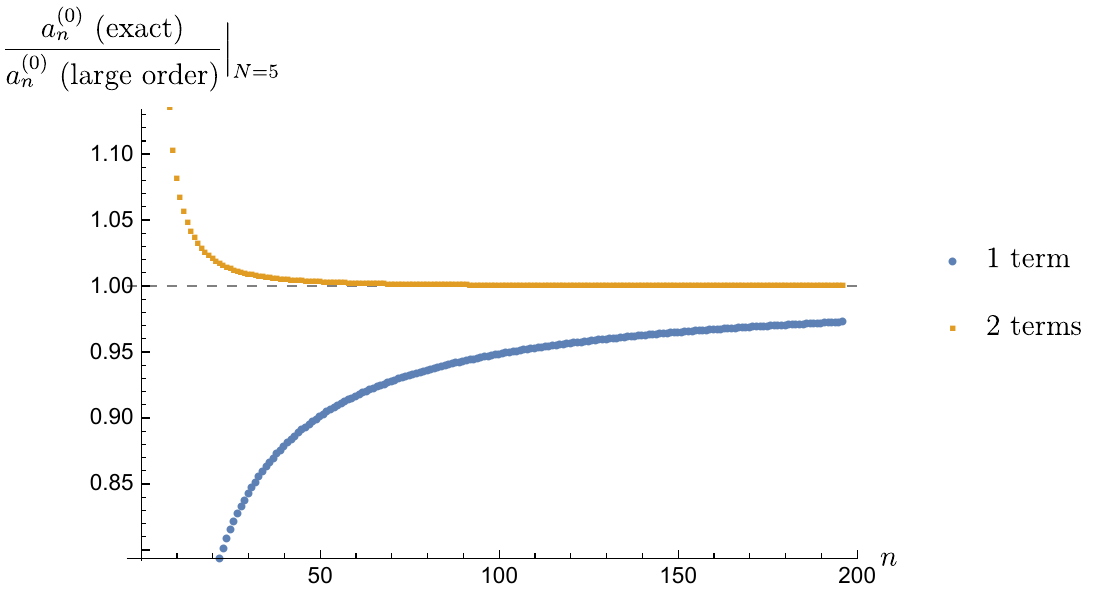}
    \caption{Ratio of numerically computed large order growth of the perturbative coefficients $a_n^{(0)}$  and the series expansion  (\ref{eq:ZGrowth0InstFiniteNWeak}) with just the first term (blue circles) and first two terms (red squares).}
    \label{fig:fig1}
\end{figure}
\begin{figure}[htb]
    \centering
    \includegraphics[scale=1.3]{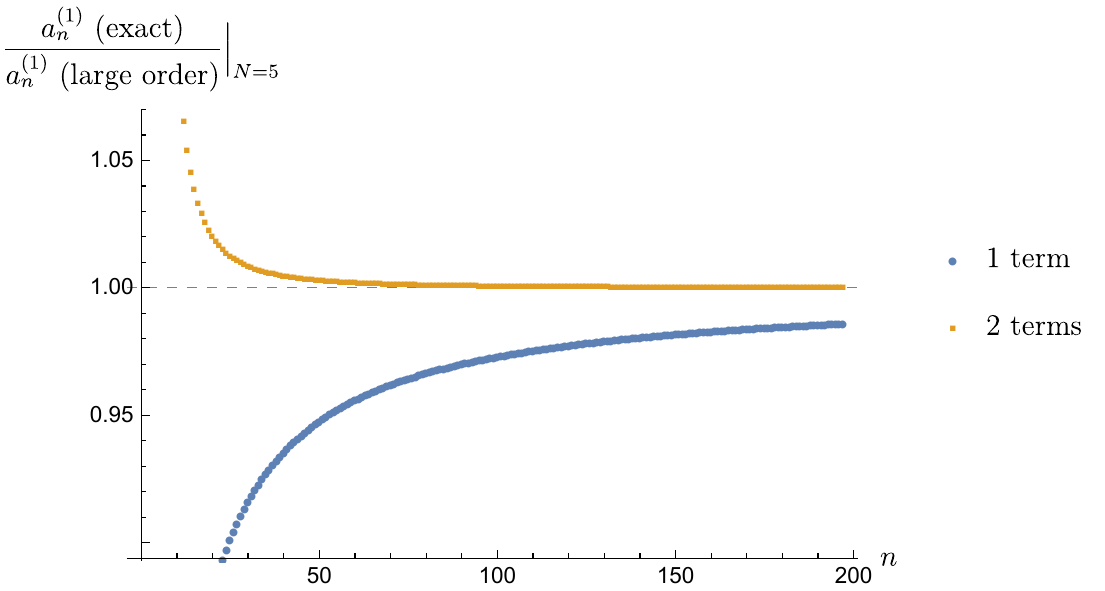}
    \caption{Ratio of numerically computed large order growth of the perturbative coefficients $a_n^{(1)}$  and the series expansion  (\ref{eq:ZGrowth1InstFiniteNWeak}) with just the first term (blue circles) and first two terms (red squares).}
    \label{fig:fig2}
\end{figure}

For a given $N$, each of the fluctuation series in (\ref{eq:zweak}) is divergent. Resurgence relations appear in the large order (in $n$) behavior of the fluctuation expansion coefficients (for a given instanton sector $k$). For example, at large order $n$ of the zero-instanton fluctuation series in (\ref{eq:zweaka}):
\begin{equation} \label{eq:ZGrowth0InstFiniteNWeak}
    \begin{aligned}
	a^{(0)}_n(N) \sim \frac{2^{N}}{\pi (N-1)!}  &\frac{ (n+N-3)!}{2^n} \bigg[1-\frac{(N-2)(2N-3)}{8}\frac{2}{(n+N-3)} \\
	&+ \frac{(4 N^4-36 N^3+129 N^2-220 N+132)}{128} \frac{2^2}{(n+N-3)(n+N-4)} + \dots \bigg]
\end{aligned}
\end{equation}
In this expression for the large order behavior of the zero-instanton series (\ref{eq:zweaka}) we recognize the low order coefficients of the one-instanton series (\ref{eq:zweakb}). See Fig. \ref{fig:fig1}. Similarly, the large-order growth of the fluctuation coefficients in the one-instanton sector is given (for $N\geq 3$) by:
\begin{eqnarray} \label{eq:ZGrowth1InstFiniteNWeak}
	a^{(1)}_n(N) \sim -\frac{2^{N-1}}{\pi (N-2)!}  \frac{ (n+N-5)!}{2^n} \left[1-\frac{(N-4)(4N-9)}{8}\frac{2}{(n+N-5)}+\dots\right]
\end{eqnarray}
with coefficients that appear in the low order expansion of the two-instanton series in (\ref{eq:zweakc}). See Fig. \ref{fig:fig2}. This pattern continues for higher instanton sectors. 

In fact, these results simply confirm well-known general results for the resurgence properties of solutions to (higher order) linear differential equations \cite{Costin:2009}, following immediately from the fact that $Z(x, N)$ satisfies a linear ODE of order $(N+1)$. 

To summarize so far: direct expansion of the Toeplitz determinant expression (\ref{eq:z1}), using the full resurgent Bessel asymptotics in (\ref{eq:bessel}), explains in elementary terms the structural form of the weak-coupling trans-series in (\ref{eq:z-transseries}) and (\ref{eq:zweak}), and confirms its resurgence properties. However, this approach becomes less practical at large $N$, so we need to develop a more powerful approach in order to study the full trans-series structure at general $N$.

\subsubsection{Strong coupling expansion of the partition function $Z(x, N)$, at fixed $N$}
\label{sec:z-strong}

The strong coupling limit is $x\to 0^+$, so we use the convergent small argument expansion \cite{dlmf:bessel-small} of the modified Bessel functions in the determinant expression (\ref{eq:z1}):
\begin{eqnarray}
I_j(x)=\left(\frac{x}{2}\right)^j \sum_{n=0}^\infty \frac{(x/2)^{2n}}{ n! \, \Gamma(n+j+1)}
\end{eqnarray}
As noted long ago by Wadia \cite{wadia}, a direct strong-coupling expansion of the partition function determinant produces a leading factor whose first $N$ terms coincide with the expansion of $e^{x^2/4}$. But there are further corrections to this, which we can write as an expansion in even powers of $x$:
\begin{eqnarray}
Z(x, N) \sim e^{x^2/4} \left[1 -  \left(\frac{(x/2)^{N+1}}{(N+1)!}\right)^2  \left(1-\frac{1}{2}\frac{(N+1)\, x^2}{(N+2)^2}
+\frac{1}{16}\frac{ (2 N+3)\, x^4}{(N+3)^2 (N+2)} +\dots \right)+\dots \right]
\label{eq:z3}
\end{eqnarray}
In contrast to the divergent weak-coupling fluctuation expansions (\ref{eq:zweaka}-\ref{eq:zweakc}), the strong coupling fluctuation expansion  in (\ref{eq:z3}) is a convergent expansion, for a given $N$. Naively, this suggests the absence of further non-perturbative terms. However, there are in fact additional non-perturbative contributions to (\ref{eq:z3}), which are difficult to deduce from this kind of brute-force determinant expansion for small $x$, with fixed $N$.\footnote{This is analogous to the non-perturbative gap-splittings in the Mathieu equation spectrum, associated with complex instantons, and for which the familiar {\it convergent} strong-coupling expansions may be re-organized in a trans-series form \cite{Basar:2015xna}. This example maps directly to the trans-series structure in the electric region of the $SU(2)$ SUSY gauge theory in the Nekrasov-Shatashvili limit
\cite{Nekrasov:2009rc,Mironov:2009uv,KashaniPoor:2012wb,Krefl:2013bsa,Basar:2015xna}.} The non-perturbative trans-series completion of this convergent strong-coupling expansion (\ref{eq:z3}) is discussed below in Sections \ref{sec:pIII} and \ref{sec:delta}. 

\subsection{Large $N$ expansions of the partition function $Z(t, N)$ in the 't Hooft limit}
\label{sec:z-weak}

The large $N$ expansion in the 't Hooft limit involves taking $N\to\infty$, keeping fixed the 't Hooft coupling 
\begin{eqnarray}
t\equiv \frac{N}{x} \equiv \frac{N g^2}{2}
\label{eq:thooft}
\end{eqnarray} 
The form of the large $N$ expansion depends on the magnitude of $t$ relative to the critical value $t_c=1$. And we will see that the large $N$ trans-series structure changes radically through this GWW phase transition.

\subsubsection{Large $N$ expansion of $Z(t, N)$ at weak coupling: $t<1$}
\label{sec:z-weak-t}

When $N\to\infty$, and $x\to\infty$, such that $t=\frac{N}{x}<1$ is fixed, we can formally rearrange the weak coupling expansion (\ref{eq:zweak}) as follows.
First, note that
\begin{eqnarray}
\ln Z_0(t, N)
&= & N^2\left(\frac{1}{2} \ln \left(\frac{t}{N}\right) +\frac{1}{t}\right) -\frac{N}{2} \ln (2\pi) + \ln G(N+1)
\label{eq:z0N}
\end{eqnarray}
which involves the well-known large $N$ asymptotics of the Barnes function \cite{dlmf:barnes}.
The zero instanton fluctuation term in (\ref{eq:zweaka}) becomes a series in inverse powers of $N^2$:
\begin{equation} \label{eq:z0a0}
\begin{aligned}
\Gamma^{(0)}(x, N)\equiv \sum_{n=0}^\infty \frac{a_n^{(0)}(N)}{x^n} &\sim \left(1+\frac{t}{8}+\frac{9 t^2}{128}+\frac{51 t^3}{1024}+\frac{1275 t^4}{32768}+\frac{8415
   t^5}{262144}+\frac{115005 t^6}{4194304} + \dots\right) \\
   & \qquad\qquad+ \frac{1}{N^2} \frac{3 t^3}{128} \left(1+\frac{25 t}{8}+\frac{825 t^2}{128}+\frac{11275
   t^3}{1024} + \dots\right) \\
   &\qquad\qquad + \frac{1}{N^4}\frac{45 t^5}{1024}  \left(1+\frac{209 t}{32} + \dots\right) \\
   & \equiv 	\sum_{n=0}^\infty \frac{f_n^{(0)}(t)}{N^{2n}}
\end{aligned}
\end{equation}
Notice that each fluctuation factor $f_n^{(0)}(t)$ has a {\it convergent} small $t$ expansion. For example, 
\begin{equation}
 f_0^{(0)}(t)=   \left(1+\frac{t}{8}+\frac{9 t^2}{128}+\frac{51 t^3}{1024}+\frac{1275 t^4}{32768}+\frac{8415
   t^5}{262144}+\frac{115005 t^6}{4194304} + \dots\right) = \frac{1}{(1-t)^{1/8}}
\label{eq:z0a0-sum}
\end{equation}
The one-instanton (and higher instanton) fluctuation terms that are obtained by re-arranging the weak-coupling fluctuation series in (\ref{eq:zweakb}-\ref{eq:zweakc}) have a different structure: they are expansions in all inverse powers of $N$ (not $N^2$). Furthermore, this rearrangement procedure produces terms with factors of $N$ in the numerator. For example, rewriting (\ref{eq:zweakb}) in the small $t$ and large $N$ limit leads to:
\begin{eqnarray}
\Gamma^{(1)}(x, N)\equiv \sum_{n=0}^\infty \frac{a_n^{(1)}(N)}{x^n} 
&\sim & \left(1-\frac{t}{8}\left(2N-7+\frac{6}{N}\right)+\dots\right)
\label{eq:z0a1}
\end{eqnarray}
This structure is an artifact of changing the order of limits, as the expansions (\ref{eq:zweakb}-\ref{eq:zweakc}) were generated at large $x$ with fixed $N$, rather than in a strict 't Hooft large $N$ limit. The proper interpretation of these extra terms is that they are associated with the small $t$ and large $N$ expansion of large $N$ instanton factors $\exp\left[-N S_{\rm weak}(t)\right]$, as illustrated below in Section \ref{sec:z-final}.\footnote{This is analogous to the situation in quantum mechanical models \cite{ZinnJustin:2004ib,Jentschura:2004jg,alvarez,Dunne:2013ada,Dunne:2014bca,Dunne:2016qix}, such as the double-well potential or the Mathieu cosine potential, where the fluctuations in the perturbative sector, as a function of $\hbar$ and $(N+\frac{1}{2})$, where $N$ is the perturbative level number, can be re-arranged into a large $(N+\frac{1}{2})$ expansion of the form in (\ref{eq:z0a0}) in inverse powers of $(N+\frac{1}{2})^2$, while the fluctuations in the one-instanton sector have a re-arranged large $(N+\frac{1}{2})$ series of the form in (\ref{eq:z0a1}) in inverse powers of $(N+\frac{1}{2})$, but also with powers of $(N+\frac{1}{2})$ in the numerators.}
In order to do this, we first need to develop a simpler and more direct approach that reveals the full trans-series structure of the large $N$ expansion.

\subsubsection{Large $N$ expansion of $Z(t, N)$ at strong coupling: $t>1$}
\label{sec:z-strong-t}

A similar rearrangement of the strong coupling expansion (\ref{eq:z3}), in the large $N$ and large $t$ limits, also produces an unusual structure:
\begin{eqnarray}
Z(t, N)\sim e^{N^2/(4t^2)}\left[1-\left(\frac{\left(N/(2t)\right)^{N+1}}{(N+1)!}\right)^2\left(1-\frac{1}{2t^2} \left(N-3+\frac{8}{N}\right)+\dots\right)+\dots\right]
\label{eq:zstrongt}
\end{eqnarray}
Again we note that the fluctuation term involves positive powers of $N$. As in the weak-coupling case (\ref{eq:z0a1}), these terms should be interpreted in terms of the expansion of a large $N$ instanton factor, $\exp\left[-N S_{\rm strong}(t)\right]$, here in the strong-coupling region. See Section \ref{sec:z-final} below for details.

\section{Tracy-Widom mapping to the Painlev\'e III equation}
\label{sec:pIII}

The previous Section showed that some features of the trans-series structure of the partition function are simply inherited from the trans-series structure (\ref{eq:bessel}) of the individual Bessel functions appearing in the Toeplitz determinant (\ref{eq:z1}), but that in the strong-coupling and large $N$ limits there are additional contributions which are rather difficult to probe using this determinant form. In order to probe the large $N$ limit in more detail we make use of classic results of Tracy and Widom from random matrix theory \cite{Tracy:1992rf,Tracy-Widom:1994,tracy-widom2,Hisakado:1996di,forrester-witte,forrester-book}, which relate the GWW partition function $Z(x, N)$ to the solution of a particular integrable non-linear ODE, the Painlev\'e III equation. This connection with the Painlev\'e III  equation holds for all $N$ and for all coupling, and so it provides a simple and direct probe of the GWW partition function $Z(x, N)$ for all $N$ and all $x$, far from the phase transition as well as close to it. Given an explicit differential equation, it is then a straightforward exercise to generate trans-series expansions \cite{Costin:2009}. This differential equation approach is complementary to the difference equation approach (based on the pre-string-equation) to trans-series developed for the GWW model in \cite{marino-matrix}, and for other matrix models in \cite{Marino:2007te}.

The Painlev\'e III equation is a differential equation with respect to the inverse coupling parameter $x$, with the unitary group index $N$ appearing as a parameter, so this can be used to define an analytic continuation in $N$ away from the positive integers, into the complex plane,\footnote{For interesting discussions of analytic continuation in $N$ in other systems, see \cite{Neuberger:2008ti,Couso-Santamaria:2015wga,Gukov:2016njj}.} which is necessary to understand fully the resurgent properties of the large $N$ expansion.

The Tracy-Widom mapping relation \cite{Tracy-Widom:1994} can be expressed by defining $s=x^2$, and
\begin{eqnarray}
E_N(s)\equiv e^{-s/4} Z\left(\sqrt{s}, N\right)
\label{eq:ez}
\end{eqnarray}
which effectively pulls out the leading exponential of the strong-coupling expansion (\ref{eq:z3}). This quantity is of interest for random matrix theory in connection with the fluctuations in the distribution of the largest eigenvalue in the Gaussian unitary ensemble (GUE) \cite{forrester-book}. Then, further defining the function $\sigma_N(s)$ via 
\begin{eqnarray}
E_N(s)\equiv \exp\left[-\int_0^s \frac{ds^\prime}{s^\prime} \sigma_N(s^\prime)\right]
\label{eq:e-sigma}
\end{eqnarray}
one finds that $\sigma_N(s)$ satisfies the Okamoto form of Painlev\'e III \cite{okamoto,okamoto2,Tracy-Widom:1994,forrester-witte,forrester-book}:
\begin{eqnarray}
\left(s\, \sigma_N^{\prime\prime}\right)^2+\sigma_N^\prime (\sigma_N-s\, \sigma_N^\prime)(4\sigma_N^\prime -1) -N^2 \left(\sigma_N^\prime\right)^2=0
\label{eq:pIII}
\end{eqnarray}
where the prime $'$ means differentiation with respect to $s$. 
This Painlev\'e III equation is satisfied for all $N$, and for all coupling (i.e. for all $s$). So, we can generate trans-series expansions for $\sigma_N(s)$ in any regime, and map them back to corresponding trans-series expansions for the partition function $Z(x, N)$ using (\ref{eq:ez}-\ref{eq:e-sigma}).

From this Painlev\'e III connection, we can then ``zoom in'' to the vicinity of the GWW phase transition region at $t_c=1$ using the known coalescence of the Painlev\'e III equation to the Painlev\'e II equation \cite{dlmf:ps}. This coalescence limit is the well-known double-scaling limit \cite{gw,wadia,Rossi:1996hs} taking $N\to \infty$, with the 't Hooft parameter $t\equiv N/x$ scaled in a particular way with $N$ (see Section \ref{sec:matching} below) close to the critical value $t_c=1$. In this case, as is also well known, the free energy is related to a particular solution (the Hastings-McLeod solution \cite{hastings,rosales}) of the Painlev\'e II equation (with zero parameter). Here we wish to study the full trans-series structure in both parameters, $x$ and $N$ (or $t$ and $N$), not just in the double-scaling region. The explicit Tracy-Widom mapping (\ref{eq:ez}-\ref{eq:e-sigma}) of $Z(x, N)$ to the Painlev\'e III equation (\ref{eq:pIII}) provides a simple way of generating the various trans-series expansions in all parameter regions. These can then be used to study how the different forms of the trans-series expansions re-arrange themselves through the GWW phase transition.

\subsection{Weak coupling expansion, for all $N$, from Painlev\'e III}

Given the explicit differential equation (\ref{eq:pIII}) for $\sigma_N(s)$, with $N$ appearing as a parameter, it is a straightforward exercise to develop weak coupling trans-series expansions in terms of the variable $s$. Tracy and Widom \cite{Tracy-Widom:1994} give the formal perturbative weak-coupling expansion for $E_N(s)$ as:
\begin{equation}
    E_N(s) = s^{-\frac{N^2}{4}} e^{-\frac{s}{4} + N\sqrt{s}} \left(1 + \frac{N}{8\sqrt{s}} + \frac{9 N^2}{128 s} + \left(\frac{3N}{128} + \frac{51N^3}{1024} \right) \frac{1}{s^\frac{3}{2}} + \left(\frac{75N^2}{1024} + \frac{1275N^4}{32768} \right) \frac{1}{s^2} + \ldots \right)
\end{equation}
Recalling that $s=x^2$, we recognize the fluctuation factor here as the perturbative fluctuation factor $\Gamma^{(0)}(x, N)$ in (\ref{eq:zweak}, \ref{eq:zweaka}). Further terms in the weak-coupling trans-series can be generated by inserting into the Painlev\'e III equation (\ref{eq:pIII}) for $\sigma_N(s)$ an weak-coupling trans-series ansatz. The perturbative weak-coupling expansion is
\begin{eqnarray}
\sigma_N^{\rm pert}(s)&\sim & \frac{s}{4}-\frac{N \sqrt{s}}{2}+\frac{N^2}{4} +\frac{N}{16 \sqrt{s}} +\frac{N^2}{16 s}+
\frac{\left(16 N^2+9\right) N}{256 s^{3/2}}+\frac{\left(4 N^2+9\right) N^2}{64 s^2} 
\nonumber\\
&&+\frac{\left(128 N^4+720 N^2+225\right) N}{2048
   s^{5/2}}+\frac{\left(4 N^4+45 N^2+54\right) N^2}{64 s^3}+ \dots
   \label{eq:sigma-weak-pert}
\end{eqnarray}
The linearized form of (\ref{eq:pIII}) reveals also non-perturbative exponentially small corrections, based on all powers of the basic exponential factor $e^{-2\sqrt{s}}$, leading to the trans-series ansatz:
\begin{eqnarray}
\sigma_N(s)\sim \frac{s}{4}-\frac{N\sqrt{s}}{2}+\frac{N^2}{4} +\sum_{k=0}^\infty (\xi_{\rm weak})^k \, d_k(\sqrt{s}, N) e^{-2 k \sqrt{s}} \sum_{n=0}^\infty \frac{\sigma_n^{(k)}(N)}{s^{n/2}}
\label{eq:sigmaweak}
\end{eqnarray}
where the weak-coupling trans-series parameter $\xi_{\rm weak}$ formally counts the ``instanton order'', and is set to 1 at the end to match the appropriate boundary conditions.
Inserting this ansatz into the differential equation (\ref{eq:pIII}) and expanding in powers of $\xi_{\rm weak}$ produces recurrence relations which iteratively determine the fluctuation coefficients, $\sigma_n^{(k)}(N)$, along with the prefactors, $d_k(\sqrt{s}, N)$, for each instanton sector $k$. The trans-series (\ref{eq:sigmaweak}) can then be mapped back to a trans-series expansion for the partition function $Z(x, N)$ using the relations (\ref{eq:ez})-(\ref{eq:e-sigma}). This confirms the general weak-coupling trans-series structure in (\ref{eq:z-transseries}).\footnote{A nontrivial point here is that the weak-coupling trans-series (\ref{eq:sigmaweak}) for $\sigma_N(s)$ has an infinite number of instanton exponentials, while as noted previously the weak-coupling trans-series (\ref{eq:z-transseries}) for $Z(\sqrt{s}, N)$ has only a finite number of instanton exponentials when $N$ is a positive integer.} For example, the leading $\frac{s}{4}$ term in (\ref{eq:sigmaweak}) corresponds to the $e^{-s/4}$ exponential factor in (\ref{eq:ez}); the subleading $-\frac{1}{2} N\sqrt{s}$ term in (\ref{eq:sigmaweak}) generates the $e^{N x}$ factor in $Z_0(x, N)$ in (\ref{eq:z0}); and the next $\frac{N^2}{4}$ term in (\ref{eq:sigmaweak}) generates the $x^{-N^2/2}$ factor in $Z_0(x, N)$ in (\ref{eq:z0}). The $N$-dependent overall normalization factor in $Z_0(x, N)$ is fixed by comparison with the explicit Toeplitz determinant expression (\ref{eq:z1}).

\subsection{Strong coupling expansion, for all $N$, from Painlev\'e III}

Tracy and Widom \cite{Tracy-Widom:1994} expressed the strong coupling (small $s$) expansion of $\sigma_N(s)$ as (here we add a few more terms)
\begin{equation}
\begin{aligned}
	\sigma_N(s) &\sim C_N s^{N+1} \bigg(1 - \frac{1}{2(N+2)}s + \frac{(2N+3) N!}{16 (N+3)!} s^2 -\frac{(2N+5) N!}{96 (N+4)!} s^3 + \frac{(2N+5)(2N+7) N!}{768 (2N+4)(N+5)!} s^4 +\ldots \bigg) \\
	&+ \frac{C^2_N s^{2N+2} }{N+1} \bigg(1 - \frac{2N+3}{2(N+2)^2} s + \frac{41 + 59 N + 27 N^2 + 
 4 N^3}{8 (2 + N)^3 (3 + N)^2}  s^2+ \ldots \bigg) + \frac{C^3_N s^{3N+3} }{(N+1)^2} \big(1 + \ldots \big) + \ldots
\end{aligned}
\label{eq:sigmastrong}
\end{equation}
where the numerical coefficient $C_N$ is defined as:
\begin{equation} 
	C_N = \frac{1}{4^{N+1} \Gamma(N+1) \Gamma(N+2)}=(N+1) \left(\frac{1}{2^{N+1} (N+1)!}\right)^2
	\label{eq:cn}
\end{equation}
As a cross-check, the reader can verify that the leading term, in the first parentheses in (\ref{eq:sigmastrong}), agrees with the first terms in (\ref{eq:z3}) when mapped back to the partition function using the relations (\ref{eq:ez})-(\ref{eq:e-sigma}). 

\begin{figure}[htb]
\centerline{
\includegraphics[scale=0.6]{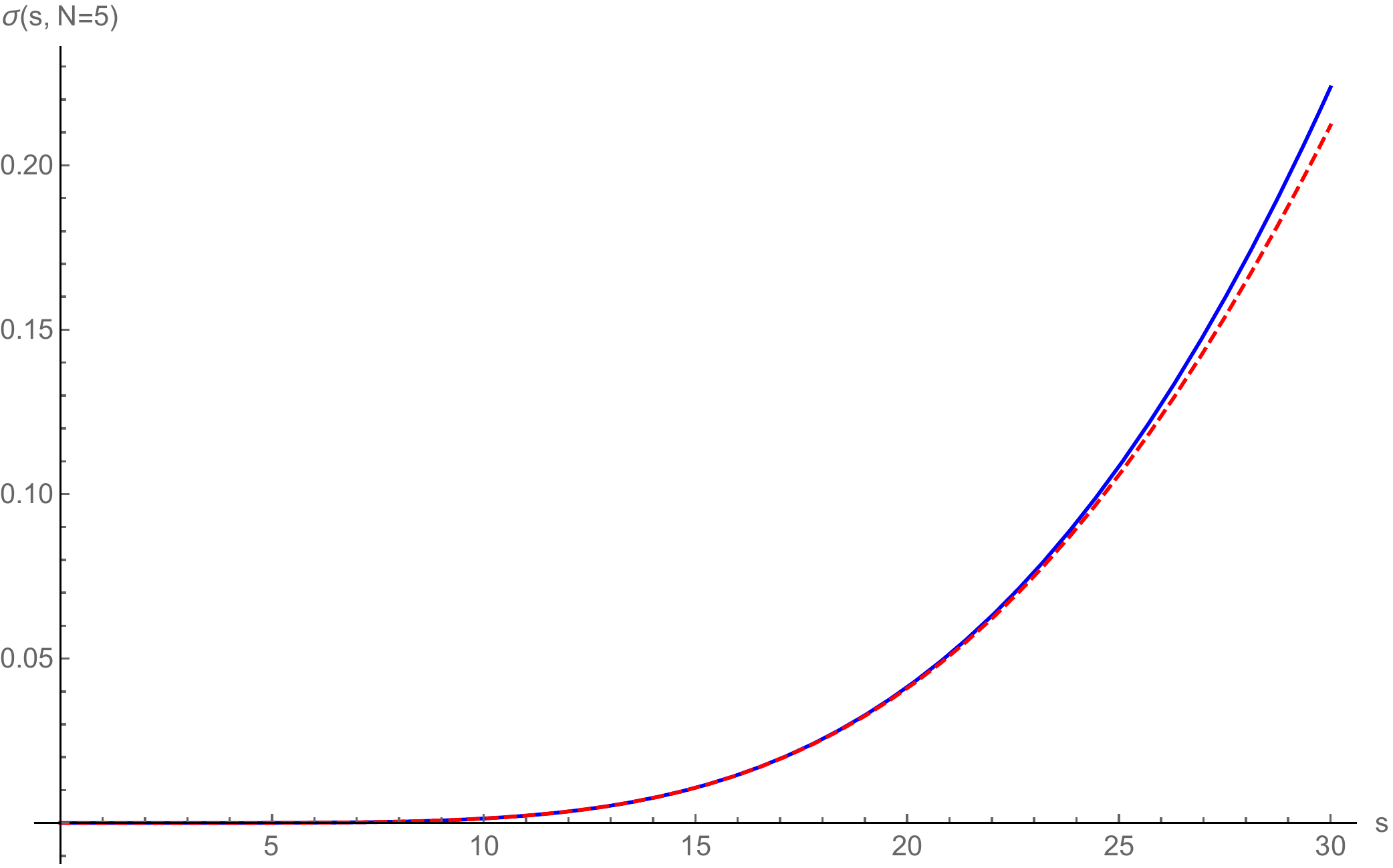}}
\caption{Comparison of the exact [solid blue curve] Tracy-Widom function $\sigma_N(s)$ in (\ref{eq:e-sigma}), for $N=5$, with the leading non-perturbative strong-coupling approximation in (\ref{eq:sigma-strong-leading-solution}). Recall that $s\equiv x^2$. The agreement is excellent at strong coupling ($s\to 0$), and also in the vicinity of the phase transition at $s=5^2$.}
\label{fig:fig3}
\end{figure}

We have found a new closed-expression for the leading term of $\sigma_N(s)$ in (\ref{eq:sigmastrong}). We differentiate (\ref{eq:pIII}) with respect to $s$ to find a simpler equation (the overall factor of $\sigma_N^{\prime\prime}$ may be dropped since the solution is monotonic for $s\geq 0$):
\begin{eqnarray}
2 s^2 \sigma_N^{\prime\prime\prime}(s)+2 s \sigma_N^{\prime\prime}(s)+
2 \left(s-N^2\right) \sigma_N^{\prime}(s)-\sigma_N(s) 
+8 \sigma_N(s) \sigma_N^\prime(s)-12 s (\sigma_N^\prime(s))^2=0
\label{eq:sigma-almost-linear}
\end{eqnarray}
The \textbf{linearized} part of this equation can be solved in closed form:
\begin{eqnarray}
&& 2 s^2\sigma_N^{\prime\prime\prime}(s) + 2s \sigma_N^{\prime\prime}(s)+ 2 \left(s-N^2\right) \sigma_N^{\prime}(s)-\sigma_N(s) = 0 
\nonumber\\
&\implies & \qquad \sigma_N^{({\rm linearized})}(s) = \text{constant}\times\frac{s}{4}\left(J_N(\sqrt{s})^2-J_{N-1}(\sqrt{s})J_{N+1}(\sqrt{s})\right)
\label{eq:sigma-strong-leading-solution}
\end{eqnarray}
Choosing the multiplicative constant to be 1, the small $s$ expansion agrees with the first parentheses term in the strong-coupling expansion (\ref{eq:sigmastrong}), to all orders. This makes it clear that this expansion is convergent, with an infinite radius of convergence. Figure \ref{fig:fig3} shows that at strong coupling (small $s$) the ``linearized'' solution in (\ref{eq:sigma-strong-leading-solution}) is an excellent approximation.
And yet, despite the fact that this expansion is convergent, there are extra non-perturbative terms in the strong-coupling trans-series (\ref{eq:sigmastrong}).\footnote{A similar phenomenon occurs in the re-arrangement of the convergent strong-coupling expressions for gap edges in the Mathieu equation, into trans-series expressions \cite{Basar:2015xna}.} 

Indeed, the remaining terms in (\ref{eq:sigmastrong}) should be understood as a strong-coupling trans-series for $\sigma_N(s)$:
\begin{eqnarray}
\sigma_N(s)
\sim 
(N+1) \sum_{k=1}^\infty (\xi^{\sigma}_{\rm strong})^k \left(\frac{C_N s^{N+1}}{N+1}\right)^{k} f^{(k)}_{\rm strong}(s,N) 
\label{eq:sigma-strong-structure}
\end{eqnarray}
in terms of strong-coupling instanton factors:
\begin{eqnarray}
\left(\frac{C_N s^{N+1}}{N+1}\right) =  \left(\frac{\left(\sqrt{s}\right)^{N+1}}{2^{N+1}\, (N+1)!}\right)^{2}
\label{eq:sigma-strong-instanton}
\end{eqnarray}
multiplied by (convergent!) fluctuation factors $f^{(k)}_{\rm strong}(s,N) $ in each instanton sector. In (\ref{eq:sigma-strong-structure}),  $\xi^{\sigma}_{\rm strong}$ is the strong-coupling trans-series parameter, whose value from the boundary conditions is determined to be $\xi^{\sigma}_{\rm strong}=1$.
We will understand more details about this unusual strong-coupling trans-series structure (\ref{eq:sigma-strong-structure}) in Section \ref{sec:delta} below, using properties of the Painlev\'e III equation.

\subsection{Large $N$ expansions in the 't Hooft limit, from Painlev\'e III}

We can convert these weak- and strong-coupling expansions into large $N$ expansions in the 't Hooft limit, as before. But a more direct way to generate such expansions is to rescale the Painlev\'e III equation (\ref{eq:pIII}) in terms of the 't Hooft parameter $t$:
\begin{eqnarray}
 \left(\left(t^3 \sigma^\prime \right)^\prime \right)^2 + 4 t \sigma^\prime \left(2 \sigma + t \sigma^\prime \right) \left(2 t^3 \sigma^\prime + N^2\right) - t^4 \left(\sigma^\prime \right)^2 = 0
 \label{eq:sigmat}
\end{eqnarray}
where the prime $'$ means differentiation with respect to $t$.
Then one can develop trans-series expansions with an appropriate ansatz form.
But we defer this step to Section \ref{sec:delta}, where we study the trans-series structure of an even simpler form of the Painlev\'e III equation, from which all these other trans-series can be derived in a much more explicit manner.

\section{Mapping to a simpler form of the Painlev\'e III equation}
\label{sec:simpler}

It turns out to be significantly easier to work with another function $\Delta(x, N)$:
\begin{eqnarray}
\Delta(x, N) \equiv \langle \det U \rangle = \frac{\det \left[ I_{j-k+1} \left(x\right)\right]_{j, k =1, \dots , N}}{\det \left[ I_{j-k} \left(x\right)\right]_{j, k =1, \dots , N}}
\label{eq:delta}
\end{eqnarray}
which is the expectation value of ${\rm det}\, U$ \cite{Rossi:1982vw}. The function $\Delta(x, N)$ is directly related to $\sigma_N(s)$ [see Equation (\ref{eq:sigma-delta1}) below], and also to the partition function $Z(x, N)$ [see Equation (\ref{eq:delta-z}) below], but it is easier to work with $\Delta(x, N)$ because it satisfies a much simpler nonlinear equation, for all $N$.

$\Delta(x, N)$ satisfies the following differential-difference and difference equations \cite{Rossi:1982vw,Rossi:1996hs} 
\begin{eqnarray}
2\Delta^\prime(x, N)&-&\left(1-\Delta^2(x, N)\right) \left(\Delta(x, N-1)-\Delta(x, N+1)\right) = 0
\label{eq:delta1a}
\\
\frac{2N}{x} \Delta(x, N)&-&\left(1-\Delta^2(x, N)\right) \left(\Delta(x, N-1)+\Delta(x, N+1)\right) = 0 
\label{eq:delta1b}
\end{eqnarray}
These two equations can be combined into a single nonlinear differential equation for $\Delta(x, N)$, which we refer to as Rossi's equation \cite{Rossi:1982vw,Rossi:1996hs}:
\begin{eqnarray}
\Delta^{\prime\prime}(x, N)+\frac{1}{x}\Delta^{\prime}(x, N) + \Delta(x, N) \left(1-\Delta^2(x, N)\right)+
\frac{\Delta(x, N)}{\left(1-\Delta^2(x, N)\right)}\left[ \left(\Delta^{\prime}(x, N)\right)^2 -\frac{N^2}{x^2}\right] =0
\label{eq:rossi}
\end{eqnarray}
Note that this Rossi equation (\ref{eq:rossi}) is valid for all $x$ and all $N$; i.e., for all $N$, and for all coupling.

From results of Tracy and Widom \cite{Tracy-Widom:1994}, 
it can be shown that there is a simple relation between $\Delta(x, N)$ and $\sigma_N(s)$, and therefore also between $\Delta(x, N)$ and the partition function $Z(x, N)$. To make this relation completely explicit, define, for all $N$ and $s$ (recall the identification: $s=x^2$): 
\begin{eqnarray}
\sigma_N(s)=\frac{N^2}{4}+\frac{x^2}{4}\left(\Delta^2(x, N)+ \frac{1}{\left(1-\Delta^2(x, N)\right)}\left[ \left(\partial_x \Delta(x, N)\right)^2-\frac{N^2}{x^2}\right]\right)
\label{eq:sigma-delta1}
\end{eqnarray}
If $\Delta(x, N)$ satisfies the Rossi equation (\ref{eq:rossi}),  then $\sigma_N(s)$ defined in (\ref{eq:sigma-delta1}) satisfies Okamoto's Painlev\'e III equation (\ref{eq:pIII}), with the identification: $s=x^2$. Therefore, a trans-series expansion for $\Delta(x, N)$ can immediately be converted into a trans-series expansion for $\sigma_N(s)$, and hence for the partition function $Z(x, N)$, for all $N$ and for any coupling.

The relation between $\Delta(x, N)$ and $\sigma_N(s)$ can also be expressed in difference form as
\begin{eqnarray}
\sigma_N(s)=\frac{x^2}{4}\left(\Delta^2(x, N)-
\left(1-\Delta^2(x, N)\right)\Delta(x, N-1)\Delta(x, N+1)\right)
\label{eq:sigma-delta2}
\end{eqnarray}
Furthermore, $\Delta(x, N)$ is also related directly to the partition function $Z(x, N)$ as:
\begin{eqnarray}
\Delta^2(x, N)=1-\frac{Z(x, N-1)\, Z(x, N+1)}{Z^2(x, N)}
\label{eq:delta-z}
\end{eqnarray}
Finally, there is yet another (related) connection to the Painlev\'e III equation: 
the Rossi equation (\ref{eq:rossi}) can be converted to the Painlev\'e V equation \cite{dlmf:ps} by defining
\begin{eqnarray}
c(s, N)=1-\frac{1}{\Delta^2(\sqrt{s}, N)}
\label{eq:c}
\end{eqnarray}
Then $c(s, N)$ satisfies the Painlev\'e V equation, with parameters $\alpha=0, \beta=-N^2/2, \gamma=1/2, \delta=0$, and for such parameters, this can be converted  to Painlev\'e III \cite{okamoto,okamoto2}.

\section{Trans-series expansions for $\Delta(x, N)\equiv \langle \det U \rangle$}
\label{sec:delta}

The main advantage of working with the function $\Delta(x, N)\equiv \langle {\rm det}\, U\rangle$ defined in (\ref{eq:delta}), instead of $\sigma_N(s)$ or the partition function $Z(x, N)$, is that the Rossi equation (\ref{eq:rossi}) is much simpler than Okamoto's Painlev\'e III equation (\ref{eq:pIII}). In particular, the second derivative term is {\it linear} in (\ref{eq:rossi}), but {\it quadratic} in (\ref{eq:pIII}). This fact leads to dramatic simplifications, especially when considering the strong-coupling and large $N$ limits. And since the partition function can be deduced immediately from $\Delta(x, N)$ using the relations (\ref{eq:ez})-(\ref{eq:e-sigma}), or from (\ref{eq:delta-z}), the trans-series structure of $Z(x, N)$  is inherited from that of $\Delta(x, N)$ in a straightforward way.

\subsection{Weak coupling expansion for $\Delta(x, N)$, at fixed $N$}

The weak coupling (large $x$) expansion of $\Delta(x, N)$ could be obtained by direct expansion of the determinants in (\ref{eq:delta}), analogous to the treatment of the partition function in Section \ref{sec:z-weak1}. However, a much simpler method is to insert an appropriate trans-series ansatz into the Rossi differential equation (\ref{eq:rossi}) and match terms. We find a trans-series
\begin{eqnarray} \label{eq:DeltaFiniteNWeakExp}
\Delta(x, N)&\sim& \sum_{k=0}^\infty \left(\xi_{\rm weak}^{\Delta}\right)^k C_k(x, N) e^{-2k x}\sum_{n=0}^\infty \frac{A_n^{(k)}(N)}{x^n} \nonumber\\
&\sim & \sum_{n=0}^\infty \frac{A_n^{(0)}(N)}{x^n} -2 i\, \xi_{\rm weak}^{\Delta}\,  \frac{(4x)^{N-1}}{(N-1)!} \, e^{-2x} \sum_{n=0}^\infty \frac{A_n^{(1)}(N)}{x^n} +\dots 
\end{eqnarray}
with fluctuation series:
\begin{eqnarray}
\sum_{n=0}^\infty \frac{A_n^{(0)}(N)}{x^n}&=& 
1-\frac{N}{2 x}-\frac{N^2}{8 x^2}  -\frac{N(N^2+1)}{16 x^3} -\frac{5 N^2 ( N^2 + 4)}{128 x^4} -\frac{N (7 N^4 + 70 N^2 + 27)}{256 x^5}  - \dots 
\label{eq:delta-a0}
\\
\sum_{n=0}^\infty \frac{A_n^{(1)}(N)}{x^n}&=& 
1-\frac{(N^2-N+1)}{4 x} + \frac{\left(N^4-4 N^3+7 N^2-12 N+5\right)}{32 x^2} 
\nonumber\\ 
&& \qquad -\frac{\left(N^6-9 N^5+34 N^4-87 N^3+202 N^2-165 N+63\right)}{384 x^3}   -\dots
\label{eq:delta-a1}
\end{eqnarray}

\begin{figure}[htb]
    \centering
    \includegraphics[scale=1.35]{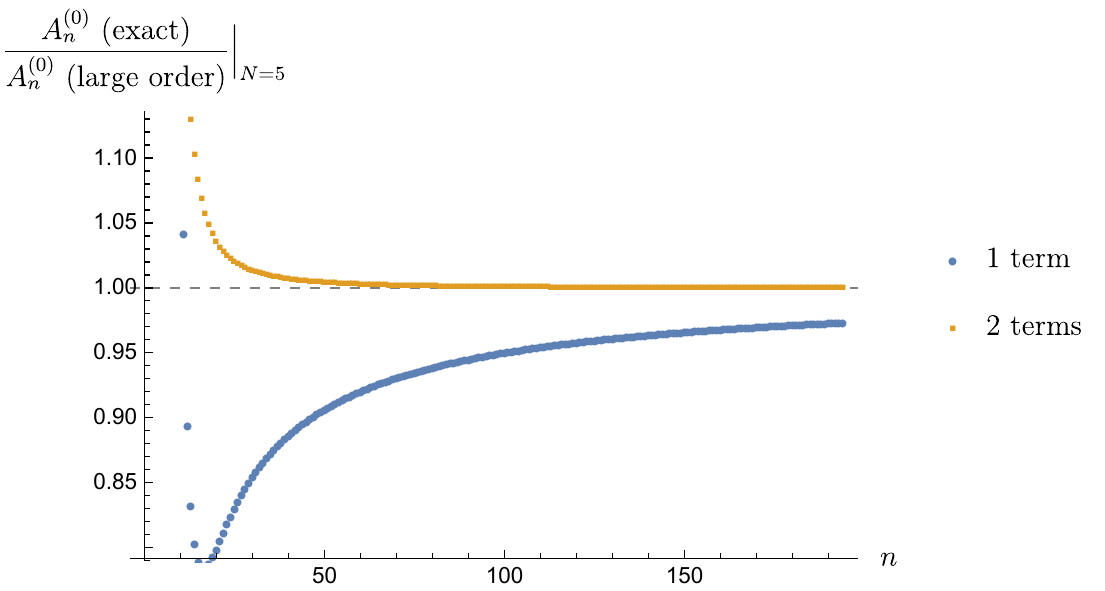}
    \caption{Ratio of the exact perturbative coefficients $A_n^{(0)}(N)$ from (\ref{eq:delta-a0}), divided by the resurgent large-order growth expression (\ref{eq:Delta0InstantonGrowth}). Both plots are for $N=5$, and include just the first term in (\ref{eq:Delta0InstantonGrowth}) [blue circles] or the first two terms [gold squares]. Both sequences have been accelerated using Aitken Extrapolation.}
    \label{fig:DeltaFiniteNWeakCouplingGrowth0Inst}
\end{figure}

This weak-coupling trans-series structure for $\Delta(x, N)$ 
clearly matches the weak-coupling trans-series form in (\ref{eq:z-transseries}), which was deduced from the Toeplitz determinant expression (\ref{eq:z1}).
As was the case with the partition function in Section \ref{sec:z-weak1}, we observe resurgence relations in the large order behavior of the fluctuation coefficients appearing in the weak-coupling trans-series (\ref{eq:DeltaFiniteNWeakExp}) for $\Delta(x, N)$. For example, for a given $N$, the perturbative coefficients in (\ref{eq:delta-a0}) have the large order growth (see Figure \ref{fig:DeltaFiniteNWeakCouplingGrowth0Inst}):
\begin{equation} \label{eq:Delta0InstantonGrowth}
		A_n^{(0)}(N) \sim - \frac{2^{N+1}}{\pi (N-1)!}\frac{(n+N-3)!}{2^n }  \left[1 - \frac{(N^2-N+1)}{4} \frac{2}{(n+N-3)} + \dots\right]
\end{equation}
We recognize typical resurgent behavior: the low order coefficients $A_n^{(1)}(N)$ of the one-instanton fluctuation series (\ref{eq:delta-a1}) appear here in this large-order behavior of the fluctuation coefficients $A_n^{(0)}(N)$ of the perturbative zero-instanton sector. This continues at higher order instanton sectors, consistent with general theorems for resurgence for nonlinear ODEs \cite{costin-duke}.

\subsection{Strong coupling expansion for $\Delta(x, N)$, at fixed $N$}
\label{sec:delta-strong}

Recall that the strong-coupling (small $x$) expansion of the partition function $Z(x, N)$, deduced in (\ref{eq:z3}) by expansion of the Toeplitz determinant expression (\ref{eq:z1}), does not have an obvious trans-series completion at strong coupling, while the Tracy-Widom result (\ref{eq:sigmastrong}) for $\sigma_N(s)$ suggests a suitable strong-coupling trans-series form (\ref{eq:sigma-strong-structure}). However, for $\Delta(x, N)$ this is a much simpler question, because $\Delta(x, N)$ satisfies a relatively simple differential equation (\ref{eq:rossi}).

In the strong coupling limit we notice from explicit computations that 
$\Delta(x, N)\sim \frac{(x/2)^N}{N!}$. Therefore, in this regime we can linearize the differential-difference and difference equations (\ref{eq:delta1a}, \ref{eq:delta1b}):
\begin{eqnarray}
2\Delta^\prime(x, N)&\approx &
\left(\Delta(x, N-1)-\Delta(x, N+1)\right) 
\label{eq:delta2a}
\\
\frac{2N}{x} \Delta(x, N)&\approx&
\left(\Delta(x, N-1)+\Delta(x, N+1)\right) 
\label{eq:delta2b}
\end{eqnarray}
We recognize these as the defining relations for the Bessel functions. Correspondingly, in this limit the nonlinear equation (\ref{eq:rossi}) linearizes to become the Bessel equation \cite{Rossi:1982vw,Rossi:1996hs}:
\begin{eqnarray}
\Delta^{\prime\prime}(x, N)+\frac{1}{x}\Delta^{\prime}(x, N)+\Delta(x, N) -\frac{N^2}{x^2} \Delta(x, N) \approx 0
\label{eq:deltaeq2}
\end{eqnarray}
Thus, the leading strong coupling behavior is given by 
\begin{eqnarray}
\Delta_{(1)}(x, N)\approx \xi_{\rm strong}^{\Delta}\, J_N(x)+\dots 
\label{eq:leading-delta-strong}
\end{eqnarray}
where the overall multiplicative constant  $\xi_{\rm strong}^{\Delta}$ must be chosen to be $\xi_{\rm strong}^{\Delta}=1$ in order to match with a direct computation of the strong-coupling expansion of the determinant ratio in (\ref{eq:delta}), as well as to match the strong-coupling expansion of the partition function in (\ref{eq:z3}), via (\ref{eq:ez}) and (\ref{eq:e-sigma}) [or via (\ref{eq:delta-z})].\footnote{This choice $\xi_{\rm strong}^{\Delta}=1$ is also precisely analogous to the Hastings-McLeod choice of unit coefficient of the Airy function in the asymptotics of the solution of Painlev\'e II \cite{rosales,hastings}, for the double-scaling limit, as discussed in Section \ref{sec:matching}.} 
Notice that already this leading-order term, $\Delta_{(1)}(x, N)$, in the strong-coupling expansion is a non-perturbative ``one-instanton" contribution: there is no ``perturbative'' contribution to $\Delta(x, N)$. 
Using the relation (\ref{eq:sigma-delta2}) we deduce from (\ref{eq:leading-delta-strong}) that the leading strong-coupling expression for $\sigma_N(s)$ agrees precisely with the expression in (\ref{eq:sigma-strong-leading-solution}), which was obtained by solving the linearized part of (\ref{eq:sigma-almost-linear}). 
We can then use the relation (\ref{eq:delta-z}) to derive a closed-form expression for the leading strong-coupling expansion contribution to $\ln Z(x, N)$:
\begin{eqnarray}
\ln Z(x, N)\Big|_{\text{leading}} &\sim & 
\frac{x^2}{4}-\frac{x^2}{4}\left[ \left(J_N(x)\right)^2
+\left(J_{N+1}(x)\right)^2
-\frac{2N}{x}J_N(x) J_{N+1}(x)\right.
\nonumber\\
   && \hskip -2cm 
   \left. -\frac{2^{-2 N} N x^{2 N}} {\Gamma (N+2)^2} \, _3F_4\left(N+\frac{1}{2},N+1,N+1;N,N+2,N+2,2 N+1;-x^2\right)\right]
 \label{eq:lnz-strong-leading}
\end{eqnarray}
The first ``perturbative'' term $\frac{x^2}{4}$ is general, simply coming from the relation (\ref{eq:ez}) between $Z(x, N)$ and $E_N(s)$ [recall that $s\equiv x^2$]. The leading correction term in (\ref{eq:lnz-strong-leading}) can be re-written as a sum, $-\sum_{l=1}^\infty l\, \left(J_{N+l}(x)\right)^2$, an expression which had been observed empirically to order $x^{4N+2}$ in \cite{Rossi:1996hs}. We now show that these higher power corrections should be understood as the non-perturbative trans-series completion of the strong-coupling expansion.

The parameter $\xi_{\rm strong}^{\Delta}$ also has an interpretation as a strong-coupling trans-series parameter, which effectively counts the instanton sectors, in addition to imposing the appropriate physical boundary condition. Therefore, returning to the full nonlinear Rossi equation (\ref{eq:rossi}), we find a strong-coupling (small $x$) trans-series ansatz of the form (with $\xi_{\rm strong}^{\Delta}$ set equal to 1 at the end):
\begin{eqnarray}
\Delta(x, N)=\sum_{k=1, 3, 5, \dots}^\infty (\xi_{\rm strong}^{\Delta})^k \Delta_{(k)}(x, N)
\label{eq:delta-strong}
\end{eqnarray}
The trans-series (\ref{eq:delta-strong}) is a sum over all odd instanton powers in terms of $\xi_{\rm strong}^{\Delta}$.
Matching these powers of $\xi_{\rm strong}^{\Delta}$ converts Rossi's nonlinear equation (\ref{eq:rossi}) into a tower of {\it linear} equations, the first of which is the Bessel equation (\ref{eq:deltaeq2}) satisfied by $\Delta_{(1)}(x, N)$, and the second of which is the inhomogeneous Bessel equation for $\Delta_{(3)}(x, N)$:
\begin{eqnarray}
\Delta_{(3)}^{\prime\prime}+\frac{1}{x} \Delta_{(3)}^\prime +\left(1-\frac{N^2}{x^2}\right)\Delta_{(3)}
&=&h_{(3)}(x, N)
\label{eq:delta-next}
\end{eqnarray}
with inhomogeneous term
\begin{eqnarray}
h_{(3)}(x, N)&=&
\left(\Delta_{(1)}(x, N)\right)^3\left(1+\frac{N^2}{x^2}\right) -\Delta_{(1)}(x, N)\left(\Delta_{(1)}^\prime(x, N)\right)^2
\nonumber\\
&=&\left(J_N(x)\right)^3\left(1+\frac{N^2}{x^2}\right) -J_N(x)\left(J_N^\prime(x)\right)^2
\label{eq:delta-next-h}
\end{eqnarray}
\begin{figure}[htb]
    \centering
    \includegraphics[scale=.6]{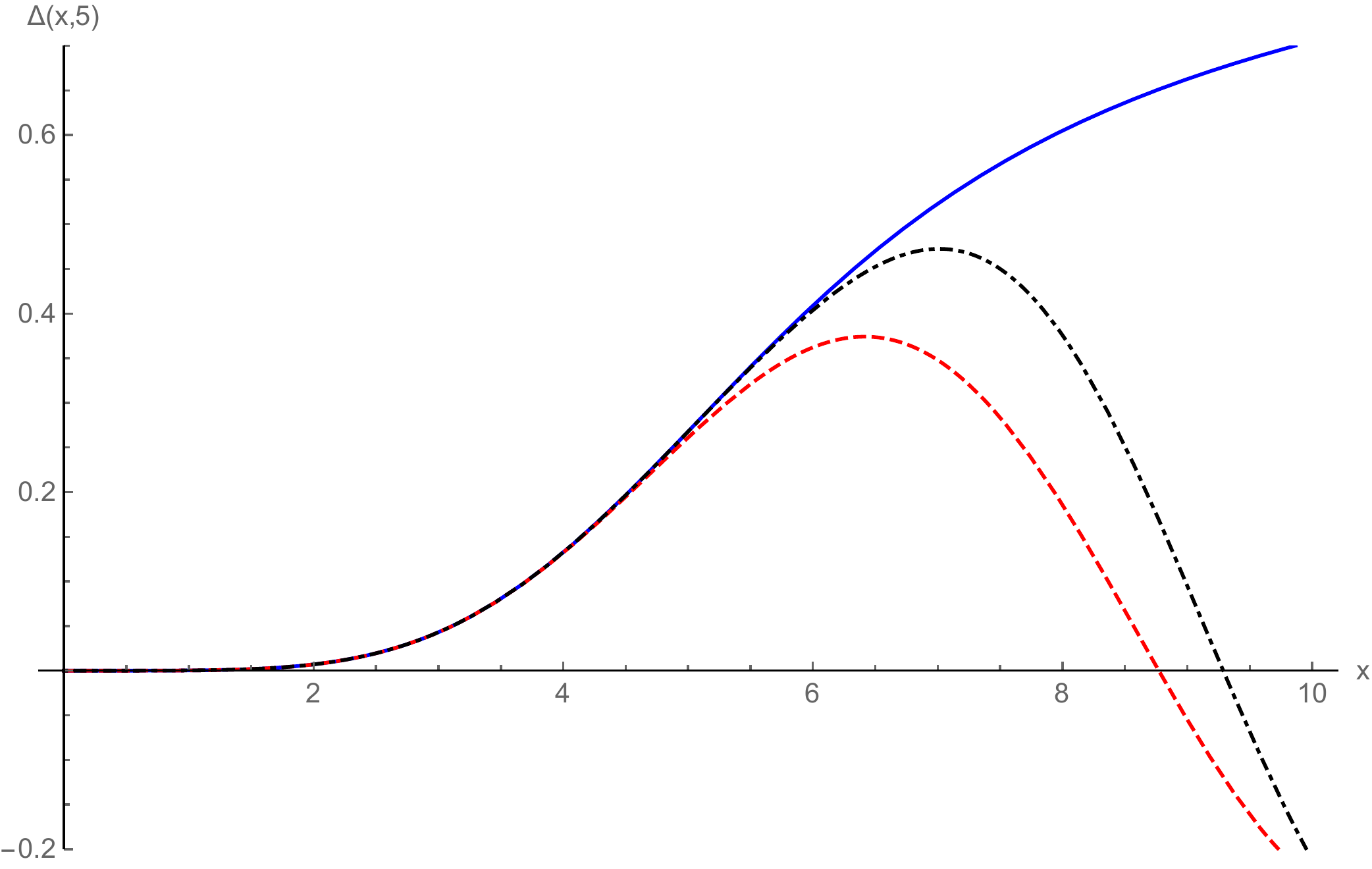}
    \caption{Comparison plots of the exact $\Delta(x, N)$ [solid blue curve] as a function of $x$, with $N=5$, compared to the inclusion of successive terms in the strong-coupling trans-series expansion (\ref{eq:delta-strong}), with $\xi_{\rm strong}^{\Delta}$ taking its physical value: $\xi_{\rm strong}^{\Delta}=1$. The agreement is excellent in the strong-coupling $x\to 0$ region, and also in the transition region $x\approx N$, and improves progressively as more terms in the instanton sum are included.}
    \label{fig:fig5}
\end{figure}
Since the homogeneous part of this equation is just the Bessel operator, we can immediately write the exact solution to (\ref{eq:delta-next}) as
\begin{eqnarray}
\Delta_{(3)}(x, N)=-\frac{\pi}{2}\, J_N(x) \int_0^x dx^\prime \, x^\prime\, Y_N(x^\prime) h_{(3)}(x^\prime, N) + \frac{\pi}{2}\,Y_N(x) \int_0^x dx^\prime \,x^\prime\, J_N(x^\prime) h_{(3)}(x^\prime, N)
\label{eq:delta-next-solution}
\end{eqnarray}
where $h_{(3)}$ is the source term in  (\ref{eq:delta-next-h}), and the factor $\frac{\pi x^\prime}{2}$ arises from the Bessel Wronskian. 
From (\ref{eq:delta-next-solution})  we deduce the (convergent) strong-coupling expansion of this next correction term:
\begin{eqnarray}
\Delta_{(3)}(x, N)
&\sim &\frac{x^{3N+2}}{2^{3N+2} \Gamma(N+1) \left(\Gamma(N+2)\right)^2}\left(1-\frac{(3N+5)}{4(N+2)^2} x^2+\dots \right) \label{eq:delta-next-strong}\\
&\sim& \left(\frac{x}{2(N+1)}\right)^2 \left(J_N(x)\right)^3 
\left(1+\frac{(4 N+7) \,x^2}{4 (N+1) (N+2)^2}+
\frac{(N (8 N+37)+41)\, x^4}{8 (N+1)^2 (N+2)^2(N+3)^2}+\dots\right) 
\nonumber
\end{eqnarray}
This expansion of $\Delta_{(3)}(x, N)$ maps to the second part of the Tracy-Widom strong coupling expansion for $\sigma_N(s)$ in (\ref{eq:sigmastrong}). Also notice that while $\Delta_{(1)}(x, N)$ is linear in the Bessel function $J_N(x)$, the next term in the trans-series, $\Delta_{(3)}(x, N)$, is essentially cubic in $J_N(x)$, multiplied by a convergent fluctuation factor. Figure \ref{fig:fig5} shows the $x$ dependence, for $N=5$, of the exact $\Delta(x, N)$ computed from (\ref{eq:delta}), compared to the leading non-perturbative expression, $\Delta_{(1)}(x, N)$ in (\ref{eq:leading-delta-strong}), and also including the the next contribution, $\Delta_{(3)}(x, N)$ in (\ref{eq:delta-next-solution}). The agreement is excellent, even in the transition region where $x\approx N$.

\begin{figure}[htb]
    \centering
    \includegraphics[scale=.6]{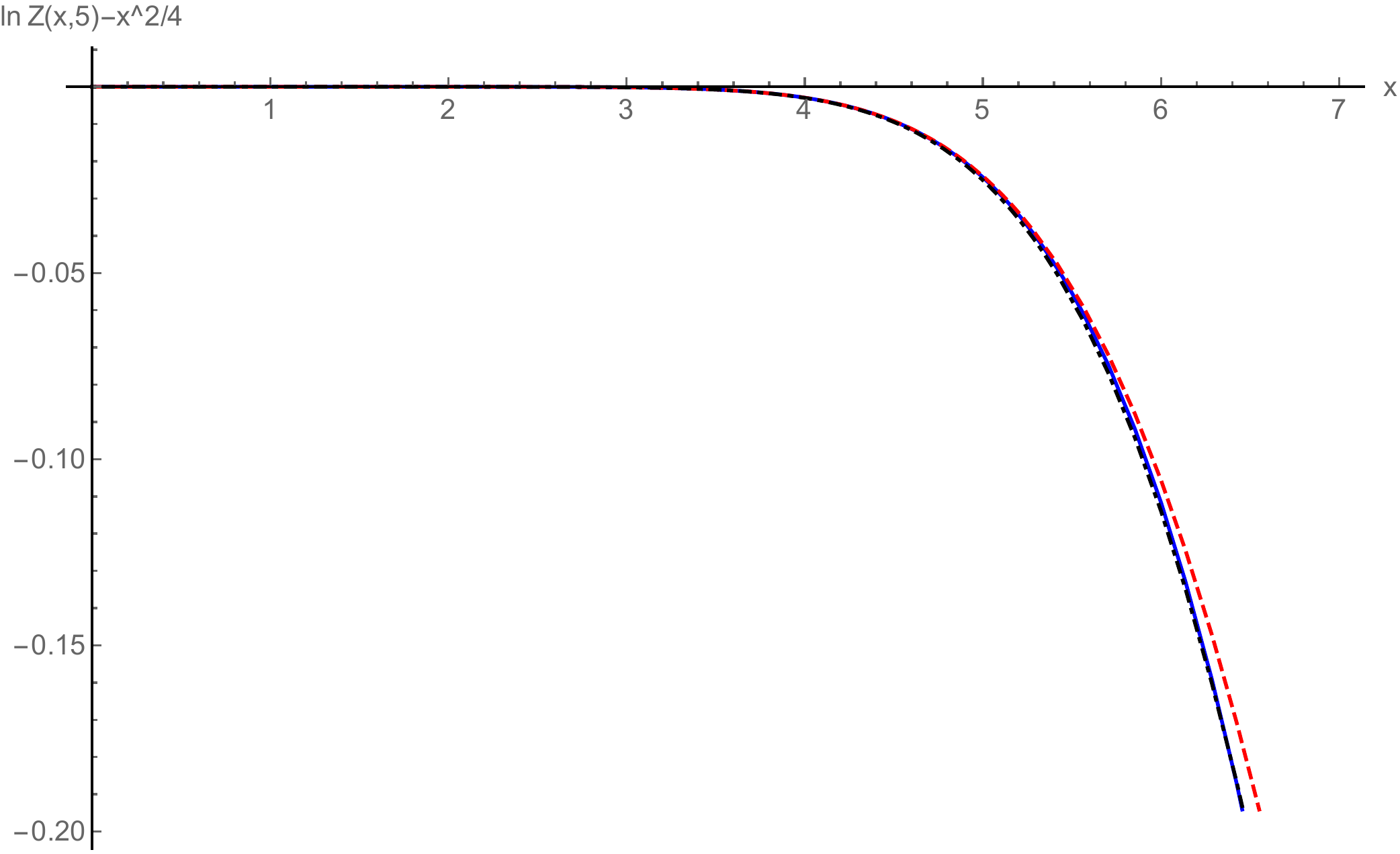}
    \caption{Comparison plots of the exact $\ln Z(x, N)$, with the trivial $x^2/4$ term subtracted [solid blue curve], as a function of $x$, for $N=5$, compared to successive terms in the strong-coupling trans-series expansion. The red dashed curve includes just the first non-trivial term, from (\ref{eq:lnz-strong-leading}), while the black dot-dashed curve includes also the next term in (\ref{eq:lnz-strong-sub-leading}). The agreement is excellent in the strong-coupling $x\to 0$ region, even in the transition region $x\approx N$, and improves progressively as more terms in the instanton sum are included.}
    \label{fig:fig6}
\end{figure}

Given the sub-leading term $\Delta_{(3)}(x, N)$ in (\ref{eq:delta-next-solution}), we can immediately deduce the associated sub-leading term for $\ln Z(x, N)$:
\begin{eqnarray}
\ln Z(x, N)\Big|_{\text{sub-leading order}} &\sim & 
\frac{x^2}{4}-\frac{x^2}{4}\left[ \left(J_N(x)\right)^2
+\left(J_{N+1}(x)\right)^2
-\frac{2N}{x}J_N(x) J_{N+1}(x)\right.
\nonumber\\
   && \hskip -1cm 
   \left. -\frac{2^{-2 N} N x^{2 N}}{\Gamma (N+2)^2} \, _3F_4\left(N+\frac{1}{2},N+1,N+1;N,N+2,N+2,2 N+1;-x^2\right)\right]
   \nonumber\\
   && \hskip -1cm 
   -\frac{1}{2} \int_0^x  dx^\prime  \, x^\prime \, \left[(J_N(x^\prime))^2 J_{N-1}(x^\prime) J_{N+1}(x^\prime) +2 J_N(x^\prime) \Delta_{(3)}(x^\prime, N)
   \right.\nonumber\\
   && \left. - J_{N-1}(x^\prime) \Delta_{(3)}(x^\prime, N+1) - J_{N+1}(x^\prime) \Delta_{(3)}(x^\prime, N-1)\right]
   \label{eq:lnz-strong-sub-leading}
\end{eqnarray}
While the first non-perturbative correction in (\ref{eq:lnz-strong-leading}) is quadratic in $J_N(x)$, the sub-leading term in (\ref{eq:lnz-strong-sub-leading}) is quartic in $J_N(x)$. Figure \ref{fig:fig6} shows the $x$ dependence, for $N=5$, of the exact $\ln Z(x, N)$ computed from (\ref{eq:z}), compared to the leading non-perturbative expression in (\ref{eq:lnz-strong-leading}), and also including the the next contribution in (\ref{eq:lnz-strong-sub-leading}). The agreement is excellent, even in the transition region where $x\approx N$.

Even though the correction $\Delta_{(3)}(x, N)$ in (\ref{eq:delta-next-strong}) looks like it contributes  innocent (and convergent) further powers of $x$ in the strong coupling expansion, this analysis shows that it is really the next term in a trans-series expansion (\ref{eq:delta-strong}). Similarly, the next term, $\Delta_{(5)}(x, N)$, in the strong-coupling trans-series (\ref{eq:delta-strong}) satisfies an inhomogeneous linear (Bessel) equation
\begin{eqnarray}
\Delta_{(5)}^{\prime\prime}+\frac{1}{x} \Delta_{(5)}^\prime +\left(1-\frac{N^2}{x^2}\right)\Delta_{(5)}=h_{(5)}(x,N)
\label{eq:delta-next-next}
\end{eqnarray}
with inhomogeneous term
\begin{equation}
    \begin{aligned}
        h_{(5)}(x,N) &=\frac{N^2 }{x^2} \Delta_{(1)}^2 \left(\Delta_{(1)}^3 + 3 \Delta_{(3)} \right) -  2 \Delta_{(1)} \Delta_{(1)}^\prime \Delta_{(3)}^\prime + 3 \Delta_{(1)}^2 \Delta_{(3)} - \left(\Delta_{(1)}^\prime \right)^2 \left(\Delta_{(1)}^3 + \Delta_{(3)} \right) 
    \end{aligned}
    \label{eq:h5}
\end{equation}
The homogeneous part involves the Bessel operator, so the solution has the same form as (\ref{eq:delta-next-solution}), but with the source term $h_{(3)}(x^\prime, N)$  replaced now by $h_{(5)}(x^\prime, N)$ from (\ref{eq:h5}). Given $\Delta_{(5)}(x, N)$, we can deduce the next contribution to the trans-series expansion for $\ln Z(x, N)$, as a nested integral.
This structure clearly continues  at higher orders: all higher terms of the strong-coupling trans-series (\ref{eq:delta-strong}) can be written as solutions to an inhomogeneous Bessel equation, with source terms expressed in terms of the previous trans-series solutions. 

In general, this suggests expressing the strong-coupling trans-series as
\begin{eqnarray}
\Delta(x, N)\sim \sum_{k=1, 3, 5, \dots}^\infty \left(\xi_{\rm strong}^{\Delta}\right)^k\,   P^{(k)}_\Delta(x, N)\, \left(J_{N}(x)\right)^{k}\, f^{(k), \Delta}_{\rm strong}(x, N)
\label{eq:delta-final}
\end{eqnarray}
in terms of odd powers of Bessel functions $J_N(x)$, multiplied by prefactors, $P^{(k)}_\Delta(x, N)$, and convergent fluctuation factors, $f^{(k), \Delta}_{\rm strong}(x, N)$. Correspondingly, 
the strong-coupling trans-series for $\ln Z(x, N)$ takes the form
\begin{eqnarray}
\ln Z(x, N)\sim \frac{x^2}{4}+\sum_{k=2, 4, 6, \dots}^\infty \left(\xi_{\rm strong}^{Z}\right)^k\,   P^{(k)}_Z(x, N)\, \left(J_{N}(x)\right)^{k}\, f^{(k), Z}_{\rm strong}(x, N)
\label{eq:lnz-final}
\end{eqnarray}
in terms of even powers of Bessel functions $J_N(x)$, multiplied by prefactors, $P^{(k)}_Z(x, N)$, and convergent fluctuation factors, 
$f^{(k), Z}_{\rm strong}(x, N)$.
This structure is particularly well-suited for studying the large $N$ 't Hooft limit at strong-coupling, as discussed below in Section \ref{sec:delta-largeN-strong}.

\subsection{Large $N$ expansions for $\Delta(t, N)$ in the 't Hooft limit}
\label{sec:delta-largeN}

\begin{figure}[htb]
\centering
\includegraphics[scale=.6]{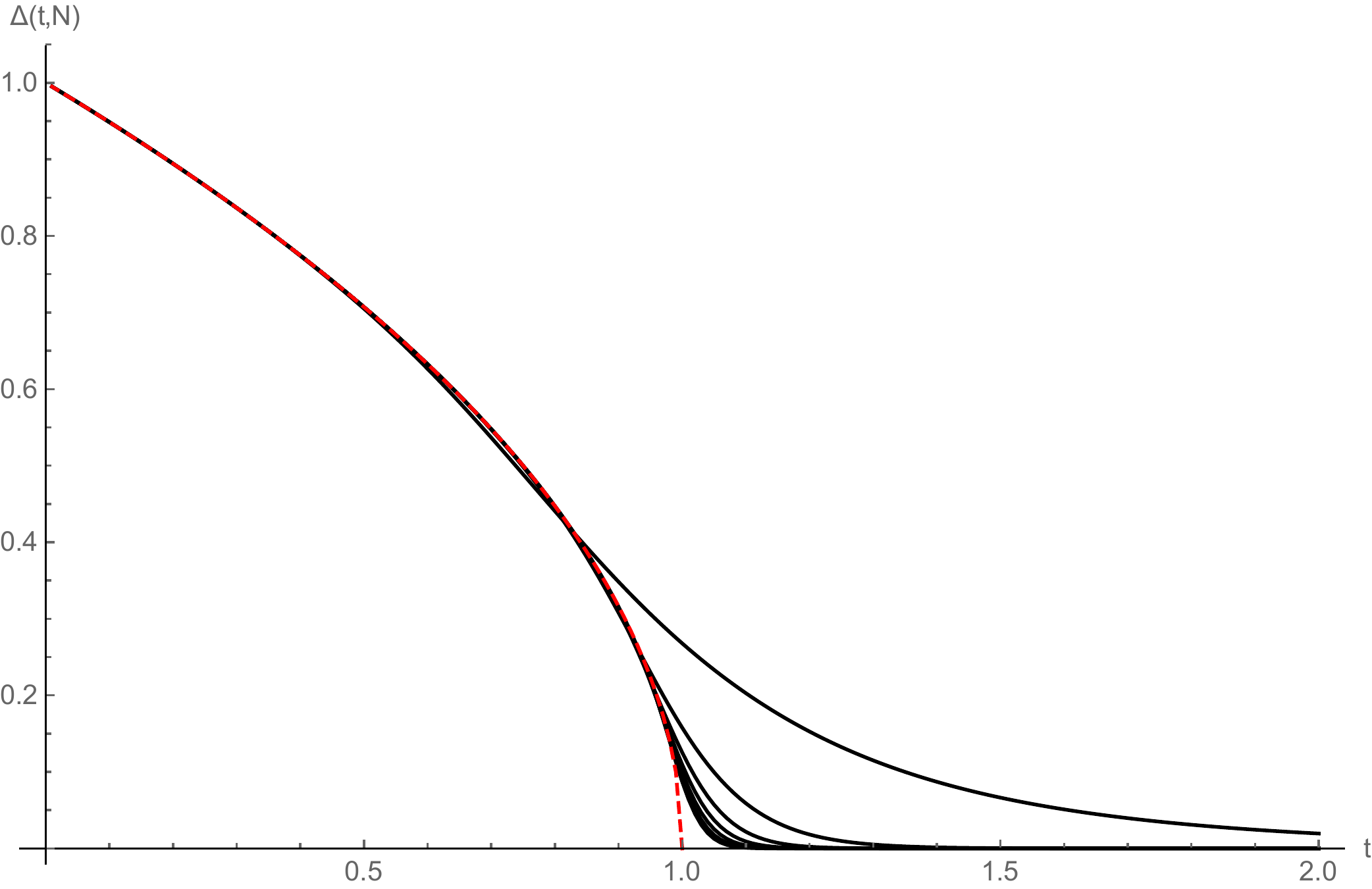}
\caption{The dependence of $\Delta(t, N)\equiv \langle \det U\rangle$ on the 't Hooft coupling $t$, for various values of $N$. The values of $N$ plotted are $N=5, 25, 50, 75, 100, 125, 150$, with the black curves lowering monotonically towards the $N=\infty$ result in (\ref{eq:kink}). The red dashed curve shows the infinite $N$ weak coupling form, while at $N=\infty$, $\Delta$ vanishes at strong coupling, for all $t\geq 1$.
}
\label{fig:kink}
\end{figure}

To generate the large $N$ expansions in the 't Hooft limit, we rescale the Rossi equation (\ref{eq:rossi}) to express it in terms of the 't Hooft parameter $t=\frac{N}{x}$:
\begin{eqnarray}
t^2 \Delta^{\prime\prime} +t \Delta^\prime + \frac{N^2 \Delta}{t^2}\left(1- \Delta^2 \right)  = \frac{\Delta}{1-\Delta^2}\left(N^2 -t^2 \left(\Delta^\prime\right)^2\right)
\label{eq:deltat}
\end{eqnarray}
The GWW phase transition in the large $N$ limit can be seen directly in the behavior of $\Delta(t, N)$ for large $N$. Figure \ref{fig:kink} shows $\Delta(t, N)$ as a function of 't Hooft coupling $t$ for various values of $N$: $N=5, 25, 50, 75, 100, 125, 150$. We see that as $N\to\infty$: 
\begin{eqnarray}
\Delta(t, N) \xrightarrow{\: N \to \infty \: } \begin{cases}
0\qquad\quad, \quad t\geq 1 \quad (\text{strong coupling})
\cr\sqrt{1-t} \quad, \quad t\leq 1 \quad (\text{weak coupling})
\end{cases}
\label{eq:kink}
\end{eqnarray}
This discontinuous derivative of $\Delta(t, N)$ at $N=\infty$ and $t=1$ is the signal of the third-order phase transition of the GWW  model. In the following sections we probe analytically this change of behavior at finite $N$, in terms of the different structure of the trans-series expansions for $\Delta(t, N)$ coming from the Rossi equation (\ref{eq:deltat}). 

\subsubsection{Large $N$ expansions for $\Delta(t, N)$ at weak coupling: 
$t < 1$}
\label{sec:delta-largeN-weak}

At weak coupling, the leading large $N$ solution for $\Delta(t, N)$ is obtained algebraically from the vanishing of the $N^2$ terms in (\ref{eq:deltat}):
\begin{eqnarray}
\frac{\Delta}{t^2}\left(1- \Delta^2 \right)  \approx \frac{\Delta}{1-\Delta^2}\qquad  \Rightarrow \qquad \Delta \sim \sqrt{1-t}
\end{eqnarray}
The fluctuations about this zero-instanton part of the weak-coupling large $N$ trans-series for $\Delta(t, N)$ are obtained by inserting an expansion of the form
\begin{eqnarray}
\Delta^{(0)}(t, N)=\sqrt{1-t} \sum_{n=0}^\infty \frac{d_n^{(0)}(t)}{N^{2n}}
\end{eqnarray}
which gives a recursive solution for the coefficient functions $d_n^{(0)}(t)$. Matching terms gives:
\begin{eqnarray}
\Delta^{(0)}(t, N)\sim \sqrt{1-t}\left(1-\frac{1}{16}\frac{t^3}{(1-t)^3}\frac{1}{N^2} -
\frac{t^5 (54 +19t)}{512 (1-t)^6}\frac{1}{N^4}
-\frac{t^7 (4500 + 5526 t + 631 t^2)}{8192(1-t)^9}\frac{1}{N^6}
-\dots \right)
\nonumber\\
\label{eq:delta0-N-weak}
\end{eqnarray}

The non-perturbative large $N$ instanton parts of the trans-series are then obtained by inserting into the differential equation (\ref{eq:deltat}) the weak-coupling trans-series ansatz
\begin{eqnarray} 
	\Delta(t, N) &\sim  & \sum_{k=0}^\infty \left(\xi_{\rm weak}^{\Delta}\right)^k\Delta^{(k)}_{\rm weak}(t, N) \\
	&\sim  &
	\sqrt{1-t}\sum_{n = 0}^\infty \frac{d^{(0)}_n(t)}{N^{2n}} -\frac{i}{\sqrt{2\pi N}}\, \xi_{\rm weak}^{\Delta}\,  f^{(1)}(t) e^{- N S_\text{weak}(t)} \sum_{n = 0}^\infty \frac{d^{(1)}_n(t)}{N^n} +\dots
	\label{eq:DeltaLargeNWeakCouplingExpansionAnsatz}
\end{eqnarray}
At leading order in $\xi_{\rm weak}^\Delta$ we obtain the following equations
\begin{equation}
\begin{aligned}
       t^4 \left( S^\prime(t) \right)^2 &= 4 (1-t), \\
       4t (1-t) \left( f^{(1)}_N(t) \right)^\prime &= (4-3t) f^{(1)}_N(t) 
\end{aligned}
\end{equation}
which determine the weak-coupling action to be
\begin{equation}
	S_\text{weak}(t) = \frac{2 \sqrt{1-t}}{t}-2 {\rm arctanh}\left(\sqrt{1-t}\right)
	\label{eq:sweakt}
\end{equation}
and the one-instanton prefactor to be
\begin{eqnarray}
f^{(1)}(t)=  \frac{1}{2} \frac{t}{(1-t)^\frac{1}{4}}
\label{eq:f1t}
\end{eqnarray}
\begin{figure}[htb]
    \centering
    \includegraphics[scale=1.35]{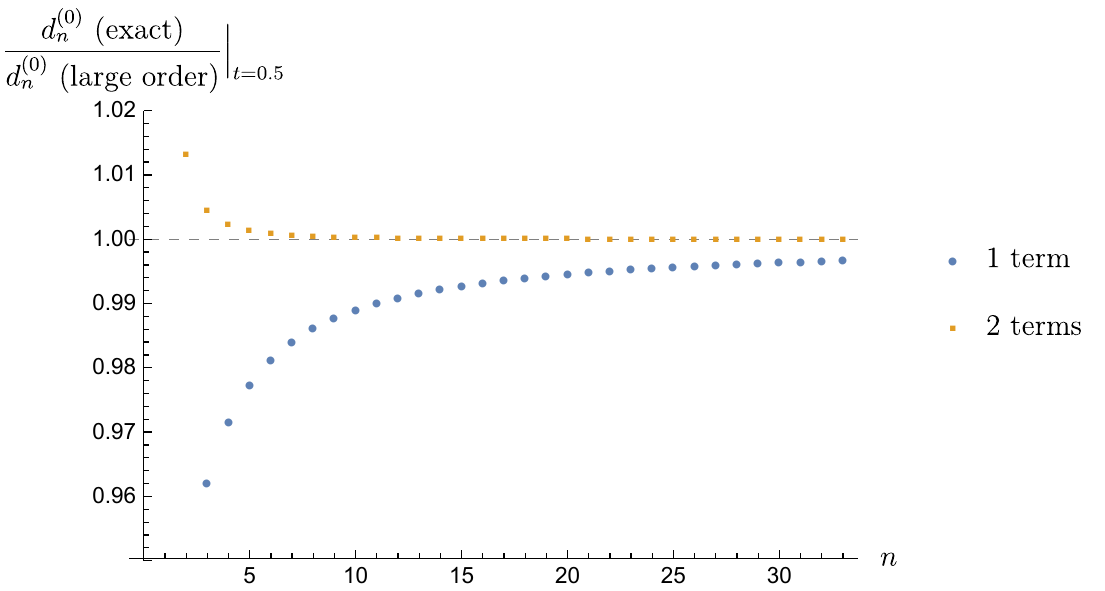}
    \caption{
    Ratio of the exact perturbative coefficients $d_n^{(0)}(t)$ from (\ref{eq:delta0-N-weak}), divided by the resurgent large-order growth expression (\ref{eq:DeltaLargeNWeak0InstantonGrowth}). Both plots are for $t=0.5$, and include just the first term in (\ref{eq:DeltaLargeNWeak0InstantonGrowth}) [blue circles] or the first two terms [gold squares]. Both sequences have been accelerated using Aitken Extrapolation.
    }
    \label{fig:fig7}
\end{figure}
The fluctuation terms in the large $N$ expansion in this one-instanton sector are:
\begin{eqnarray}
\sum_{n = 0}^\infty \frac{d^{(1)}_n(t)}{N^n} = 1 +\frac{(3 t^2-12 t-8)}{96 (1-t)^{3/2}} \frac{1}{N} +
\frac{(81 t^4-324 t^3-300 t^2+2892 t-836)}{18432 (1-t)^3}\frac{1}{N^2}+
\dots 
\label{eq:delta-d1}
\end{eqnarray}
Note that the perturbative fluctuation series (\ref{eq:delta0-N-weak}) is a series in inverse powers of $N^2$, while the one-instanton fluctuation series (\ref{eq:delta-d1}) [and also all higher order instanton fluctuation series] is a series in all inverse powers of $N$. This explains the structure found in Section \ref{sec:z-weak-t} for the partition function $Z(t, N)$.

The divergent weak-coupling large $N$ expansions in (\ref{eq:delta0-N-weak}, \ref{eq:delta-d1}) also display characteristic large order/low order resurgent behavior. 
For example, for a given $t<1$, the perturbative coefficients in (\ref{eq:delta0-N-weak}) grow with order $n$ as:
\begin{equation} \label{eq:DeltaLargeNWeak0InstantonGrowth}
		\sqrt{1-t} \text{ } d^{(0)}_n(t) \sim - \frac{\sqrt{2}f^{(1)}(t) }{\pi^{3/2}}    \frac{\Gamma(2n-\frac{5}{2})}
		{(S_{\rm weak}(t))^{2n-\frac{5}{2}}}
		\left[1+\frac{(3 t^2-12 t-8)}{96 (1-t)^{3/2}} \frac{S_{\rm weak}(t)}{(2n-\frac{7}{2})}+\dots\right]
\end{equation}
In this large-order behavior of the perturbative fluctuation coefficients (\ref{eq:delta0-N-weak}) we recognize the low-order coefficients of the one-instanton fluctuations, from (\ref{eq:delta-d1}). See Figure \ref{fig:fig7} for an illustrative plot. Similar relations connect the fluctuations in higher order instanton sectors.

\subsubsection{Large $N$ expansions for $\Delta(t, N)$ at strong coupling: $t > 1$.}
\label{sec:delta-largeN-strong}

The leading strong-coupling contribution to $\Delta(x, N)$ is (\ref{eq:leading-delta-strong}) [recall $t\equiv N/x$]:
\begin{eqnarray}
\Delta_{(1)}(t, N)\sim J_N\left(\frac{N}{t}\right)
\label{eq:delta-N1}
\end{eqnarray}
Thus, the large $N$ 't Hooft limit at strong coupling is based on the Debye expansion \cite{dlmf:debye}
\begin{eqnarray}
J_{N}\left(N\, {\rm sech}\alpha\right)\sim
\frac{e^{-N(\alpha-\tanh\alpha)}}{(2\pi N\tanh\alpha)^{\frac{1}{2}}}
\sum_{n=0}^{\infty}\frac{U_{n}(\coth\alpha)}{N^{n}},
\label{eq:debye}
\end{eqnarray}
where we identify $\frac{1}{t}\leftrightarrow {\rm sech}\,\alpha$, and the polynomials $U_n(p)$ are generated by the following recursion relation (with $U_0(p)\equiv 1$):
\begin{eqnarray}
U_{n+1}(p)=\tfrac{1}{2}p^{2}(1-p^{2})
U_{n}^{\prime}(p)+\frac{1}{8}\int_{0}^{p}
(1-5t^{2})U_{n}(t)\mathrm{d}t
\end{eqnarray}
In the strong coupling regime $t>1$, and from (\ref{eq:debye}) we identify the strong-coupling large $N$ action as $(\alpha-\tanh \,\alpha)$:\footnote{This differs by a factor of 2 from the strong coupling large $N$ action normalization in  \cite{marino-matrix}, because this is the action for the strong-coupling trans-series for $\Delta$. In the corresponding strong-coupling trans-series for $Z$ or $\ln Z$, these non-perturbative factors are squared [see, for example, (\ref{eq:lnz-final})], so the strong-coupling action is doubled. Our normalization convention is the same as in \cite{Okuyama:2017pil,Alfinito:2017hsh}. }
\begin{eqnarray}
S_{\rm strong}(t) = {\rm arccosh(t)} - \sqrt{1-1/t^2}
\label{eq:strong-t}
\end{eqnarray}
In fact, the Debye expansion (\ref{eq:debye}) is derived by inserting a  trans-series ansatz 
\begin{eqnarray}
J_N(N/t) \sim P(t, N)\, e^{-N S_{\rm strong}(t)} \sum_{n=0}^{\infty}\frac{1}{N^n} U_n\left(q(t)\right)
\label{eq:delta-strong-t-leading}
\end{eqnarray}
into the Bessel equation, and matching terms in order to determine the functional form of $S_{\rm strong}(t)$, the prefactor $P(t, N)$, and a recursive expression for the fluctuation coefficients $U_n\left(q(t)\right)$, where $q(t) = \frac{t}{\sqrt{t^2-1}}$.

We observe resurgent large order behavior in the coefficients $U_{n}(\coth\alpha)\equiv U_n\left(q(t)\right)$, of the large $N$ expansion (\ref{eq:debye}), as they have large order growth as $n\to\infty$ (for any given $t>1$):
\begin{eqnarray} \label{eq:DebyeGrowth}
U_n\left(q(t)\right) \sim \frac{1}{2\pi} \frac{(-1)^n\, (n-1)!}{(2 S_{\rm strong}(t))^n}
\left(1+ U_1\left(q(t)\right) \frac{(2 S_{\rm strong}(t))}{(n-1)}
+ U_2\left(q(t)\right) \frac{(2 S_{\rm strong}(t))^2}{(n-1)(n-2)}+\dots \right)
\end{eqnarray}
\begin{figure}[htb]
\centering
\includegraphics[scale=1.35]{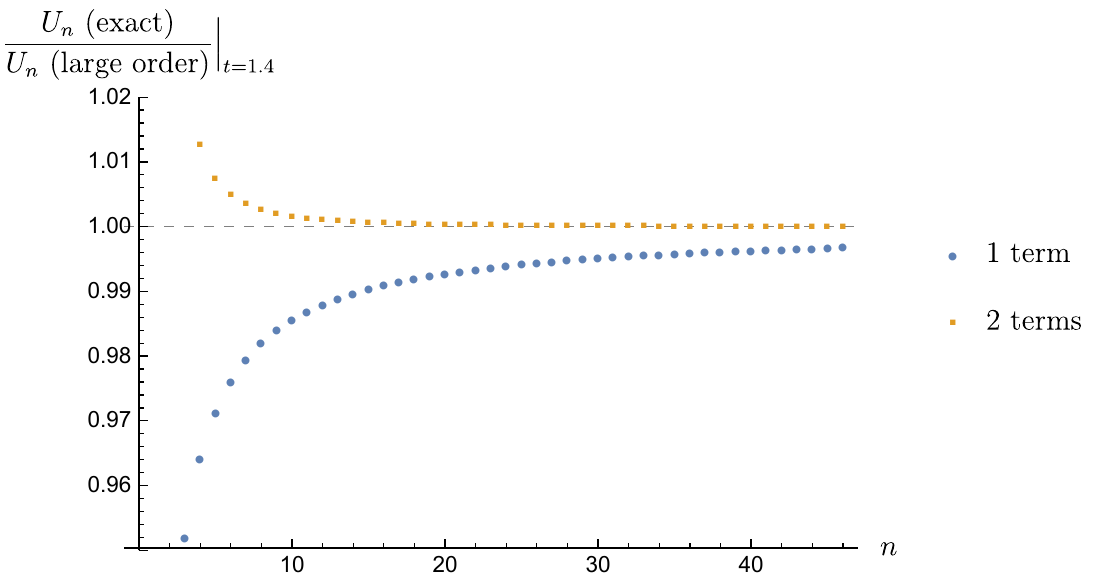}
\caption{Ratio of the exact perturbative coefficients $U_n(q(t))$ from the Debye expansion (\ref{eq:debye}), divided by the resurgent large-order growth expression (\ref{eq:DebyeGrowth}). Both plots are for $t=1.4$, and include just the first term in (\ref{eq:DebyeGrowth}) [blue circles] or the first two terms [gold squares]. 
}
\label{fig:debye}
\end{figure}
Analogous to the situation of resurgence in the Bessel function asymptotics in (\ref{eq:self-resurgence}), we observe the phenomenon of ``self-resurgence'', whereby the large-order growth (\ref{eq:DebyeGrowth}) of the coefficients $U_n(q(t))$ in the large $N$ Debye expansion (\ref{eq:debye}) involves the same low-order coefficients: $U_0(q(t))$, $U_1(q(t))$, $U_2(q(t))$, \dots, for any given $t>1$.
This is illustrated in Figure \ref{fig:debye}. 

The factor of $2$ multiplying $S_{\rm strong}(t)$ in the large-order behavior (\ref{eq:DebyeGrowth}) implies that the next term in the large $N$ strong-coupling trans-series has exponential factor $\exp(-3N S_{\rm strong}(t))$. Matching terms in the differential equation (\ref{eq:deltat}), we find:
\begin{eqnarray}
\Delta(t, N) &\sim& \frac{\sqrt{t}\, e^{-N S_{\rm strong}(t)}}{\sqrt{2\pi N}\, (t^2-1)^{1/4}}  \sum_{n=0}^{\infty}\frac{U_n\left(q(t)\right)}{N^{n}}
+ \frac{1}{4(t^2-1)} \left( \frac{\sqrt{t} e^{-N S_{\rm strong}(t)}}{\sqrt{2\pi N} \, (t^2-1)^{1/4}} \right)^3 \sum_{n=0}^{\infty}\frac{U^{(1)}_n\left(q(t)\right)}{N^{n}}
\nonumber\\
&&+ O\left(e^{-5 N S_{\rm strong}(t)}\right) 
\label{eq:delta-large-n-strong}
\end{eqnarray}
where
\begin{equation}
 U^{(1)}_0(t)=1\qquad ,  \qquad  U^{(1)}_1(t) = \frac{t-6 t^3}{8 \left(t^2-1\right)^{3/2}}\qquad, \quad \dots
\end{equation}
Given the differential equation (\ref{eq:deltat}), it is straightforward to generate the fluctuations about higher instanton sectors, producing a strong-coupling large $N$ trans-series of the form
\begin{eqnarray}
\Delta(t, N) \sim \sum_{k=0}^\infty  P_{(k) \text{strong}}(t) \left(\frac{\sqrt{t} e^{-N S_\text{strong}(t)}}{\sqrt{2\pi N} (t^2-1)^{1/4}}\right)^{2 k+1} f_{(k) \text{strong}}(t, N)
\label{eq:general-delta-t}
\end{eqnarray}
in odd powers of the leading instanton factor, with prefactors and fluctuations at each instanton order. This is consistent with the strong coupling form in (\ref{eq:delta-final}).

\begin{figure}[htb]
\centering
\includegraphics[scale=.6]{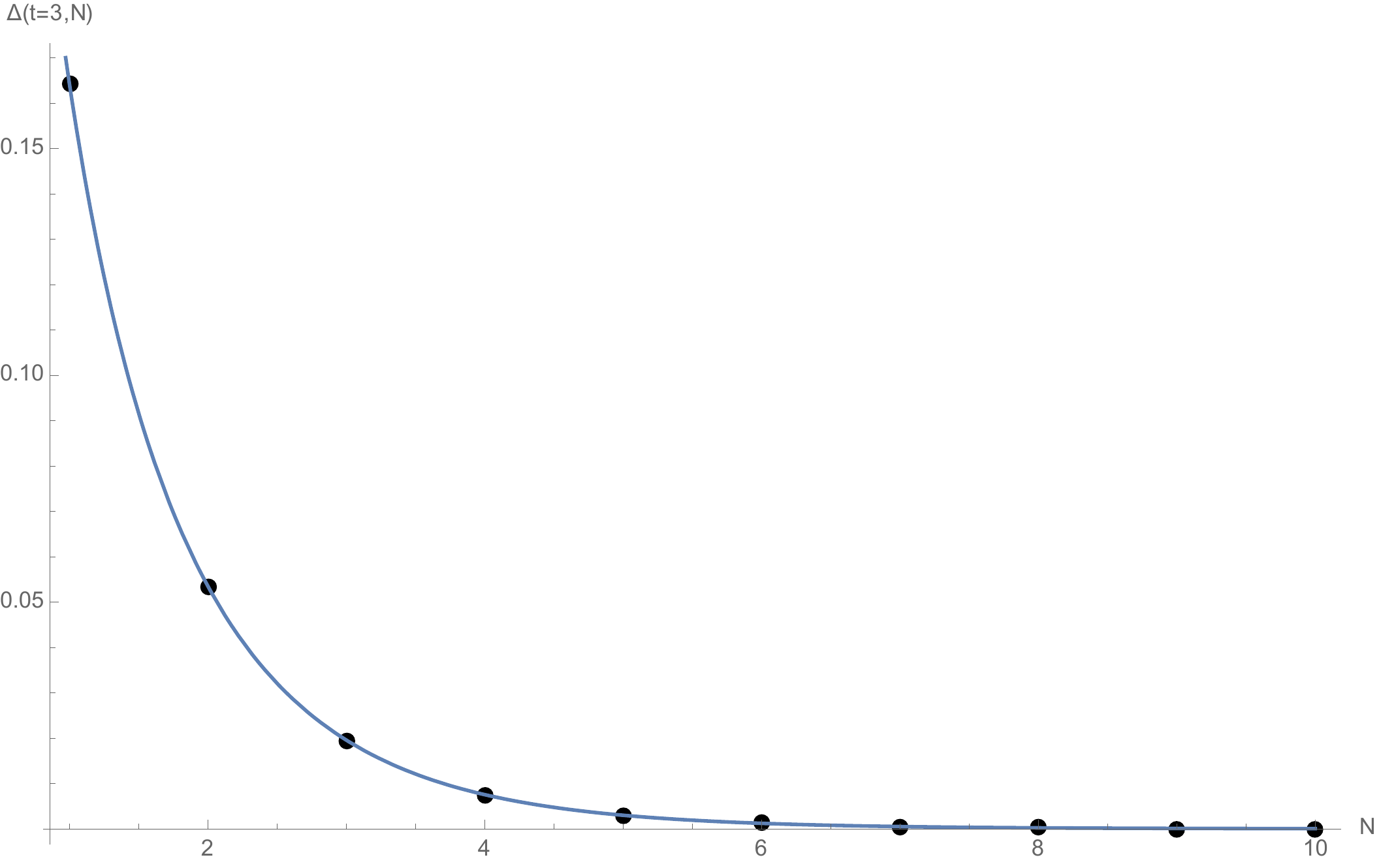}
\caption{The $N$ dependence of the leading large $N$ contribution $J_N(N/t)$ in (\ref{eq:delta-N1}) in the strong coupling regime, plotted here as the solid blue curve for $t=3$. The black dots are the exact values from the determinant expression (\ref{eq:delta}) for $\Delta(t, N)$. Notice that the agreement appears to be excellent, but in fact there are tiny non-peerturbative corrections, shown in Figure \ref{fig:fig10}, coming from the next term $\Delta_{(3)}$ in the strong-coupling trans-series (\ref{eq:delta-strong}).}
\label{fig:fig9}
\end{figure}

\begin{figure}[htb]
\centering
\includegraphics[scale=.6]{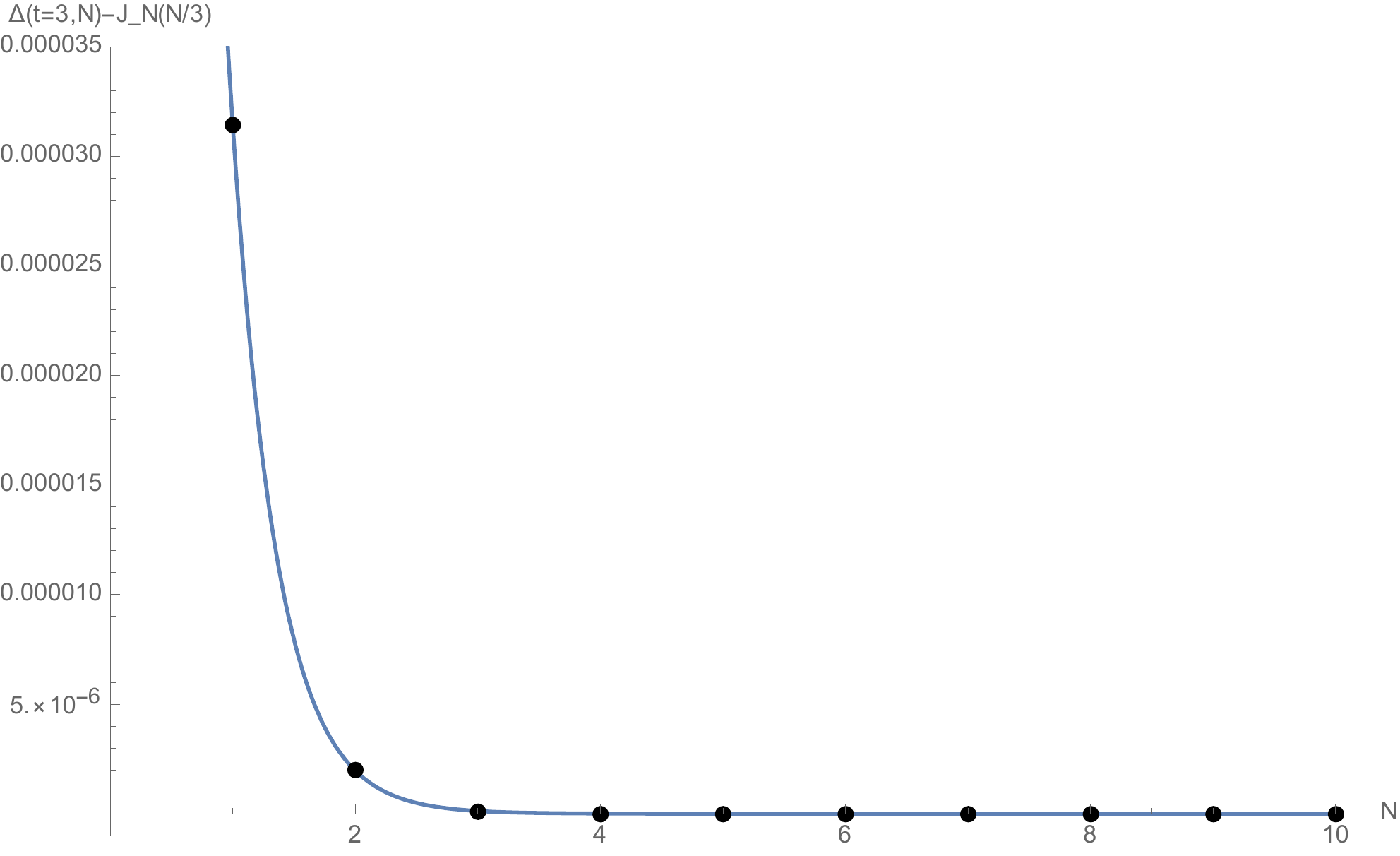}
\caption{The $N$ dependence of the {\it difference} between $\Delta(t,N)$ and the leading large $N$ contribution $J_N(N/t)$ in the strong coupling regime, plotted here as the solid blue curve for $t=3$ using the next trans-series term $\Delta_{(3)}\left(x=\frac{N}{3}, N\right)$ in (\ref{eq:delta-next-solution}). The black dots are the exact values for the difference: $\Delta(t,N)-J_N(N/t)$, with $t=3$. Even at $N=1$ the difference is tiny once we include this next non-perturbative correction.}
\label{fig:fig10}
\end{figure}

Another interesting comparison consists of plotting the $N$ dependence of the first non-perturbative terms of the strong-coupling expansion (\ref{eq:delta-strong}), converted to expressions as functions of $t$ and $N$, rather than $x$ and $N$. In Figure \ref{fig:fig9} we plot the leading strong-coupling term $\Delta_{(1)}(t, N)=J_N(N/t)$ as a function of $N$, compared with the exact $\Delta(t, N)$ from (\ref{eq:delta}), for a fixed strong-coupling value: $t=3$. There is excellent agreement, even at $N=1$, between the exact values (the black dots) and the function $\Delta_{(1)}(3, N)=J_N(N/3)$ (solid blue curve). But in fact, this agreement is not perfect: there is a tiny non-perturbative difference. Figure \ref{fig:fig10} compares the difference, $\Delta_{(1)}(3, N)-J_N(N/3)$, with the next strong-coupling trans-series term, $\Delta_{(3)}(3, N)$, from (\ref{eq:delta-next-solution}), with the conversion: $x=N/3$. The deviation is very small, and again there is excellent agreement. This procedure can be continued to higher order in the strong-coupling large $N$ trans-series expansion.

\subsection{Uniform Large $N$ Strong Coupling Expansion for $\Delta(t, N)$}
\label{sec:uniform}
The strong-coupling large $N$ expansion in (\ref{eq:delta-large-n-strong}), based on the Debye expansion (\ref{eq:debye}) of the Bessel function $J_N(N/t)$, has the disadvantage that when truncated at any finite order it diverges at $t=1$, which is the $N=\infty$ critical point. This is an unphysical divergence, associated with the Debye approximation itself, and it persists for very large $N$. Physically, the full contribution to $\Delta(t, N)$ is finite and smooth at $t=1$, at any instanton order. But the fact that the strong coupling trans-series (\ref{eq:delta-final},\ref{eq:lnz-final}) are expansions in powers of Bessel functions means that we can interpret the phase transition as the coalescence of saddle points. Therefore, this deficiency of the conventional large $N$ expansion can be overcome by using a {\it uniform} large $N$ expansion \cite{dlmf-uniform} for the Bessel functions:
\begin{eqnarray}
J_{N}\left(N/t\right)\sim
\left(
\frac{4\zeta(1/t)}{1-1/t^{2}}\right)^{\frac{1}{4}}
\, \left(\frac{\mathrm{Ai}\left(
N^{\frac{2}{3}}\zeta(1/t)\right)}{N^{\frac{1}{3}}} \sum_{n=0}^{\infty}\frac{A_{n}(\zeta(1/t))}{N^{2n}}+
\frac{\mathrm{Ai}'\left(N^{\frac{2}{3}}
\zeta(1/t)\right)}{N^{\frac{5}{3}}}
\sum_{n=0}^{\infty}\frac{B_{n}(\zeta(1/t))}{N^{2n}}\right)
\nonumber\\
\label{eq:bessel-uniform}
\end{eqnarray}
This large $N$ expansion is valid {\it uniformly} in $t$, for all $t\in (0, \infty)$, including both small $t$ (weak coupling, $0< t\leq 1$) and large $t$ (strong coupling, $1\leq t<\infty$). In (\ref{eq:bessel-uniform}), $\zeta(z)$ is defined as \cite{dlmf-uniform}:
\begin{eqnarray}
\zeta(z)=\begin{cases} 
\left(\frac{3}{2}\left(\ln\left(\frac{1+\sqrt{1-z^{2}}}{z}\right)-\sqrt{1-z^{2}}
\right)\right)^{2/3} \qquad, \quad 0<z<1
\\ ~\\
-\left(\frac{3}{2}\left(\sqrt{z^{2}-1}-\operatorname{arcsec}z\right)\right)^{2/3} \qquad, \quad z >1
\end{cases}
\label{eq:zeta-uniform}
\end{eqnarray}
and $A_k(\zeta)$ and $B_k(\zeta)$ are functions defined in \cite{dlmf-uniform}, with $A_0(\zeta)\equiv 1$. Note that $\zeta(z)$ is related to the strong coupling action (\ref{eq:strong-t}) as:
\begin{eqnarray}
S_{\rm strong}(t)\equiv \frac{2}{3} \, \left(\zeta\left(\frac{1}{t}\right)\right)^{3/2}\quad, \quad t\geq 1
\label{eq:s-zeta}
\end{eqnarray}
Also note that for large argument, ${\rm Ai}^\prime(x)\sim -\sqrt{x}\, {\rm Ai}(x)$, so the second sum term in (\ref{eq:bessel-uniform}) could be re-expanded amd combined with the first sum, giving inverse odd powers of $N$ in the large $N$  expansion.
\begin{figure}[htb]
\center\includegraphics[scale=0.6]{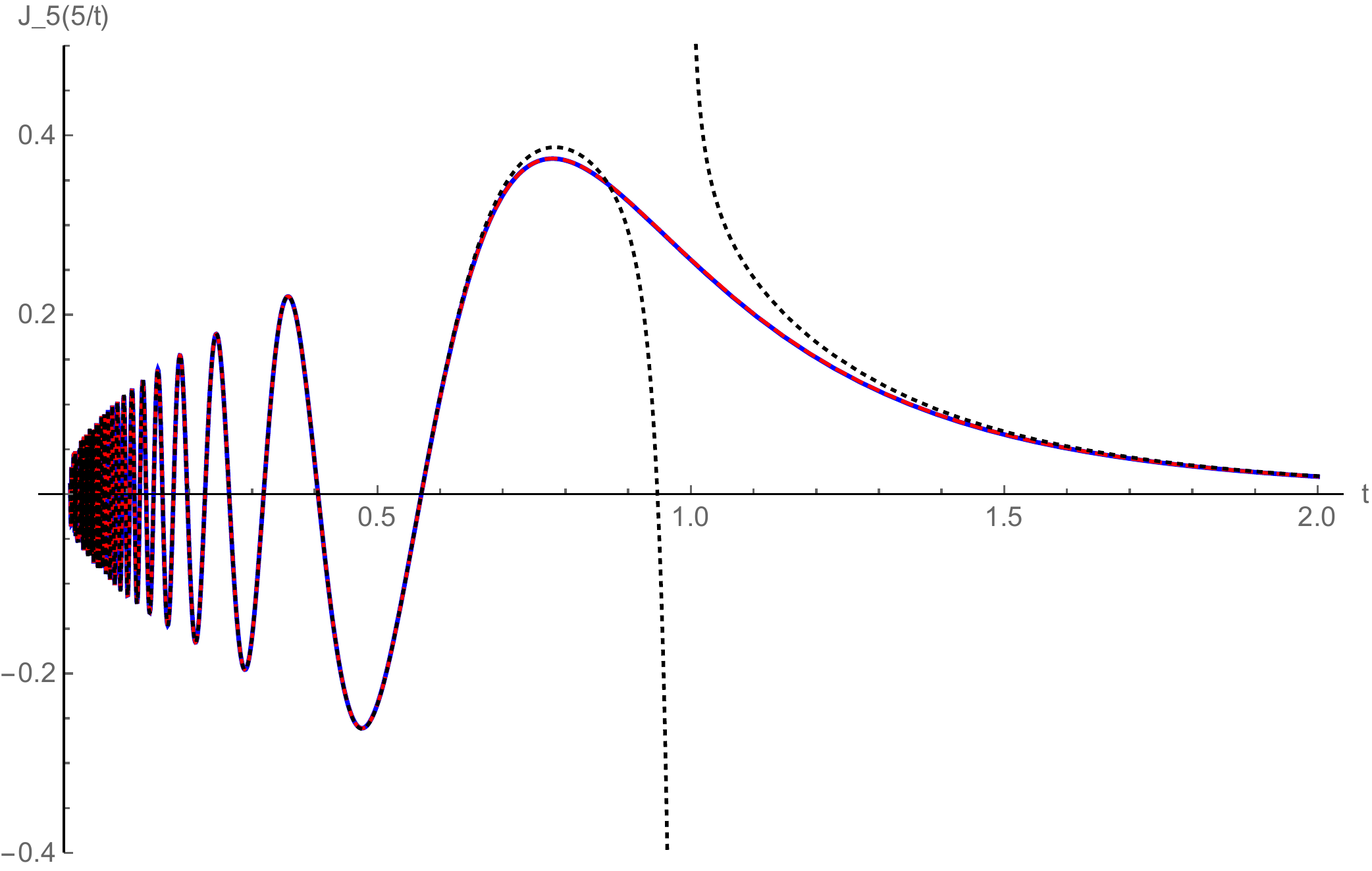}
\caption{The exact Bessel function $J_5(5/t)$ as a function of $t$ [blue solid curve], compared with the leading term of the uniform expansion (\ref{eq:bessel-uniform}) [red dashed curve]. The curves are indistinguishable in this plot. The black dotted curves show  the leading terms of the Debye expansion \cite{dlmf:debye}, which diverge at $t=1$. This divergence is an artifact of the Debye approximation and is responsible for the unphysical divergence of the conventional large $N$ approximation at $t=1$, shown in Fig. \ref{fig:uniform-delta}. }
\label{fig:uniform-j}
\end{figure}

\begin{figure}[htb]
\center\includegraphics[scale=0.6]{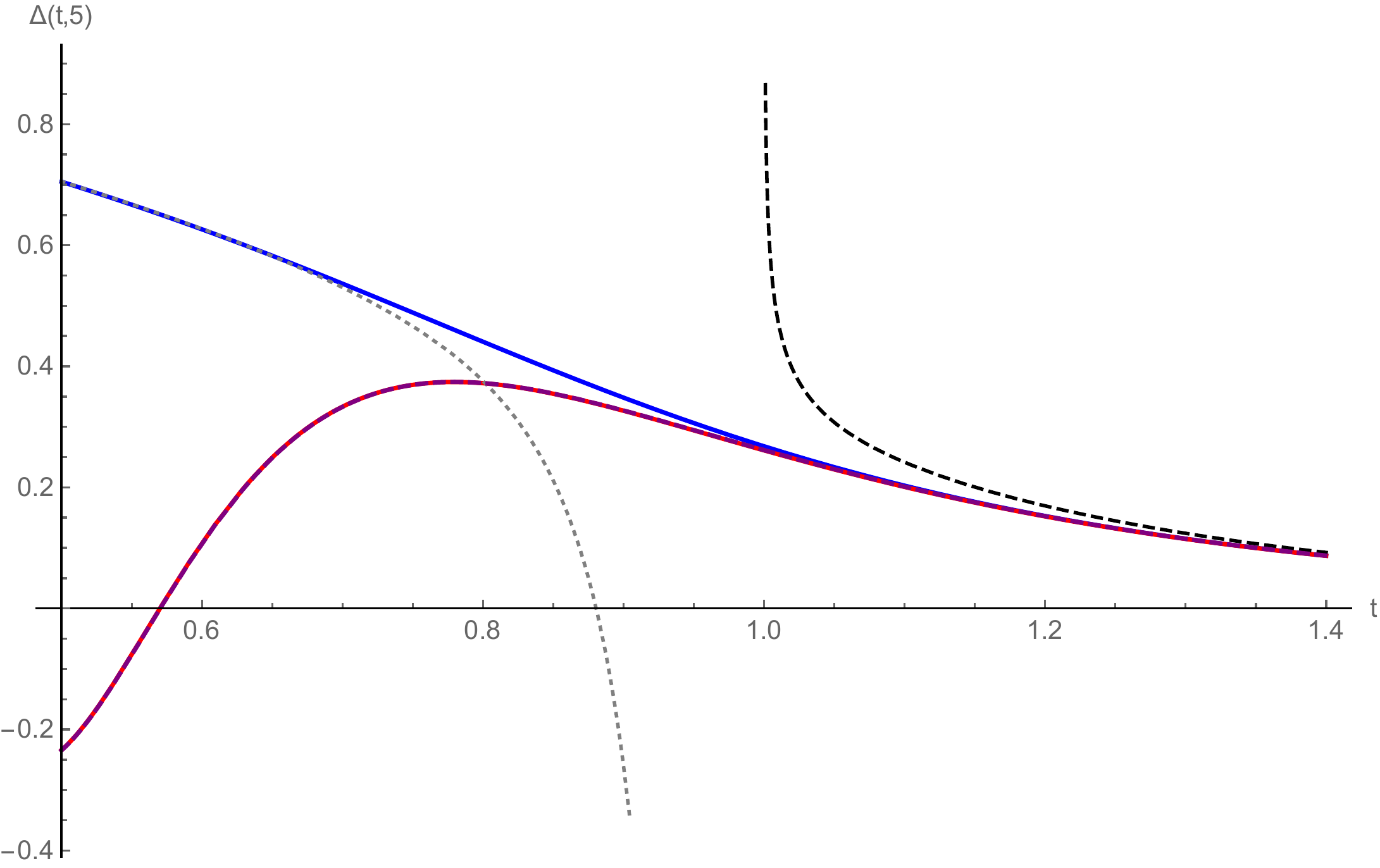}
\caption{Plot of $\Delta(t, N)$ for $N=5$, as a function of the 't Hooft coupling $t$, showing the unphysical divergence of the usual large $N$ expansion approximations at the critical value of $t=1$. The solid blue curve is the exact result. The full strong-coupling one-instanton approximation, $\Delta_{(1)}(t, N)=J_N(N/t)$, and its uniform large $N$ approximation, are shown in the blue and red dashed curves, which are indistinguishable. Note that the strong-coupling one instanton approximation does not diverge at the critical point $t=1$, and is very accurate in the vicinity of the GWW transition point at $t=1$. }
\label{fig:uniform-delta}
\end{figure}

Figure \ref{fig:uniform-j} shows the Bessel function $J_5(5/t)$ as a function of the 't Hooft coupling $t$, compared with the leading term ($A_0(\zeta)=1$) of the uniform large $N$ expansion (\ref{eq:bessel-uniform}), and with the leading terms of the Debye large $N$ expansion \cite{dlmf:debye}. The leading uniform approximation is indistinguishable from the exact function, while the Debye large $N$ expansion diverges at $t=1$. And this plot is for the very small value of $N=5$. In fact, the agreement of the uniform large $N$ approximation to the Bessel function is remarkably good even for $N=1$ or $N=2$. On the other hand, the divergence of the Debye expansion at $t=1$ is responsible for an un-physical divergence as $t\to 1^+$ for $\Delta(t, N)$, as plotted in Figure \ref{fig:uniform-delta}.  Figure \ref{fig:uniform-delta} shows that the weak-coupling large $N$ expansion for $\Delta(t, N)$ also diverges as $t\to 1^-$, from the weak-coupling side, where we have plotted the leading large $N$ contribution  from the weak-coupling expression (\ref{eq:delta0-N-weak}). These divergences persist at large $N$. In contrast, the uniform approximation is smooth through the transition, and even the leading term is remarkably accurate at $t=1$.

These un-physical divergences (at $t=1$) of the conventional large $N$ approximations for $\Delta(t, N)$ are inherited by the analogous conventional large $N$ approximations for physical quantities such as the free energy, $\ln Z(t, N)$, and the specific heat, which are discussed below in the Section \ref{sec:large-n-physical}. In Section \ref{sec:wilson} we show how the uniform large $N$ approximation (\ref{eq:bessel-uniform}) can be applied to Wilson loop expectation values.

\section{Matching finite $N$ trans-series to the double-scaling limit}
\label{sec:matching}

\subsection{Coalescence of Painlev\'e III to Painlev\'e II}

In standard form \cite{dlmf:ps}, the Painlev\'e II and III equations read:
\begin{eqnarray}
\text{Painlev\'e II:}\qquad \frac{d^2 W}{d\chi^2}&=&
2 W^3+\chi \, W +a
\label{eq:p2}\\
\text{Painlev\'e III:}\qquad \frac{d^2 w}{dz^2}&=&\frac{1}{w}\left(\frac{d w}{dz}\right)^2 -\frac{1}{z}\frac{d w}{d z} +\frac{\alpha w^2+\beta}{z}+\gamma\, w^3+\frac{\delta}{w}
\label{eq:p3}
\end{eqnarray}
The coalescence of the Painlev\'e III equation to the Painlev\'e II equation is achieved by rescaling the variable $z$, the function $w(z)$, and the parameters $(\alpha, \beta, \gamma, \delta)$ and $(a)$ in a correlated manner \cite{dlmf:ps}:
\begin{eqnarray}
z=1+\epsilon^2 \chi \quad; \quad w=1+2 \epsilon W\quad; \quad \alpha=-\frac{1}{2\epsilon^6}\quad; \quad \beta=\frac{1}{2\epsilon^6}+\frac{2a}{\epsilon^3}\quad ;\quad  \gamma=\frac{1}{4\epsilon^6}=-\delta
\label{eq:coal}
\end{eqnarray}
Then in the limit $\epsilon\to 0$, the Painlev\'e III equation (\ref{eq:p3}) reduces to  the Painlev\'e II equation (\ref{eq:p2}).
Since the Rossi equation (\ref{eq:rossi}) for $\Delta(x, N)$ is directly related to the Painlev\'e III equation, we can use this same coalescence rescaling (\ref{eq:coal}) in the Rossi equation (\ref{eq:rossi}), in order to probe more finely the vicinity of the GWW phase transition at $t_c=1$. We define a new new variable $\kappa$ and a new function $W(\kappa)$ by: 
\begin{eqnarray}
t=1+\frac{1}{N^{2/3}}\kappa\quad;\quad \Delta(t, N)=\frac{t^{1/3}}{N^{1/3}} W(\kappa)
\label{eq:ds}
\end{eqnarray}
which effectively uses the $N\to\infty$ limit as the $\epsilon\to 0$ limit. The parameter $\kappa$ measures the deviation from the critical 't Hooft coupling, $t_c=1$, rescaled by $N^{-2/3}$. Then it is straightforward to check that the $t$ form (\ref{eq:deltat}) of Rossi's equation  reduces as $N\to\infty$ to the Painlev\'e II equation (\ref{eq:p2}) with parameter $a=0$, and with the argument $\kappa$ and function $W(\kappa)$ each rescaled by a factor of $2^{1/3}$:
\begin{eqnarray}
(\ref{eq:deltat}) \quad \longrightarrow \quad \frac{d^2 W}{d\kappa^2}=2\, W^3(\kappa)+2\kappa\, W(\kappa)
\label{eq:p2-scaled}
\end{eqnarray}
This coalescence limit is precisely the double-scaling limit \cite{gw,wadia,forrester-book}, in which we zoom in on the region close to the critical point $t_c=1$, scaled by the factor $\frac{1}{N^{2/3}}$.

With our conventions here, the weak-coupling ($t<1$) side of the double-scaling limit is $\kappa<0$, and the strong-coupling ($t>1$) side is $\kappa>0$, with the GWW phase transition occurring at $\kappa=0$.\footnote{This differs in sign from the convention in \cite{marino-matrix}, but matches the NIST convention for the Painlev\'e II equation \cite{dlmf:p2-graphics}.} It is well known that the character of the Hastings-McLeod solution to Painlev\'e II changes at $\kappa=0$ \cite{hastings,rosales,marino-matrix}. See for example the illustrative plots at \cite{dlmf:p2-graphics}.

\subsection{Matching to the Double-scaling region from the weak-coupling side: $t\to 1^-$}
\label{sec:ds-weak}

Approaching the GWW phase transition from the weak-coupling side, as $t\to 1$ from below, in the double-scaling limit (\ref{eq:ds}) with $N\to\infty$, the leading term (\ref{eq:delta0-N-weak}) of the large $N$ trans-series for $\Delta(t, N)$ scales as (recall that $\kappa<0$ at weak coupling):
\begin{eqnarray}
\left(\frac{N}{t}\right)^{1/3} \Delta^{(0)}_{\rm weak}(t, N) &\sim & \left(\frac{N}{t}\right)^{1/3} \sqrt{1-t}\left( 1-\frac{1}{16}\frac{t^3}{(1-t)^3}\frac{1}{N^2}-\frac{1}{512}\frac{t^5(54+19t)}{(1-t)^6}\frac{1}{N^4}+O\left(\frac{1}{N^6}\right)\right)
\nonumber \\
&\sim & \sqrt{-\kappa}\left(1-\frac{1}{16\, (-\kappa)^3}-\frac{73}{512 \, (-\kappa)^6}-\dots\right)
\label{eq:delta-ds-weak1}
\end{eqnarray}
This agrees (see for example equation (4.1) in \cite{Tracy:1992rf}, or equation (4.45) in \cite{marino-matrix}) with the leading perturbative term in the weak-coupling  trans-series expansion of the double-scaling limit solution to the Painlev\'e II equation (\ref{eq:p2-scaled}). Similarly, the first non-perturbative term in the weak-coupling  trans-series expansion (\ref{eq:DeltaLargeNWeakCouplingExpansionAnsatz}) scales as
\begin{eqnarray}
\left(\frac{N}{t}\right)^{1/3} \Delta^{(1)}_{\rm weak}(t, N) &\sim & -\left(\frac{N}{t}\right)^{1/3} \frac{i}{2\sqrt{2\pi N}} \frac{t}{(1-t)^{1/4}}\, e^{-N S_{\rm weak}(t)}\left( 1+\frac{1}{96}\frac{(3t^2-12t-8)}{(1-t)^{3/2}}\frac{1}{N} \right.\nonumber\\
&& \qquad \left. 
+\frac{1}{18432}
\frac{(81t^4-324t^3-300t^2+2892t-836)}{(1-t)^3}\frac{1}{N^2}+\dots \right)
\nonumber\\
&\sim & -\frac{i}{2\sqrt{2\pi}}\frac{1}{(-\kappa)^{1/4}}\, e^{-\frac{4}{3} (-\kappa)^{3/2}}\left(1-\frac{17}{96 \, (-\kappa)^{3/2}}+\frac{1513}{18432 \, (-\kappa)^3}-\dots\right)
\label{eq:delta-ds-weak2}
\end{eqnarray}
Here we have used the fact that in the $t\to 1^-$ and $N\to\infty$ double-scaling limit the weak-coupling action (\ref{eq:sweakt}) behaves as
\begin{eqnarray}
S_{\rm weak}(t)&\sim& 2\sqrt{1-t}+\ln\left(\frac{1-\sqrt{1-t}}{1+\sqrt{1-t}}\right)+\dots \qquad,\quad t\to1^-\nonumber\\
&\sim & \frac{4}{3} \frac{(-\kappa)^{3/2}}{N}+\dots
\label{eq:sweak-ds}
\end{eqnarray}
Expression (\ref{eq:delta-ds-weak2}) agrees with the leading one-instanton term in the weak-coupling trans-series expansion of the (appropriately scaled) double-scaling solution to the Painlev\'e II equation: see for example equation (4.2) in \cite{Tracy:1992rf}, or equation (4.52) in \cite{marino-matrix}.

In fact, the correspondence of these and all higher trans-series terms follows immediately from the coalescence reduction of the Rossi equation (\ref{eq:deltat}) to the Painlev\'e II equation (\ref{eq:p2-scaled}). Thus, we see that the weak-coupling trans-series expansion (\ref{eq:DeltaLargeNWeakCouplingExpansionAnsatz}) for $\Delta(t, N)$, which is valid for all 't Hooft coupling $t<1$, and for all $N$, matches smoothly to the weak-coupling trans-series expansion for the double-scaling solution $W(\kappa)$ to the Painlev\'e II equation (\ref{eq:p2-scaled}), in the weak-coupling $\kappa<0$ region.

\subsection{Matching to the Double-scaling region from the strong-coupling side: $t\to 1^+$}
\label{sec:ds-strong}
On the strong-coupling side of the GWW phase transition, the leading term (\ref{eq:leading-delta-strong}) of the trans-series expression for $\Delta(t, N)$ is a Bessel function: $\Delta(t, N)\sim J_N\left(\frac{N}{t}\right)$. In the large $N$ double-scaling limit (\ref{eq:ds}) this reduces to a (re-scaled) Airy function due to the identity \cite{dlmf:bessel-to-airy}:
\begin{eqnarray}
\lim_{N\to\infty} J_N(N-N^{1/3} \kappa) = \left(\frac{2}{N}\right)^{1/3} {\rm Ai}\left(2^{1/3} \kappa \right)
\label{eq:bessel-to-airy}
\end{eqnarray}
Allowing for the rescaling of the argument and function by $2^{1/3}$ in (\ref{eq:p2-scaled}), this shows that the choice of multiplicative constant $\xi_{\rm strong}^\Delta=1$ in (\ref{eq:leading-delta-strong}) coincides precisely with the choice of multiplicative constant to be the identity in the Hastings-McLeod solution of the standard Painlev\'e II equation \cite{hastings,rosales,dlmf:p2-graphics}. 

This is also consistent with the large $N$ expansion for $\Delta(t, N)$ in (\ref{eq:delta-large-n-strong}). Expanding the strong-coupling action $S_{\rm strong}(t)$ in (\ref{eq:strong-t}) near the transition region $t\to 1^+$ (note that $\kappa>0$ in the strong-coupling region):
\begin{eqnarray}
S_{\rm strong}(t)&\sim & \frac{2\sqrt{2}}{3}\left(t-1\right)^{3/2} 
+\dots 
\nonumber \\
&\sim & \frac{1}{N} \frac{2\sqrt{2}}{3}\kappa^{3/2} -\dots 
\label{eq:sstrong-ds}
\end{eqnarray}
Therefore, the leading large $N$ behavior in (\ref{eq:delta-large-n-strong}) yields
\begin{eqnarray}
\left(\frac{N}{t}\right)^{1/3} \Delta_{(1)}(t, N) & \sim & \left(\frac{N}{t}\right)^{1/3} 
\frac{1}{\sqrt{2\pi N}}\frac{\sqrt{t}}{(t^2-1)^{1/4}} e^{-N S_{\rm strong}(t)}
\nonumber \\
&\sim & 
\frac{1}{\sqrt{2\pi } (2\kappa)^{1/4}}\exp\left[-\frac{2\sqrt{2}}{3} \, \kappa^{3/2}\right] 
\nonumber \\
& \sim & 2^{1/3}\, {\rm Ai}\left(2^{1/3}\, \kappa\right)
\label{eq:Delta-ds-strong1}
\end{eqnarray}
which is the leading large $\kappa$ behavior of the Hastings-McLeod solution to the scaled Painlev\'e II equation (\ref{eq:p2-scaled}).
Furthermore, the sub-leading strong-coupling trans-series term in (\ref{eq:delta-large-n-strong}) yields
\begin{eqnarray}
\left(\frac{N}{t}\right)^{1/3} \Delta_{(3)}(t, N) 
&\sim & \left(\frac{N}{t}\right)^{1/3} \frac{1}{4 (t^2-1)}\left(\frac{1}{\sqrt{2\pi N}}\frac{\sqrt{t}}{(t^2-1)^{1/4}} e^{-N S_{\rm strong}(t)}\right)^3 
\nonumber \\
&\sim &
\frac{1}{4\,(2\pi)^{3/2} (2\kappa)^{7/4}}\exp\left[-2\sqrt{2} \, \kappa^{3/2}\right] 
\nonumber \\
& \sim & \frac{1}{4\, \kappa} \left({\rm Ai}\left(2^{1/3}\, \kappa\right)\right)^{3}
\label{eq:Delta-ds-strong2}
\end{eqnarray}
in agreement with the next term in the trans-series expansion of the Hastings-McLeod solution \cite{hastings,rosales} on the strong-coupling side.

Thus the strong-coupling trans-series for $\Delta(t, N)$, valid for all $N$ and all $t>1$ on the strong-coupling side of the GWW phase transition,  matches smoothly to the double-scaling limit form of the strong-coupling trans-series expansion for the solution to Painlev\'e II.

\subsection{Instanton condensation}
\label{sec:condensation}

As we approach the GWW phase transition, either from the strong-coupling or weak-coupling side, the large $N$ approximation trans-series expansions  (\ref{eq:DeltaLargeNWeakCouplingExpansionAnsatz}) and (\ref{eq:delta-large-n-strong}) break down numerically. This is in part because the relevant instanton factors, $e^{-N\, S_{\rm strong}(t)}$ and $e^{-N\, S_{\rm weak}(t)}$, approach 1 and so are no longer exponentially small. Physically, this is the phenomenon of ``instanton condensation''
\cite{Neuberger:1980qh,Neuberger:1980as}, whereby the one-instanton approximation is no longer a good approximation, requiring the treatment of all instanton orders. But the breakdown of the large $N$ expansion at $t=1$ is also because the prefactor terms diverge as $t\to 1$. This means we must have a better treatment of the fluctuations about the instanton factors, as well as the full effect of multi-instanton interactions at higher instanton order. In the case of the GWW model we can probe this condensation in great detail on the strong coupling side of the transition because we have simple closed-form expressions, in terms of Bessel functions, for the various orders of the instanton expansion. While (\ref{eq:delta-large-n-strong}) is indeed the structure of the strong-coupling large $N$ trans-series as $t\to\infty$, we also know the structure of the strong-coupling large $N$ trans-series for all $t\geq 1$. The leading one-instanton term (\ref{eq:delta-N1}) is exactly $\Delta_{(1)}(t, N)=J_N(N/t)$, while the next three-instanton term is exactly $\Delta_{(3)}(t, N)$ given in (\ref{eq:delta-next-solution}), with the replacement $x\to N/t$.
\begin{figure}[htb]
\centering
\includegraphics[scale=.6]{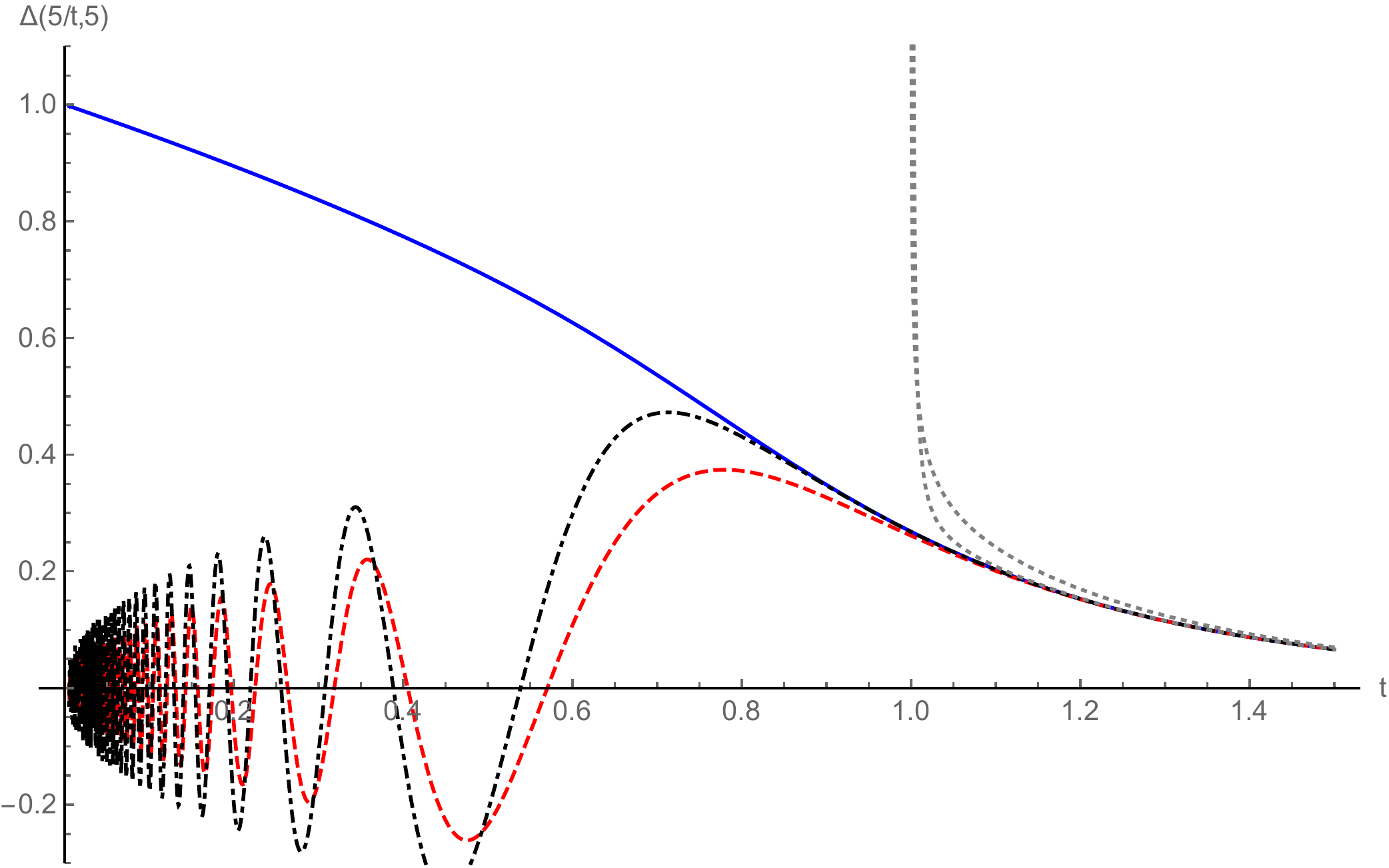}
\caption{The $t$ dependence of $\Delta(t, N)$ [blue curve] and the leading strong-coupling large $N$ contribution
$\Delta_{(1)}(t, N)=J_N(N/t)$ [red dashed curve] in  the  strong  coupling  regime at $N=5$. The  black dot-dashed curve  includes  the  three-instanton
contribution $\Delta_{(3)}(t, N)$ in (\ref{eq:delta-next-strong}). The agreement is excellent, even at the $N=\infty$ GWW transition point $t=1$. The dotted grey curves, which diverge at the transition point, show the leading instanton terms in the conventional large $N$ strong-coupling expansion in (\ref{eq:delta-large-n-strong}), which has effectively used the Debye expansion for the Bessel functions appearing in (\ref{eq:leading-delta-strong}) and  (\ref{eq:delta-next-strong}). }
\label{fig:fig17}
\end{figure}
This is illustrated in Figure \ref{fig:fig17}, where we see that the one-instanton and three-instanton terms $\Delta_{(1)}(t, N)$, and $\Delta_{(3)}(t, N)$, give an excellent approximation to $\Delta(t, N)$, all the way down to the transition point at $t=1$, and even below [compare with Figure \ref{fig:uniform-delta} which shows the leading instanton terms]. In fact, the agreement at the transition point $t=1$ is better than $3.4\%$ with just the leading term, and better than $0.1\%$ when the next term is included. See Figure \ref{fig:fig18}. On the other hand, the corresponding terms from the usual large $N$ expansion (\ref{eq:delta-large-n-strong}) diverge as we approach $t=1$ from above, as shown in Figure \ref{fig:fig17}. Thus we can view $\Delta_{(1)}(t, N)$ as the full one-instanton expression, including fluctuations, and $\Delta_{(3)}(t, N)$ as the full three-instanton expression, including fluctuations and interactions. 
\begin{figure}[htb]
\centering
\includegraphics[scale=.6]{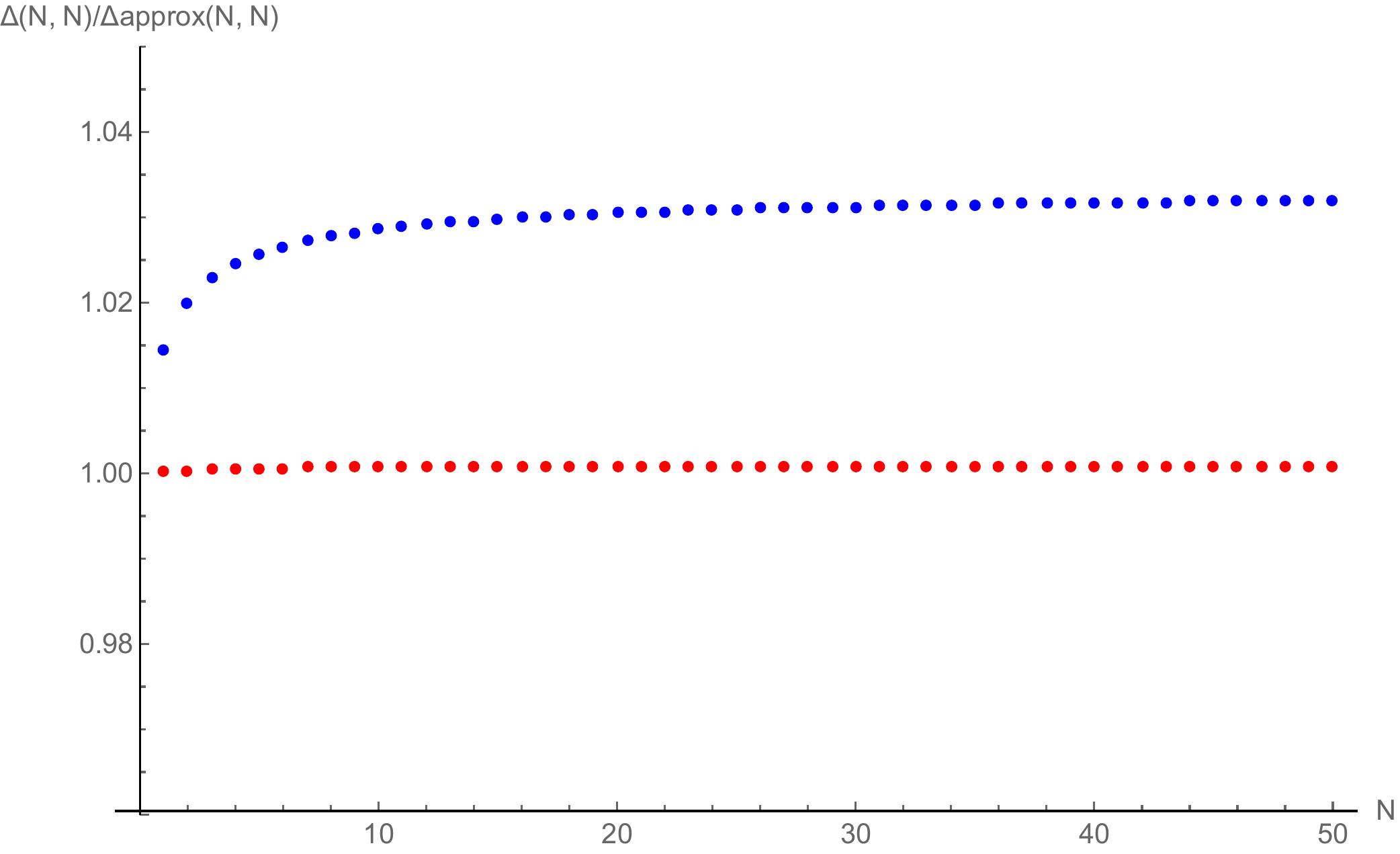}
\caption{The $N$ dependence of the ratio of $\Delta(N, N)$, the exact value at the GWW transition point, 
to the leading one-instanton strong-coupling large $N$ contribution $\Delta_{(1)}(N, N)=J_N(N)$ [blue circles], and then including also the three-instanton contribution $\Delta_{(3)}(N, N)$ [red circles] in (\ref{eq:delta-next-solution}). The agreement is excellent, even at small values of $N$.}
\label{fig:fig18}
\end{figure}
A non-trivial consistency check is that we can evaluate $\Delta(x=N, N)$ from the explicit determinant expressions (\ref{eq:delta}), and verify that at the GWW transition point, $t=1$, the limit
\begin{eqnarray}
\left(\frac{N}{2}\right)^{1/3} \Delta(N, N) \to 0.36706155 ...
\end{eqnarray}
which is the numerical value of the Hastings-McLeod solution $W(\kappa)$ at $\kappa=0$, scaled by the appropriate factor of $2^{-1/3}$ \cite{fornberg,hastings,rosales,spohn}. 

This condensation phenomenon can be probed even more precisely very close to the transition point using the double-scaling limit. Analogous to the above interpretation for $\Delta(t, N)$, we can consider $W(\kappa)$ in the vicinity of the transition point at $\kappa=0$. The standard weak-coupling and strong-coupling trans-series expansions diverge at $\kappa=0$ for the same reason: the exponential factors are no longer small, and also the prefactors diverge. On the strong-coupling side we can uniformize this behavior by regarding the Airy functions Ai($\kappa$) as the basic trans-series building blocks, much as the Bessel functions $J_N(N/t)$ are the basic trans-series building blocks at finite $N$ [see equation (\ref{eq:delta-final})]. 

In order to facilitate comparison with numerical results, we consider the Painlev\'e II equation with its conventional normalization (\ref{eq:p2}), and with $a=0$. Then, write a trans-series expansion in the strong-coupling ($\chi >0$) region:
\begin{eqnarray}
W(\chi) \sim \sum_{k=1, 3, 5, \dots} \left(\xi_{\rm strong}^W \right)^k W_{(k)}(\chi)
\label{eq:W-strong}
\end{eqnarray}
where the physical Hastings-McLeod solution has the trans-series parameter $\xi_{\rm strong}^W=1$. Expanding in instanton order, we find:
\begin{eqnarray}
W_{(1)}(\chi)&=& {\rm Ai}(\chi) \nonumber\\
W_{(3)}(\chi)&=& 2\pi\left({\rm Ai}(\chi)\int_{\chi}^\infty {\rm Bi}(\chi^\prime) {\rm Ai}^3(\chi^\prime)\, d\chi^\prime 
-{\rm Bi}(\chi)\int_{\chi}^\infty  {\rm Ai}^4(\chi^\prime)\, d\chi^\prime\right)
\label{eq:p2-W}
\end{eqnarray}
Remaining terms can be generated by iterating the integral equation form of the Painlev\'e II equation:
\begin{eqnarray}
W(\chi)={\rm Ai}(\chi)+2\pi\int_{\chi}^\infty\left[
{\rm Ai}(\chi) {\rm Bi}(\chi^\prime)
-{\rm Ai}(\chi^\prime) {\rm Bi}(\chi)\right] W^3(\chi^\prime)\, d\chi^\prime
\label{eq:p2-integral}
\end{eqnarray}
\begin{figure}[htb]
\centering
\includegraphics[scale=.6]{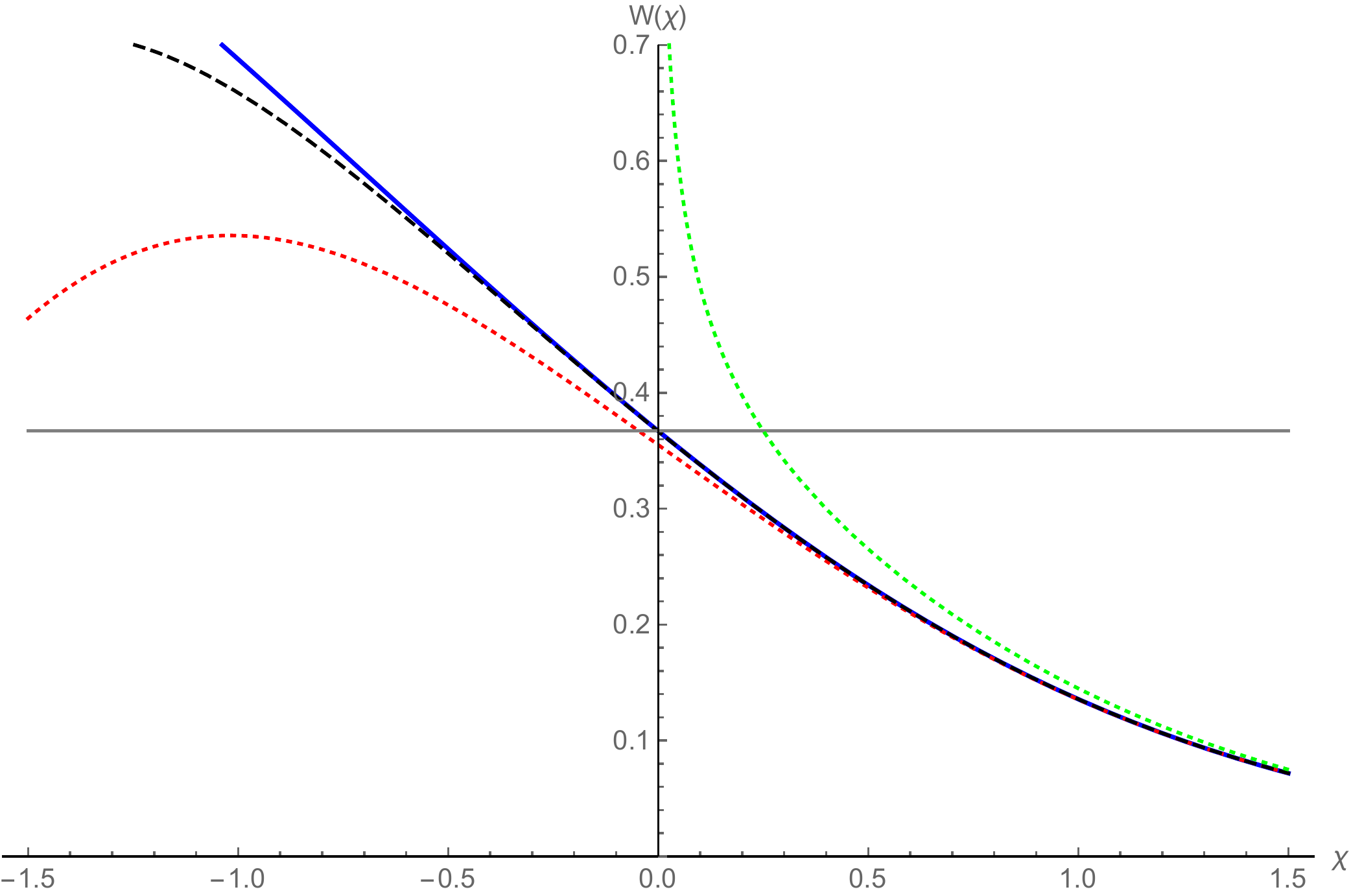}
\caption{Illustration of the instanton condensation phenomenon for the Painlev\'e II equation, which characterizes the immediate vicinity of the GWW phase transition. The solid blue curve is the exact (numerical) solution of the Hastings-McLeod solution to the Painlev\'e II equation (\ref{eq:p2}), (with parameter $a=0$). The red dotted curve is the uniformized leading instanton contribution $W_{(1)}(\chi)={\rm Ai}(\chi)$, and the black dashed curve includes also the uniformized three-instanton contribution in (\ref{eq:p2-W}). The horizontal grey line is the numerical value \cite{hastings,rosales,spohn,fornberg} at the transition point $\chi=0$: $W(0)=0.36706155$, which agrees to $0.1\%$ precision with the first two terms of the trans-series (\ref{eq:W-strong}). The dotted green line, which diverges at the transition point, is the conventional leading one-instanton approximation, from the leading large $\chi$ behavior of the Airy function.}
\label{fig:fig13}
\end{figure}
In Figure \ref{fig:fig13}
we plot the contributions of the first two instanton terms, $W_{(1)}(\chi)$ and 
$W_{(3)}(\chi)$ in (\ref{eq:p2-W}), compared to the exact numerical solution. The agreement is excellent throughout the strong coupling, $\chi>0$, region, all the way down to the transition point at $\chi=0$. Indeed, we can evaluate at $\chi=0$, to find\footnote{We have used the integrals: $\int_0^\infty {\rm Ai}^4(\chi)\,d\chi=\frac{\ln 3}{24\pi^2}$, and
$\int_0^\infty {\rm Bi}(\chi) {\rm Ai}^3(\chi)\,d\chi=\frac{1}{24\pi}$.}
\begin{eqnarray}
W_{(1)}(0)&=&{\rm Ai}(0)=\frac{1}{3^{2/3} \, \Gamma\left(\frac{2}{3}\right)} \approx 
0.355028054...
\\
W_{(3)}(0)&=&2\pi\left(\frac{1}{24\pi}{\rm Ai}(0)-\frac{\ln 3}{24\pi^2}{\rm Bi}(0)\right)=\frac{3^{1/3}\pi-3^{5/6}\ln 3}{36 \pi\, \Gamma\left(\frac{2}{3}\right)} \approx 
0.011665728...
\end{eqnarray}
Therefore, taking into account the first two instanton orders of the trans-series (\ref{eq:W-strong}) yields an approximation:
\begin{eqnarray}
W(0)\approx W_{(1)}(0)+
W_{(3)}(0)= 0.366693782...
\end{eqnarray}
which agrees to better than $0.1\%$ precision with the actual numerical Hastings-McLeod solution value \cite{spohn,fornberg}, 
$W(0)=0.367061552...
$ . Similarly, the derivative at $\chi=0$ is approximated well by the first two terms of the uniformized instanton (trans-series) expansion (\ref{eq:W-strong}):
\begin{eqnarray}
W^\prime(0)&\approx& {\rm Ai}^\prime(0)+2\pi\left(\frac{1}{24\pi}{\rm Ai}^\prime(0)-\frac{\ln 3}{24\pi^2}{\rm Bi}^\prime(0)\right) \nonumber\\
&=&
-\frac{13\times 3^{2/3} \pi +{3}^{7/6} \ln (3)}{36 \pi  \Gamma \left(\frac{1}{3}\right)}
 \approx 
-0.293451526...
\label{eq:slope}
\end{eqnarray}
compared to the precise numerical value $W^\prime(0)=-0.295372105...$ \cite{spohn,fornberg}.

\section{Uniform Large $N$ Expansion for Wilson Loops}
\label{sec:wilson}

Wilson loop expectation values in the GWW model can also be expressed in terms of the function $\Delta(x, N)\equiv \langle \det\, U\rangle$ defined in (\ref{eq:delta}). For example, the (normalized) one-winding Wilson loop is
\begin{eqnarray}
{\mathcal W}_1(x, N)&\equiv &\frac{1}{N}\frac{\partial}{\partial x}\, \ln Z(x, N)\\
&=& \frac{x}{2N}-\frac{2}{N x}\, \sigma_N\\
&=& \frac{x}{2N}-\frac{x}{2N}\left(\Delta^2(x,N)-\left(1-\Delta^2(x,N)\right)\Delta(x, N-1) \, \Delta(x, N+1)\right)
\label{eq:w1}
\end{eqnarray}
\begin{figure}[htb]
\center\includegraphics[scale=0.5]{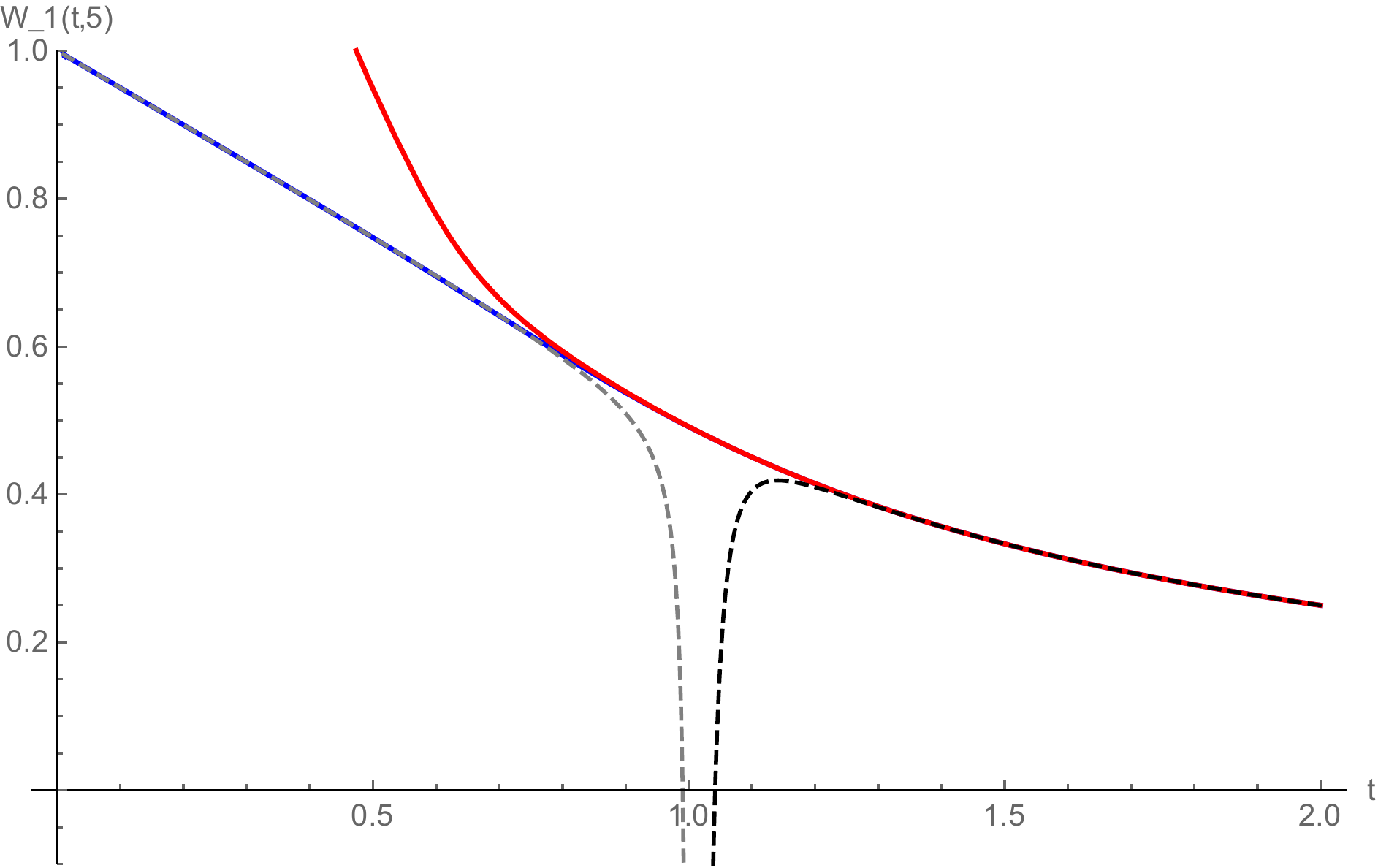}
\caption{The solid blue curve shows the Wilson loop ${\mathcal W}_1(t, N)$ as a function of 't Hooft coupling $t$, with $N=5$. The red solid curve is the leading strong coupling approximation in (\ref{eq:w1-leading}), for which the uniform strong coupling approximation (\ref{eq:w1-uniform}) is indistinguishable. The grey and black dashed lines show the leading terms of the conventional large $N$ approximations for the weak and strong coupling regimes, respectively. Note that the conventional large $N$ approximations diverge at $t=1$, while the uniform approximation is smooth and in excellent agreement with the exact result, even at this small value of $N$.}
\label{fig:wilson1}
\end{figure}
Here we have used the relations (\ref{eq:ez}, \ref{eq:e-sigma}) between the partition function $Z(x, N)$ and $\sigma_N$, and the relation (\ref{eq:sigma-delta2}) between $\sigma_N$ and $\Delta(x, N)$. This means we can immediately deduce trans-series expansions for the Wilson loop from the trans-series expansions for $\Delta(x, N)$, described in Section \ref{sec:delta}. 
For example, using the leading instanton strong coupling expression for $\Delta(x, N)\approx J_N(x)$ in (\ref{eq:leading-delta-strong}), we deduce a simple strong coupling approximation for the Wilson loop (recall $t=g^2 N/2=N/x$)
\begin{eqnarray}
{\mathcal W}_1(t, N)\Big |^{\rm strong}&\approx& \frac{1}{2t}-\frac{1}{2t}\left(J_N^2(N/t)-J_{N-1}(N/t)J_{N+1}(N/t)\right) \nonumber\\
&=& \frac{1}{2t}-\frac{1}{2t}\left((1-t^2)J_N^2(N/t)
+\left(J_{N}^\prime(N/t)\right)^2 \right)
\label{eq:w1-leading}
\end{eqnarray}
The first term is the familiar perturbative expression at strong coupling [recall that the strong coupling large $N$ expansion truncates after just one term]. The second term is the full leading instanton contribution, with all fluctuation factors resummed to all orders. Expanding the Bessel functions according to the Debye expansion (\ref{eq:debye}) leads to the usual leading instanton expression at large $N$ (see \cite{Okuyama:2017pil,Alfinito:2017hsh}),  which diverges at $t=1$ when truncated at any fluctuation order:
\begin{eqnarray}
{\mathcal W}_1(t, N)\Big |^{\rm strong}_{{\rm large}\, N}\approx \frac{1}{2t}+\frac{e^{-2N S_{\rm strong}(t)}}{4\pi N^2} \left(\frac{t}{1-t^2}+\frac{1}{12N}\frac{t^2(3+14t^2)}{(t^2-1)^{5/2}}+O\left(\frac{1}{N^2}\right)\right)
\label{eq:w1-largeN-leading}
\end{eqnarray}
On the other hand, the uniform large $N$ approximation introduced in Section \ref{sec:uniform} leads to a uniform large $N$ approximation for $\mathcal{W}_1(t,N)$ (we use also the uniform approximation for $J_N^\prime(N/t)$ at \cite{dlmf-uniform}):
\begin{equation} \label{eq:w1-uniform}
\begin{aligned}
	\mathcal{W}_1(t,N)&\Big |^\text{strong}_\text{uniform large N} \approx \frac{1}{2t} + \sqrt{t^2-1}\sqrt{\zeta(1/t)} \Bigg[ \frac{\left(\text{Ai} \left( N^{2/3} \zeta(1/t) \right) \right)^2}{N^{2/3}} - \frac{1}{N^{4/3}} \frac{\left(\text{Ai}^\prime \left( N^{2/3} \zeta(1/t) \right) \right)^2}{\zeta(1/t)} \\
	&+ \frac{\text{Ai}\left( N^{2/3} \zeta(1/t) \right) \text{Ai}^\prime \left( N^{2/3} \zeta(1/t) \right)}{2N^2} \left(\frac{2t}{(t^2-1)^{3/2} \sqrt{\zeta(1/t)}}  - \frac{1}{\zeta^2(1/t)}\right) + \dots\Bigg]
\end{aligned}
\end{equation}

This uniform approximation is indistinguishable from the Bessel expression (\ref{eq:w1-leading}) even for very small $N$, and both are in excellent agreement with the exact result even through the transition point. This is illustrated in Figure \ref{fig:wilson1}. Furthermore, it is a non-trivial check that the uniform approximation (\ref{eq:w1-uniform}) reduces to the Debye large $N$ expansion in (\ref{eq:w1-largeN-leading}) if we expand the Airy functions at large $N$, in which case all the $\zeta$ dependence in the fluctuation terms cancels. 

This analysis of Wilson loops in terms of the function $\Delta(t, N)$ extends also to higher winding Wilson loops.
It is interesting to compare with the recent large $N$ instanton computations in \cite{Okuyama:2017pil,Alfinito:2017hsh} for these higher winding Wilson loops. 
Using the results of \cite{Green:1980bg} higher winding Wilson loops are related to ${\mathcal W}_1$.
For example (see \cite{Alfinito:2017hsh}), for the double-winding Wilson loop:
\begin{eqnarray}
{\mathcal W}_2(x, N)&\equiv &
\frac{1}{N} \langle {\rm tr}\left( U^2\right)\rangle
= 1-\frac{2N}{x}{\mathcal W}_1(x, N) 
\label{eq:w2w1}
\end{eqnarray}
Therefore, (\ref{eq:w1}) implies that ${\mathcal W}_2$ can also be expressed in terms of $\Delta$:  
\begin{eqnarray}
{\mathcal W}_2(x, N)
= \frac{4}{x^2}\sigma_N
=\Delta^2(x,N)- \left(1-\Delta^2(x,N)\right)\Delta(x, N-1) \, \Delta(x, N+1)
\end{eqnarray}
\begin{figure}[htb]
\center\includegraphics[scale=0.5]{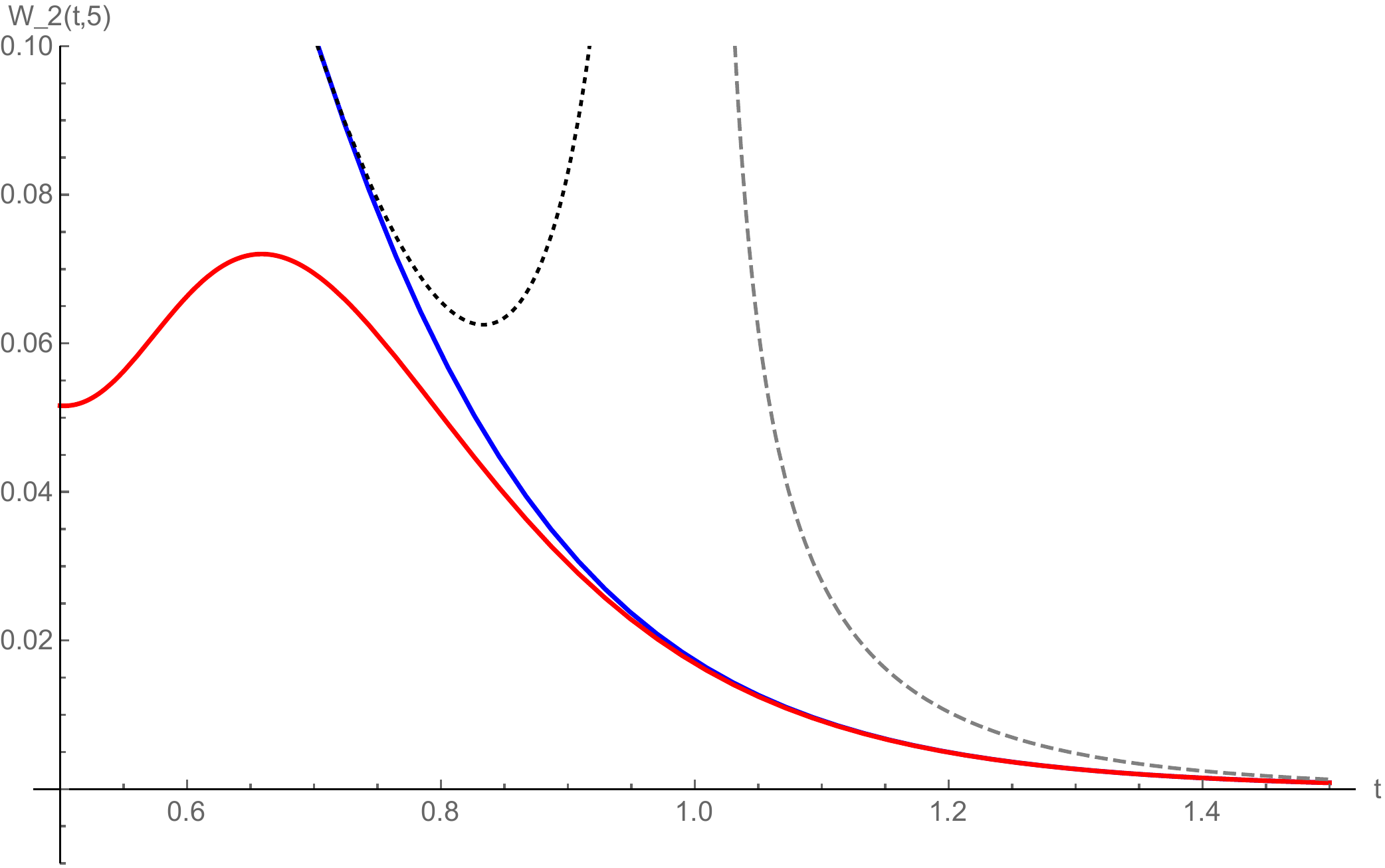}
\caption{The solid blue curve shows the Wilson loop ${\mathcal W}_2(t, N)$ as a function of 't Hooft coupling $t$, with $N=5$. The red solid curve is the leading strong coupling approximation in (\ref{eq:w1-leading}), for which the uniform strong coupling approximation is indistinguishable. The grey and black dashed lines show the leading terms of the conventional large $N$ approximations for the weak and strong coupling regimes, respectively. Note that the conventional large $N$ approximations diverge at $t=1$, while the uniform approximation is smooth and in excellent agreement with the exact result, even at this small value of $N$.}
\label{fig:wilson2}
\end{figure}
Thus, the leading strong coupling approximation for ${\mathcal W}_2(x, N)$ is already an instanton effect:
\begin{eqnarray}
{\mathcal W}_2(t, N)\Big |^{\rm strong} 
\approx J_N^2(N/t)-J_{N-1}(N/t)J_{N+1}(N/t)
\label{eq:w2-leading}
\end{eqnarray}
This is the full leading instanton contribution at strong coupling, with all fluctuation factors resummed in closed form. This is plotted in Figure \ref{fig:wilson2}, showing excellent agreement, in contrast to the standard large $N$ approximations which diverge at the transition point. The leading uniform large $N$ approximation, using the uniform approximation for the Bessel functions in the two-instanton contribution (\ref{eq:w2-leading}), is indistinguishable from the Bessel function expression.

\begin{figure}[htb]
\center\includegraphics[scale=1.2]{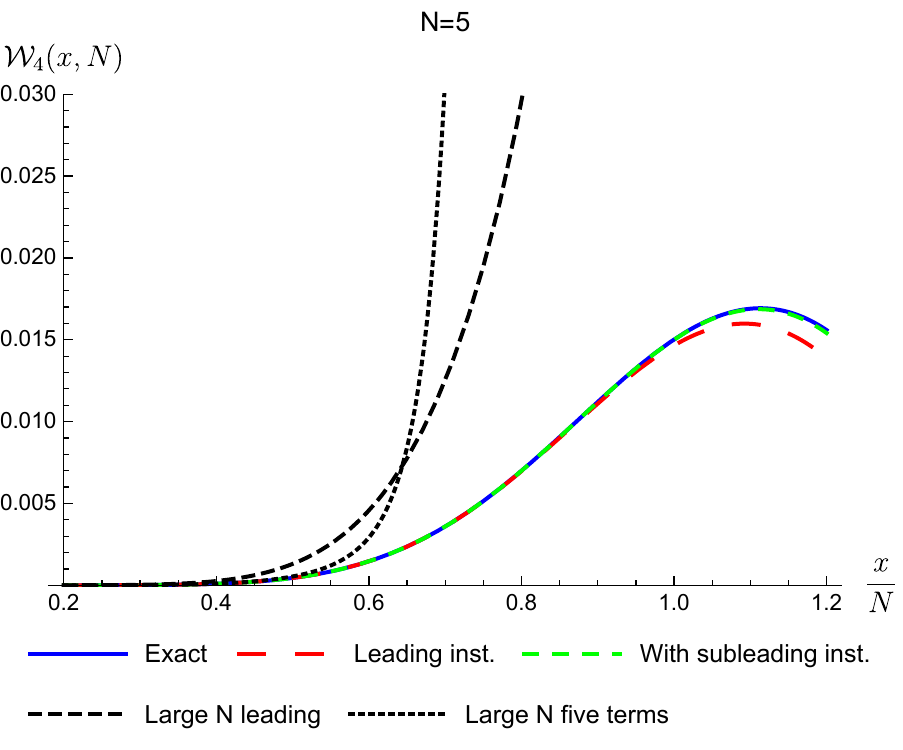}
\caption{The solid blue curve shows the exact numerically calculated Wilson loop $\mathcal W_4$ as a function of the inverse 't Hooft coupling $1/t \equiv x/N$, compared to  $\mathcal{W}_4$ calculated from the strong coupling expansion for $\Delta$ (\ref{eq:delta-strong}). The large dashed red curve (with $\Delta \approx \Delta_{(1)}$) and medium dashed green curve (with $\Delta \approx \Delta_{(1)} + \Delta_{(3)}$), show the leading and also subleading strong coupling instanton contributions. These agree well with the exact result all the way to the transition point at $x=N$. The black curves with medium and small dashes show $\mathcal{W}_4$ calculated from the one-instanton series in the large N strong coupling expansion (\ref{eq:delta-large-n-strong}), with just the leading term and the first five terms (see equation (4.8) in \cite{Alfinito:2017hsh}), respectively. All quantities are calculated at $N=5$.} 
\label{fig:w4a}
\end{figure}

\begin{figure}[htb]
\center\includegraphics[scale=1.2]{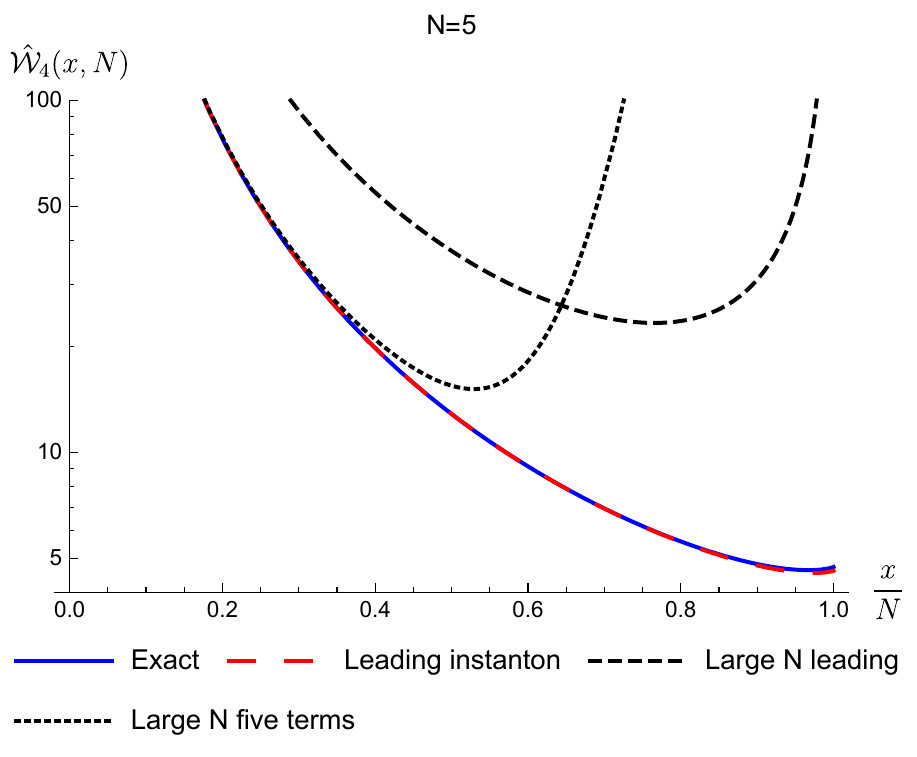}
\caption{The solid blue curve shows the exact numerically calculated Wilson loop $\hat{\mathcal{W}}_4 \equiv 4\pi N^2 e^{2N S_{\rm strong}}\, {\mathcal W}_4$ as a function of the inverse 't Hooft coupling $1/t \equiv x/N$. Compare with Figure 1 of \cite{Alfinito:2017hsh}. The large dashed red curve shows $\hat{\mathcal{W}}_4$ calculated from the leading approximation for $\Delta $ in (\ref{eq:delta-strong}) ($\Delta \approx \Delta_{(1)}$). This leading approximation is excellent all the way to the transition point at $x=N$. The black curves with medium and small dashes show $\hat{\mathcal{W}}_4$ calculated from the one-instanton series in the large N strong coupling expansion (\ref{eq:delta-large-n-strong}) with just the leading term and the first five terms, respectively (see for example, expression (4.8) in \cite{Alfinito:2017hsh}). The latter plot agrees well at very small $x$ (very strong coupling) but diverges well before the transition point. All quantities are calculated at $N=5$.} 
\label{fig:w4b}
\end{figure}

Similarly, all higher ${\mathcal W}_p(x, N)$ may be expressed in terms of $\Delta(x, N)$, and the trans-series expressions for $\Delta(x, N)$ in Section \ref{sec:delta} produce corresponding trans-series expressions for the winding Wilson loops. 
The Wilson loop ${\mathcal W}_4(x, N)$ has been studied in \cite{Alfinito:2017hsh}, where it is expressed in terms of ${\mathcal W}_1(x, N)$ as:
\begin{eqnarray}
{\mathcal W}_4(x, N)=1+\frac{4N^2}{x^2}+\frac{8}{x^2}-
\left(\frac{8N}{x}+\frac{8N^3}{x^3}+\frac{28N}{x^3}\right){\mathcal W}_1
+\frac{12N^2}{x^2} {\mathcal W}_1^2+\frac{12 N}{x^2}\frac{\partial}{\partial x} {\mathcal W}_1
\label{eq:w4}
\end{eqnarray}
Using the leading strong coupling instanton approximation (\ref{eq:w1-leading}) for ${\mathcal W}_1$ we obtain the leading instanton approximation for ${\mathcal W}_4$
\begin{eqnarray}
{\mathcal W}_4(x, N)\Big |^{\rm strong} \approx 
\frac{\left(4 N^2-2 x^2+8\right) J_N(x){}^2+2 \left(x^2-2 \left(N^2+5\right)\right) J_{N-1}(x) J_{N+1}(x)}{x^2}
\label{eq:w4-leading}
\end{eqnarray}
which can be written in terms of the 't Hooft coupling as:
\begin{eqnarray}
\hskip-.5cm {\mathcal W}_4(t, N)\Big |^{\rm strong} \approx 
-2\left(1-2t^2-\frac{4t^2}{N^2}\right) J_N^2\left(\frac{N}{t}\right)
+2 \left(1-2 t^2-\frac{10t^2}{N^2}\right) J_{N-1}\left(\frac{N}{t}\right) J_{N+1}\left(\frac{N}{t}\right)
\label{eq:w4-leadingt}
\end{eqnarray}
These expressions includes all fluctuations in this instanton sector, resummed to all orders. In Figure \ref{fig:w4a} we plot ${\mathcal W}_4(x, N)$ for $N=5$, comparing the exact result, from equations (\ref{eq:w4}) and (\ref{eq:w1}), with the leading instanton contribution  (\ref{eq:w4-leading}). We also plot the expression including the effect of the next instanton term $\Delta_{(3)}(x, N)$ from the strong-coupling trans-series (\ref{eq:delta-strong}) for $\Delta(x, N)$. 
The black dotted and dashed curves show the conventional large $N$ approximation, including just the leading fluctuation term, or also including five fluctuation terms: see expression (4.8) in \cite{Alfinito:2017hsh}.
The approximation (\ref{eq:w4-leading}), and its uniform large $N$ approximation, are much better, all the way to the transition point $x=N$. We can also compare directly to Figure 1 of \cite{Alfinito:2017hsh}: in Figure \ref{fig:w4b} we show a log plot of  the Wilson loop ${\mathcal W}_4(x, N)$ with its exponential prefactor removed: $\hat{{\mathcal W}}_4(x, N)\equiv 4\pi N^2 e^{2N S_{\rm strong}}\, {\mathcal W}_4(x, N)$. Notice that including the higher $\frac{1}{N}$ corrections improves the agreement at strong coupling (small $x/N$) but yields worse agreement near the transition point $x=N$. On the other hand, using just the leading instanton expression (\ref{eq:w1-leading}) for ${\mathcal W}_1(x, N)$ in the expression  (\ref{eq:w4}) for ${\mathcal W}_4(x, N)$ produces much better agreement all the way to $x=N$. The leading large $N$ uniform approximation is indistinguishable, already for $N=5$, with even better accuracy at larger $N$.

\section{Conclusions}

In this paper we have studied in detail how the trans-series structure in the Gross-Witten-Wadia (GWW) unitary matrix model changes as the coupling $g^2$ and gauge index $N$ are varied. The trans-series has a very different form in the different parameter regions, and in different limits. The appearance of trans-series expansions are often thought to be a `consequence' of divergent perturbative expansions. However, the fixed $N$ strong coupling expansion in the GWW model is convergent and still has a non-perturbative instanton expansion as a trans-series. The fact that resurgent relations, connecting the fluctuation expansions about different non-perturbative sectors in the trans-series, appear at finite $N$ is a simple consequence of the fact that the partition function $Z(x, N)$ (where $x=2/g^2$ is the inverse coupling variable) satisfies an $(N+1)^{\rm th}$ order linear differential equation with respect to $x$. It is, nevertheless, instructive to see how these resurgent relations work in practice. To probe the transition to the large $N$ limit we have used the Tracy-Widom mapping to connect the GWW model partition function to solutions of simple nonlinear ordinary differential equations with respect to $x$, with $N$ appearing as a parameter in the equation. These nonlinear equations are of Painlev\'e III type, and the well-known double-scaling limit of the GWW model is the coalescence cascading of Painlev\'e III to Painlev\'e II. This is a non-linear analogue of the reduction of the Bessel equation to the Airy equation. Having an explicit differential equation permits straightforward algorithmic generation of trans-series expansions.  The large $N$ trans-series have a different form, and different actions, in the weak coupling and strong coupling phases, but in each phase we again observe resurgent relations between different non-perturbative sectors. The Painlev\'e III differential equation leads to a uniform large $N$ approximation which connects accurately and smoothly to the exact solution even in the immediate vicinity of the transition point. We explored implications of this uniform large $N$ approximation for winding Wilson loops. The uniform large $N$ approximation identifies the GWW phase transition as the merging of saddle points, and it would be interesting to understand this in more physical detail in terms of semiclassical saddle configurations, especially in associated finite $N$ gauge theories.

\bigskip

{\noindent\bf Acknowledgments:}
We thank Carl Bender, Ovidiu Costin and Pavel Buividovich for helpful discussions and comments. This material is based upon work supported by the U.S. Department of Energy, Office of Science, Office of High Energy Physics under Award Number DE-SC0010339.

\section{Appendix: Large $N$ trans-series expansions for physical quantities}
\label{sec:large-n-physical}

In this Appendix we record trans-series expansions for the partition function, the free energy and the specific heat, which can all be derived directly from the trans-series expansions for the function $\Delta(t, N)$ studied in Section \ref{sec:delta}.

\subsection{Large $N$ trans-series expansions for $Z(t, N)$}
\label{sec:z-final}
To obtain a large $N$ expansion for the partition function $Z(t, N)$, we take the expansion for the function $\Delta(t, N)$ in   (\ref{eq:DeltaLargeNWeakCouplingExpansionAnsatz}) and plug it into (\ref{eq:sigma-delta1}). This gives an expansion for $\sigma(x^2, N)$, which can then be converted to an expansion for $Z(t, N)$ via equations (\ref{eq:ez}) and (\ref{eq:e-sigma}). 

At weak coupling we find the large $N$ trans-series expansion
\begin{equation}
\begin{aligned}
 Z(t,N) \sim \frac{G(N+1)}{(2\pi)^{N/2}} & \left(\frac{t}{N}\right)^{N^2/2} e^{N^2/t} \bigg[ \frac{1}{(1-t)^{1/8}} \left( 1 + \frac{3 t^3 }{128 (1-t)^3 N^2} 
 +\frac{45 t^5 (13 t+32)}{32768 (t-1)^6 N^4}
 + \dots \right)\\
& + \frac{i}{4 \sqrt{2 \pi N}} \frac{t}{(1-t)^\frac{7}{8}} e^{- N S_\text{weak}(t)} \left( 1 + \frac{(3 t^2-60 t-8)}{96 (1-t)^{3/2} N}  +\dots \right) +\dots  \bigg]
\end{aligned}
\label{eq:zweak-t}
\end{equation}
Expanding the first perturbative term at weak coupling (small $t$), we obtain
precisely the large $N$ expansion in (\ref{eq:z0a0}). To analyze the first non-perturbative term in (\ref{eq:zweak-t})
we use the small $t$ expansion of $S_{\rm weak}(t)$
\begin{eqnarray}
S_{\rm weak}(t)\sim \frac{2}{t}-1+\ln\left(
\frac{t}{4}\right)+\frac{t}{4}+\dots \qquad, \quad t\to 0
\label{eq:sweak-t}
\end{eqnarray}
to find (suppressing the obvious prefactor terms)
\begin{eqnarray}
&&\frac{i}{4 \sqrt{2 \pi N}} \frac{t}{(1-t)^\frac{7}{8}} e^{- N S_\text{weak}(t)} \left( 1 + \frac{(3 t^2-60 t-8)}{96 (1-t)^{3/2} N}  +\dots \right)  \nonumber\\
&& \sim 
\frac{i}{4\sqrt{2\pi N}}\left(\frac{t}{4}\right)^N e^N e^{-2N/t} e^{-N t/4}\, t\left(1+\frac{7t}{8}+\dots\right)\left(1-\frac{1}{N}\left(\frac{3t}{4}+
\frac{1}{12}+\dots\right)+\dots\right)
\label{eq:zweak-comp1}
\end{eqnarray}
Compare this with the first instanton term in (\ref{eq:zweak}, \ref{eq:zweakb}), which was found from the large $x$ expansion at fixed $N$, again suppressing the obvious prefactor terms:
\begin{eqnarray}
&&i\frac{(4N/t)^{N-1}}{(N-1)!} e^{-2N/t} \left(1-\frac{t}{8}\left(2N-7+\frac{6}{N}\right)+\dots\right) \nonumber\\
&& 
\sim \frac{i}{4\sqrt{2\pi N}}\left(\frac{t}{4}\right)^N e^N e^{-2N/t} t\left(1-\frac{1}{12N}+\dots\right)\left(1-\frac{t}{8}\left(2N-7+\frac{6}{N}\right)+\dots\right)
\label{eq:zweak-comp2}
\end{eqnarray}
Comparing the expressions (\ref{eq:zweak-comp1}) and (\ref{eq:zweak-comp2}), we see that the $-\frac{1}{12N}$ term in (\ref{eq:zweak-comp1}) arises in (\ref{eq:zweak-comp2}) from the large $N$ expansion of the factorial; the $-\frac{N t}{4}$ term in (\ref{eq:zweak-comp2}) is generated in (\ref{eq:zweak-comp1}) from the $e^{-N t/4}$ factor coming from expanding $S_{\rm weak}(t)$ at small $t$; the $\frac{7t}{8}$ term in (\ref{eq:zweak-comp2}) is generated by the prefactor in (\ref{eq:zweak-comp1}); and the $-\frac{3t}{4N}$ term in (\ref{eq:zweak-comp1}) is generated from the fluctuation factor in (\ref{eq:zweak-comp2}). 

The large $N$ strong-coupling expansion for $Z(t, N)$ can be derived from the corresponding large $N$ strong-coupling expansion for $\Delta(t, N)$ in (\ref{eq:delta-large-n-strong}). We find:
\begin{equation}
\begin{aligned}
    Z(t,N) e^{-\frac{N^2}{4t^2}} \sim 1 - & \frac{t}{8 \pi N \left(t^2-1\right)^{3/2}} e^{- 2 N S_\text{strong}(t)} \left(1 - \frac{t \left(26 t^2+9\right )}{12 N \left(t^2-1\right)^{3/2}} +\frac{t^2 \left(964 t^4+2484 t^2+297\right)}{288 N^2 \left(t^2-1\right)^3}+\dots \right) \\
    & -3 \left(t^2 + \frac{1}{2} \right) \left(\frac{t}{8 \pi N \left(t^2-1\right)^{3/2}} e^{- 2 N S_\text{strong}(t)} \right)^2 \left( 1 - \frac{(372 t^5 + 500 t^3 - 39 t)}{96 N \left(t^2-1\right)^{3/2} \left(2 t^2+1\right)} + \dots \right)
\end{aligned}
\end{equation}
The leading non-perturbative term behaves in the large $N$ and large $t$ limits as:
\begin{eqnarray}
&& -  \frac{t}{8 \pi N \left(t^2-1\right)^{3/2}} e^{- 2 N S_\text{strong}(t)} \left(1 - \frac{t \left(26 t^2+9\right )}{12 N \left(t^2-1\right)^{3/2}} +\dots \right) \nonumber\\
&&\sim \frac{1}{8\pi N}\frac{e^{2N}}{(2t)^{2N}}\, e^{-\frac{N}{2t^2}}\, \frac{1}{t^2}\left(1+\frac{3}{2t^2}+\dots\right)\left(1-\frac{1}{N}\left(\frac{13}{6}+\frac{4}{t^2}+\dots\right)+\dots\right)
\label{eq:zstrong-comp1}
\end{eqnarray}
where we have used the fact that the strong-coupling action (\ref{eq:strong-t}) behaves at large 't Hooft coupling as:
\begin{eqnarray}
S_{\rm strong}(t)\sim -1 + \ln(2t)+\frac{1}{4t^2}+\frac{1}{32t^4}+\dots  \quad, \quad t\to\infty 
\label{eq:strong-t2}
\end{eqnarray}
The expression (\ref{eq:zstrong-comp1}) should be compared with the strong coupling result from (\ref{eq:zstrongt}), in the large $N$ limit: 
\begin{eqnarray}
Z(t, N)e^{-\frac{N^2}{4t^2}} \sim 1-\frac{1}{2\pi N} \frac{e^{2N}}{(2t)^{2N+2}}\left(1-\frac{13}{6N}+\dots\right)\left(1-\frac{1}{2t^2} \left(N-3+\frac{8}{N}\right)+\dots\right)+\dots
\nonumber\\
\label{eq:zstrong-comp2}
\end{eqnarray}
Once again we see that these leading terms match, coming from very different origins: either from the expansion of the action, from prefactors, from combinatorial large $N$ factors, or from fluctuation terms.

\subsection{Large $N$ trans-series expansions for the free energy}
The free energy per degree of freedom  is given by
\begin{equation}\label{eq:FreeEnergyDefinition}
	F(t, N) = \frac{1}{N^2} \ln Z(t, N)
\end{equation}
At weak coupling  the large $N$ trans-series expansion has the form
\begin{equation}
\begin{aligned}
    F(t,N) \sim \frac{1}{t} + &\frac{1}{2} \log t - \frac{1}{2 N} \log (2 \pi) - \frac{1}{2} \log N  - \frac{1}{8N^2} \log(1-t) + \frac{1}{N^2} \log G(N+1)\\
    & +	\sum_{k=0}^\infty P^{(k)}_\text{weak}(t,N) e^{-N k S_\text{weak}(t) } \sum_{n=0}^\infty \frac{f^{(k)}_{\text{weak},n}(t)}{N^n}
\end{aligned}
\end{equation}
with the perturbative and one-instanton terms being
\begin{equation}
\begin{aligned}
    P^{(0)}_\text{weak}(t,N) \sum_{n=0}^\infty\frac{f^{(0)}_{\text{weak},n}(t)}{N^n} &= \frac{3 t^3}{128 N^4 (1-t)^3}\left(1+\frac{3 t^2 (5+2 t)}{8 N^2 (1-t)^3} +\dots \right) \\
    P^{(1)}_\text{weak}(t,N) \sum_{n=0}^\infty\frac{f^{(1)}_{\text{weak},n}(t)}{N^n} &=\frac{i}{\sqrt{\pi} (2N)^\frac{5}{2}} \frac{t}{(1-t)^\frac{5}{8}} \left(1 + \frac{3 t^2-60 t-8}{96 N (1-t)^{3/2}} + \dots\right)
\end{aligned}
\end{equation}
Structurally, these have the same form as the weak coupling large $N$ trans-series expansions for $\Delta(t, N)$ in (\ref{eq:DeltaLargeNWeakCouplingExpansionAnsatz}). 

At strong coupling  the large $N$ trans-series expansion has the form
\begin{equation}
    F(t,N) \sim \frac{1}{4t^2}  -	\sum_{k=0}^\infty P^{(k)}_\text{strong}(t,N) e^{-2 N (k+1) S_\text{strong}(t) } \sum_{n=0}^\infty \frac{f^{(k)}_{\text{strong},n}(t)}{N^n}
    \label{eq:free-largen-strong}
\end{equation}
with leading terms
\begin{equation}
\begin{aligned}
    P^{(0)}_\text{strong}(t,N) \sum_{n=0}^\infty\frac{f^{(0)}_{\text{str},n}(t)}{N^n} &=  \frac{t}{8 \pi N^3 \left(t^2-1\right)^{3/2}} \left(1 - \frac{t \left(26 t^2+9\right )}{12 N \left(t^2-1\right)^{3/2}} +\dots \right)\\
    P^{(1)}_\text{strong}(t,N) \sum_{n=0}^\infty\frac{f^{(1)}_{\text{str},n}(t)}{N^n} &= \frac{t^2 \left(3 t^2+2\right)}{64 \pi ^2 N^4 \left(t^2-1\right)^3} \left(1 -\frac{t \left(1116 t^4+1916 t^2+27\right)}{192 N \left(t^2-1\right)^{3/2} \left(3 t^2+2\right)} +\dots \right)
\end{aligned}
\end{equation}
Structurally, these have the same form as the strong coupling large $N$ trans-series expansions for $\Delta(t, N)$ in (\ref{eq:delta-large-n-strong}). Recall that the factor of $2$ in the exponent in  (\ref{eq:free-largen-strong}) explains why the strong coupling action in (\ref{eq:strong-t}) differs by a factor of $2$ from the strong-coupling action for the partition function and free energy \cite{marino-matrix}.

\subsection{Large $N$ trans-series expansions for the specific heat}
The specific heat is given by \cite{gw,Rossi:1996hs}
\begin{equation}
    C = \frac{x^2}{2} \frac{\partial^2 F}{\partial x^2} 
    \label{eq:specific-heat1}
\end{equation}
This can be expressed in terms of the function $\sigma_N$ defined in Section \ref{sec:pIII}:
\begin{equation}
    C  =\frac{x^2}{4N^2}+\frac{1}{N^2}\, \sigma_N(x)-\frac{x}{N^2}\frac{\partial}{\partial x} \sigma_N
    \label{eq:specific-heat2}
\end{equation}
And since $\sigma_N$ is expressed in terms of $\Delta(x, N)$ via the explicit mapping (\ref{eq:sigma-delta1}), we can use the trans-series expansions for $\Delta$ to write trans-series expressions for the specific heat.

For example, in the weak coupling regime the large N expansion is 
\begin{equation}
    C(t,N) \sim \sum_{k=0}^\infty C^{(k)}_\text{weak}(t,N) e^{-N k S_\text{weak}(t) } \sum_{n=0}^\infty \frac{c^{(k)}_{\text{weak},n}(t)}{N^n}
\end{equation}
with the perturbative and one-instanton terms being
\begin{equation}
    \begin{aligned}
       C^{(0)}_\text{weak}(t,N) \sum_{n=0}^\infty \frac{c^{(0)}_{\text{weak},n}(t)}{N^n} &=   \frac{1}{4} + \frac{(2-t) t}{16 N^2 (1-t)^2} + \frac{9 t^3}{64 N^4 (1-t)^5}  + \frac{27 t^5 (24 t+25)}{1024 N^6 (1-t)^8} + \dots \\
       C^{(1)}_\text{weak}(t,N) \sum_{n=0}^\infty \frac{c^{(1)}_{\text{weak},n}(t)}{N^n} &= \frac{i (1-t)^\frac{3}{8}}{2 \sqrt{2 \pi N} t} \left(1 - \frac{57 t^2-36 t+8}{96 N (1-t)^{3/2}} + \dots  \right)
    \end{aligned}
\end{equation}
In the strong coupling regime the large N expansion is
\begin{equation}
\begin{aligned}
     C(t,N) \sim  \frac{1}{4t^2} + \sum_{k=0}^\infty C^{(k)}_\text{strong}(t,N) e^{-2 N (k+1) S_\text{strong}(t) } \sum_{n=0}^\infty \frac{c^{(k)}_{\text{strong},n}(t)}{N^n}
\end{aligned}
\end{equation}
with
\begin{equation}
    \begin{aligned}
       C^{(0)}_\text{strong}(t,N) \sum_{n=0}^\infty \frac{c^{(0)}_{\text{strong},n}(t)}{N^n} &= -\frac{1}{4 \pi  N t \sqrt{t^2-1}}  \left(1-\frac{t \left(8 t^2-3\right)}{12 N \left(t^2-1\right)^{3/2}} + \dots \right)\\
       C^{(1)}_\text{strong}(t,N) \sum_{n=0}^\infty \frac{c^{(1)}_{\text{strong},n}(t)}{N^n} &= -\frac{3 t^2+2}{8 \pi ^2 N^2 \left(t^2-1\right)^2} \left(1 + \frac{t \left(684 t^4+92 t^2-357\right)}{192 N \left(t^2-1\right)^{3/2} \left(3 t^2+2\right)} + \dots\right)
    \end{aligned}
\end{equation}
The leading strong coupling expression, following from the leading strong coupling approximation for $\Delta(x, N)\approx J_N(x)$, reads:
\begin{eqnarray}
C(x, N)\Big|_{\rm strong}^{\rm leading}\approx
\frac{x^2}{4 N^2}+
\frac{\left(x^2-4 (N-1) N\right) J_{N-1}(x){}^2-x^2 J_N(x){}^2+2 N x J_{N-2}(x) J_{N-1}(x)}{4 N^2}
\label{eq:c-strong}
\end{eqnarray}
This expression resums, in closed form, all fluctuations about the two-instanton sector.
Further, it can be converted to a uniform large $N$ expression using  uniform approximations for the Bessel functions.

\end{document}